\documentclass [11pt,a4paper]{article}
\usepackage{jheppub}

\usepackage{hyperref}%[breaklinks]
\usepackage[english]{babel}
\usepackage[utf8]{inputenc}
\usepackage[OT2,T1]{fontenc}
\usepackage{graphicx}
\usepackage{color}
\usepackage{amsmath,amssymb,amsfonts,amsthm}
\usepackage{esint}
\usepackage{bm,bbm}
\usepackage{pifont} %for ding symbols
\usepackage{changepage}

\newcommand{\be}{\begin{equation}}
\newcommand{\ee}{\end{equation}}
\newcommand{\ba}{\begin{eqnarray}}
\newcommand{\ea}{\end{eqnarray}}
\def\bs{\begin{subequations}}
\def\es{\end{subequations}}
\renewcommand{\leq}{\leqslant}
\renewcommand{\geq}{\geqslant}
\def\a{\alpha}
\def\b{\beta}
\def\de{\delta}
\def\g{\gamma}
\def\la{\lambda}
\def\k{\kappa}
\def\e{\epsilon}

\def\Om{\Omega}
\def\om{\omega}
\def\De{\Delta}
\def\G{\Gamma}
  
\def\s{\sigma}
\def\vr{\varrho}
\def\vp{\varphi}
\def\N{\nabla}
\def\cA{\mathcal{A}}

\def\cC{\mathcal{C}}
\def\cD{\mathcal{D}}

\def\cF{\mathcal{F}}

\def\cH{\mathcal{H}}

\def\cK{\mathcal{K}}
\def\cL{\mathcal{L}}
\def\cM{\mathcal{M}}

\def\cP{\mathcal{P}}

\def\cR{\mathcal{R}}
\def\cS{\mathcal{S}}
\def\cT{\mathcal{T}}
\def\cV{\mathcal{V}}
\def\cX{\mathcal{X}}
\def\bE{\mathbbm{e}}

\def\ds{d_\textsc{s}}
\def\dh{d_\textsc{h}}
\def\dw{d_\textsc{w}}
\def\dc{d_\textsc{c}}
\def\dcs{d_{\textsc{cs},m}}
\def\p{\partial}

\def\B{\Box}
\newcommand{\Eq}[1]{(\ref{#1})}
\def\com{\color{magenta}}
\def\cob{\color{blue}}
\newcommand{\book}[5]{\emph{#1}, #2, #3, #4 (#5)}
\newcommand{\books}[4]{\emph{#1}, #2, #3 (#4)}
\newcommand{\oarX}[1]{\href{http://arxiv.org/abs/#1}{{\ttfamily\com arXiv:#1}}}
\newcommand{\arX}[1]{\href{http://arxiv.org/abs/#1}{{\ttfamily\com arXiv:#1}}}
\newcommand{\doin}[6]{\href{http://dx.doi.org/#1}{{\cob {\it #2} {\bf #3 #4} (#6) #5}}}
\newcommand{\doinn}[5]{\href{http://dx.doi.org/#1}{{\cob {\it #2} {\bf #3} (#5) #4}}}
\newcommand{\doij}[5]{\href{http://dx.doi.org/#1}{{\cob {\it #2} {\bf #3} (#5) #4}}}
\newcommand{\ndoin}[6]{\href{#1}{{\cob {\it #2} {\bf #3 #4} (#6) #5}}}
\newcommand{\ndoinn}[5]{\href{#1}{{\cob {\it #2} {\bf #3} (#5) #4}}}
\newcommand{\procsinm}[5]{in \emph{#1}, #2 eds., #3, #4 (#5)}
\newcommand{\procm}[6]{in \emph{#1}, #2  eds., #3, #4, #5 (#6)}
\newcommand{\tia}[1]{\textit{#1},}

\def\Pl{{\rm Pl}}
\def\lp{\ell_{\rm Pl}}

\def\mpl{m_{\rm Pl}}
\def\ep{E_{\rm Pl}}
\def\rme{e}
\def\rmd{d}
\def\rmi{i}
\def\bd{\mathbbm{d}}
\newcounter{listcounter}
\newcommand*{\lolabel}[1]{\noindent$\blacktriangleright${\it \textbf{0\refstepcounter{listcounter}\thelistcounter\label{#1}}}$\blacktriangleleft$}
\newcommand*{\loref}[1]{\textit{\cob 0\ref{#1}}}
\newcommand*{\llabel}[1]{\noindent$\blacktriangleright$\textbf{\refstepcounter{listcounter}\thelistcounter\label{#1}}$\blacktriangleleft$}
\newcommand*{\lref}[1]{\textit{\ref{#1}}}

\begin{document}

\title{Multifractional theories: an unconventional review}

\author{Gianluca Calcagni,}
\emailAdd{calcagni@iem.cfmac.csic.es}
\affiliation{Instituto de Estructura de la Materia, CSIC, Serrano 121, 28006 Madrid, Spain}

\date{December 18, 2016}

\abstract{We answer to 72 frequently asked questions about theories of multifractional spacetimes. Apart from reviewing and reorganizing what we already know about such theories, we discuss the physical meaning and consequences of the very recent flow-equation theorem on dimensional flow in quantum gravity, in particular its enormous impact on the multifractional paradigm. We will also get new theoretical results about the construction of multifractional derivatives and the symmetries in the yet-unexplored theory $T_\gamma$, the resolution of ambiguities in the calculation of the spectral dimension, the relation between the theory $T_q$ with $q$-derivatives and the theory $T_\gamma$ with fractional derivatives, the interpretation of complex dimensions in quantum gravity, the frame choice at the quantum level, the physical interpretation of the propagator in $T_\gamma$ as an infinite superposition of quasiparticle modes, the relation between multifractional theories and quantum gravity, and the issue of renormalization, arguing that power-counting arguments do not capture the exotic properties of extreme UV regimes of multifractional geometry, where $T_\gamma$ may indeed be renormalizable. A careful discussion of experimental bounds and new constraints are also presented.}

% 04.30.-w Gravitational waves
% 04.50.-h Higher-dimensional gravity and other theories of gravity
% 04.60.Bc Phenomenology of quantum gravity
% 05.45.Df Fractals

%\pacs{04.30.-w, 04.50.-h, 04.60.Bc, 05.45.Df}

\preprint{\doij{10.1007/JHEP03(2017)138}{JHEP}{1703}{138}{2017} [\arX{1612.05632}]}
%\preprint{\doin{10.1103/PhysRevD.88.084017}{Phys.\ Rev.}{D}{88}{084017}{2013} \hspace{10.5cm} \arX{1307.6122}}

\maketitle

\tableofcontents

%%%%%%%%%%%%%%%%%%%%%%%%%%%%%%%%%%%%%%%%%%%%%%%%%%%%%%%%%%%%%%%%%%%%%%%%%%%%%
%%%%%%%%%%%%%%%%%%%%%%%%%%%%%%%%%%%%%%%%%%%%%%%%%%%%%%%%%%%%%%%%%%%%%%%%%%%%%

\section{Introduction}\label{intro}

The unprecedented convergence of experiments in particle physics (LHC), astrophysics (LIGO) and cosmology (\textsc{Planck}) has led to discoveries that confirmed the standard knowledge of quantum interactions and classical gravity, either through the observation of phenomena predicted by the theories (the Higgs boson of the Standard Model \cite{Aad12,Cha12,CFKVZ} and general-relativistic gravitational waves from black-hole binary systems \cite{Abb16,LIG41,Ab16b}) or the gradual refinement of models of the early universe \cite{P1513,P1520}. New physics involving supersymmetry, effects of quantum gravity, or an explanation of the cosmological constant are the next \emph{desirata}, which many scenarios beyond standard predict to be in the range of our current or next-generation instruments. Some of these scenarios, such as string theory \cite{Pol98,BBSb,Zwi09}, loop quantum gravity (LQG) \cite{rov07,thi01}, spin foams \cite{Per13}, noncommutative spacetimes \cite{Sza01,Con04,ADKLW,BIMM}, and effective quantum gravity \cite{Don12,DoHo}, are very well known by theoreticians and phenomenologists of various extractions. Others, which include asymptotic safety \cite{NiR,CPR,RSnax}, causal dynamical triangulations (CDT) \cite{AGJL4}, causal sets \cite{Sur11,Dow13}, and group field theory (GFT) \cite{Ori09,Fousp,GiSi}, are perhaps that famous in the more restricted community of quantum gravity, while nonlocal quantum gravity \cite{Tom97,BKM3,Mod1,BGKM,Mod3,CaMo2,MoRa,TaBM} and multifractional spacetimes \cite{fra4,frc1,frc2,ACOS,frc3,fra6,frc4,frc5,fra7,frc6,frc7,frc8,frc9,frc10,frc11,frc12,frc13,trtls,qGW,frc14,CaRo1,first,CaRo2a,CaRo2b,frc15} have just begun to make their appearance on the scene (despite some older precedents), both as theoretical foundations of new paradigms of exotic geometry and as producers of novel phenomenology. 

It is part of the game that new proposals may meet some resistance at first and, in fact, multifractional theories have been considered in two rather radical ways: either welcomed as a fresh insight into several aspects of quantum gravity or rejected \emph{tout court} with a wide range of qualifications, from trivial to uninteresting to outright inconsistent. The first purpose of this paper is to collect the most frequent questions and criticism the author came across in the last few years and to give them a hopefully clear answer. Rather than concluding the debate, this contribution will probably fuel it further, either because some of the answers might not satisfy everybody or because new questions or objections can arise. The reader is free to make their own judgment on the matter or even to contribute to the debate actively in the appropriate channels. The recent formulation of two theorems \cite{first} showing how a universal multiscale measure of geometry naturally emerges whenever the dimension of spacetime changes with the scale (as in all quantum gravities) provides the perhaps most powerful justification to the choice of measure in multifractional theories, and an answer to many of the questions we will see below. 

The remarks are presented in an order that permits to introduce the basic ingredients of multifractional theories in a self-contained way. Therefore, the present work is an updated review on the subject, which was long due. The most recent one \cite{AIP} dates back to 2012 and it does not cover any of the major advancements regarding the motivations of the theory, several conceptual points about the measure, the field-theory and cosmological dynamics, and observational constraints. We divide the topics in a preliminary but necessary setting of the terminology (section \ref{termi}, 3 items), general motivations (section \ref{moti}, 3 items), basic aspects of the geometry and symmetry of multifractional spacetimes (section \ref{geo}, 15 items), frames and their physical interpretation (section \ref{frafi}, 9 items), field theory (section \ref{qft}, 9 items), classical gravity and cosmology (section \ref{cosmo}, 6 items), quantum gravity (section \ref{qg}, 5 items), observational and experimental constraints (section \ref{phen}, 18 items), and a final perspective (section \ref{concl}, 4 items). See table \ref{tab1}.
\begin{table}[h]
\begin{center}
\begin{tabular}{|llcc|}\hline
Sec. & Topic & Items & No.\ of items \\\hline
\ref{termi} & Terminology & \loref{01}--\loref{03} & 3 \\
\ref{moti} & Motivations & \loref{04}--\loref{06} & 3 \\
\ref{geo} & Geometry and symmetry & \loref{07}--\lref{19} & 15 \\
\ref{frafi} & Frames and physics & \lref{20}--\lref{28} & 9 \\
\ref{qft} & Field theory & \lref{29}--\lref{36} & 9 \\
\ref{cosmo} & Classical gravity and cosmology & \lref{37}--\lref{42} & 6 \\
\ref{qg} & Quantum gravity & \lref{43}--\lref{47} & 5\\
\ref{phen} & Observations and experiments & \lref{48}--\lref{65} & 18 \\
\ref{concl} & Perspective & \lref{66}--\lref{69} & 4\\\hline
\end{tabular}
\caption{\label{tab1} Summary of the questions per topic.}
\end{center}
\end{table}

The questions are the subsection titles in the table of content (actual text of the questions adapted). For each answer, bibliography is given where one can find more technical details. The question-answer format should both facilitate the search for specific topics and make an easier reading than the traditional review article. We also note that this is a ``review plus plus'' because it contains a number of novel results that augment the theory by new elements: 
\begin{enumerate}
\item a more thorough discussion about the physical meaning and consequences of the very recent flow-equation theorems, succinctly presented in ref.\ \cite{first}, which have repercussions both in general quantum gravity and on the theories of multifractional spacetimes (questions \loref{04}, \loref{07}, \lref{09}, \lref{11}, \lref{14}, \lref{27}, \lref{42}, \lref{45}, and \lref{47});
\item advances in the theory $T_\g$ with fractional derivatives, incompletely formulated in refs.~\cite{frc2,frc4}, regarding its symmetries (question \lref{11}), a proposal for a multiscale fractional derivative (question \lref{11}), a multiscale line element generalizing the no-scale one of ref.\ \cite{frc1} (question \lref{11}), the recasting of the propagator as a superposition of quasiparticle modes with a characteristic mass distribution (question \lref{35}), and its renormalizability (question \lref{47});
\item a clarification of the unit conversion of the scales of these geometries, previously assumed without an explanation (question \loref{07b});
\item the formulation of an important approximation of $T_\g$, that we will denote by $T_{\g=\a}\cong T_q$, with the theory with $q$-derivatives, carried through a comparison of their critical behavior (question \loref{07b}), a comparison and mutual approximation of their differential calculus (question \lref{11}), a comparison of their propagators (question \lref{34}) and of their renormalization properties (question \lref{47}); 
\item the dissipation of some ambiguities \cite{frc7} in the calculation of the spectral dimension (question \lref{13});
\item a discussion on complex dimensions in quantum gravity and fractal geometry (question \lref{14});
\item some remarks clarifying that the frame choice in multifractional theories and in scalar-tensor theories is made, respectively, at the classical and at the quantum level (question \lref{26});
\item the recognition, of utmost importance for this class of theories, that the second flow-equation theorem fixes the presentation of the geometry measure in an elegant way, which eventually leads to an unexpected solution of the presentation problem (question \lref{27});
\item a detailed summary of results on the renormalization in multifractional theories and discussions on the new perspectives opened by the stochastic view and on the inadequacy of the usual power-counting argument (question \lref{47});
\item new experimental bounds on the theory with $q$-derivatives, approximated in the stochastic view, coming from general dispersion relations (questions \lref{49} and \lref{55}) and from vacuum Cherenkov radiation (questions \lref{49} and \lref{56}).
\end{enumerate}

%%%%%%%%%%%%%%%%%%%%%%%%%%%%%%%%%%%%%%%%%%%%%%%%%%%%%%%%%%%%%%%%%%%%%%%%%%%%%
%%%%%%%%%%%%%%%%%%%%%%%%%%%%%%%%%%%%%%%%%%%%%%%%%%%%%%%%%%%%%%%%%%%%%%%%%%%%%

\section{Terminology}\label{termi}

\lolabel{01} \textbf{\emph{What is the dimension of spacetime?}}\addcontentsline{toc}{subsection}{\texorpdfstring{\loref{01}}{} What is the dimension of spacetime?}

There are several definitions of dimension. The most used in theoretical physics is that of topological dimension $D$, which is simply the total number of spatial and time directions. In a spacetime with Lorentzian signature and one time direction, $D=4$ means that there are three space directions. Other important geometric indicators are the Hausdorff dimension $\dh$, the spectral dimension $\ds$, and the walk dimension $\dw$. In all these cases and by a convention accepted by physicists and mathematicians, the dimension of spacetime is defined after Euclideanizing the time direction.\footnote{The reader uneasy with this convention can limit the discussion in the text to spatial slices and time separately. Little changes about the main results.}

The \emph{Hausdorff dimension} is defined as the scaling of the Euclideanized volume $\cV(\ell)$ of a $D$-ball of radius $\ell$ or of a $D$-hypercube of edge size $\ell$. There is no difference in scaling between the ball and the hypercube. On a classical continuum spacetime, this reads
\be\label{dh}
\dh(\ell):=\frac{\rmd\ln\cV(\ell)}{\rmd\ln\ell}\,.
\ee
Since the volume is the integral $\cV=\int\rmd\vr(x)$ of the spacetime measure $\vr(x)=\vr(x^0,x^1,$ $\dots,x^{D-1})$ in a given region, an approximately constant $\dh$ is nothing but the scaling of the measure under dilations of the coordinates, $\vr(\la x)=\la^{\dh}\vr(x)$. On a quantum geometry, the volume $\cV$ may be replaced by the expectation value $\langle\hat\cV\rangle$ of the volume operator $\hat\cV$ on a superposition of quantum states of geometry \cite{COT3,Thu15}. By using an embedding space, $D$-balls can be defined also on a discrete geometry or on a pre-geometric combinatorial structure (for instance, LQG and GFT), as well as on totally disconnected or highly irregular sets such as fractals \cite{Fal03}. In the latter case, the definition of $\dh$ is more complicated than eq.\ \Eq{dh} but it conveys essentially the same information, in particular about the scaling of the measure defining the set \cite{frc2}. Moreover, a continuous parameter $\ell$ exists in all discrete settings or quantum gravities with a notion of distance, even in the absence of a fundamental notion of continuous spacetime \cite{COT2,COT3}. In such settings, $\ell$ is measured in units of a lattice spacing or of the labels of combinatorial complexes.

The \emph{spectral dimension} $\ds$ is the scaling of the return probability in a diffusion process (see \cite{CMNa} for a review). Let $\bar\cK(\p)$ be the Laplacian on a smooth manifold. Placing a pointwise test particle at point $x'$ on the manifold and letting it diffuse, its motion will obey the nonrelativistic diffusion equation $(\p_\s-\kappa_1\bar\cK)P(x,x';\s)=0$ with initial condition $P(x,x';0)=\de(x-x')/\sqrt{g}$, where $\kappa_1$ is a diffusion coefficient, $\s$ is an abstract diffusion time parametrizing the process, and $g$ is the determinant of the metric. Integrating the heat kernel $P$ for coincident points over all points of the geometry, one obtains a function $\cP(\s):=Z/\cV=\int\rmd^D x\sqrt{g}P(x,x;\s)/\cV$ called return probability (the volume factor makes the normalization finite). In an alternative interpretation \cite{CMNa}, the diffusion process is replaced by a probing of the geometry with a resolution $\sim 1/\ell$, where $\ell=\sqrt{\kappa_1\s}$ is the characteristic length scale detectable by the apparatus. Adding also a quantum-field-theory twist to the story, the diffusion equation is reinterpreted as the running equation of the transition amplitude $P$ defined by the Green function $G(x,x')=-\int_0^{+\infty}\rmd(L^2)\,P(x,x';L)$, corresponding in momentum space to the Schwinger representation 
\be\label{Sch}
\tilde G(k)=-\frac{1}{\tilde\cK(k)}=-\int_0^{+\infty}\rmd(L^2)\,\exp[-L^2\tilde\cK(k)]\,.
\ee
Here $L$ is a parameter related to the probed scale $\ell$ and $\tilde\cK$ is the Fourier transform of the kinetic operator $\cK(\p)$ in the field action (not necessarily equal to $\bar\cK$, in general; see question \lref{13}). The propagator $G$ governs the quantum propagation of a particle from $x'$ to $x$ and $\cP[L(\ell)]$ is the probability of finding the particle in a neighborhood of $x$ of size $\ell$.

Whatever the interpretation of $\cP$, the spectral dimension is the scaling of the return probability:
\be\label{ds}
\ds(\ell):=-\frac{\rmd\ln\cP(\ell)}{\rmd\ln\ell}\,.
\ee
Using $\s$ instead, one gets the more common form $\ds=-2\rmd\ln\cP(\s)/\rmd\ln\s$. For a set with approximately constant spectral dimension, $\cP(\ell)\sim \ell^{-\ds}$. As in the case of the Hausdorff dimension, a continuous parameter $\ell$ can always be defined. In quantum geometries, the return probability in eq.\ \Eq{ds} may be replaced by the expectation value $\langle\hat\cP\rangle$ of a certain operator $\hat\cP$ on a superposition of quantum states of geometry \cite{COT3}. 

The \emph{walk dimension} is the scaling of the mean-square displacement of a random walker $X(\s)$ (a stochastic motion $X$ over the manifold):
\be\label{dw}
\dw:=2\left(\frac{\rmd\ln\langle X^2(\s)\rangle}{\rmd\ln\s}\right)^{-1},
\ee
where $\langle X^2(\s)\rangle=\int\rmd^Dx\,\sqrt{g}\,x^2\,P(x,0;\s)$. For a set with approximately constant walk dimension, $\langle X^2(\s)\rangle\sim \s^{2/\dw}$. More information on $\dw$ can be found in refs.\ \cite{frc7,trtls}. 

In a continuous space, there is a relation between the three dimensions we just introduced. Simply by scaling arguments, one notes that\footnote{In this chain of relations, a small typo in ref.\ \cite{trtls} is corrected.} $\s^{-\ds/2}\sim \cP=Z/\cV\sim \cV^{-1} \sim \ell^{-\dh} \sim X^{-\dh} \sim \s^{-\dh/\dw}$, hence $\ds=2\dh/\dw$. We will comment on this equation in the next question. For Euclidean space or imaginary-time Minkowski spacetime ($\cK=\N^2$), it is immediate to check that $\dh=D=\ds$ and $\dw=2$. Other definitions of dimension, much less frequently used in theoretical physics, can be found in refs.\ \cite{frc2,Fal03}. In footnote \ref{fu1} and questions \loref{04} and \loref{07b}, we will invoke one such definition, called \emph{capacity} of a set.

For continuous manifolds and in the presence of very simple but nontrivial dispersion relations $\cK(\p)\to\tilde\cK(k)\neq-k^2$, it is easy to show that the spectral dimension $\ds$ is nothing but the Hausdorff dimension $\dh^{(k)}$ of momentum space \cite{AAGM3,Ron16}. For fractals, this identification is conjectured but not yet proved \cite{Akk2,Akk12}. In general, it is not true that $\ds=\dh^{(k)}$ for the most general multiscale geometry, as already recognized in ref.\ \cite{AAGM3}. Consider the case where $\tilde\cK(k)$ [a function almost always such that $\tilde\cK(0)=0$ and $\tilde\cK(\infty)=\infty$] depends on $k=\sqrt{k_\mu k^\mu}$ and the measure in $k$-momentum space is the usual one, $\rmd^Dk=\rmd k\,k^{D-1}\rmd\Om_D$, where $\rmd\Om_D$ is the angular measure. All the other cases, including multifractional spacetimes, can be derived from this straightforwardly. Calling $K^2:=\tilde\cK(k)$, we have $2K\rmd K=\tilde\cK'(k)\rmd k$, where a prime denotes a derivative with respect to $k$.
Therefore, up to an angular prefactor the measure in $K$-momentum space is $\rmd k\,k^{D-1}=\rmd K\,w(K)$, where
\be
w(K)=\frac{2K[\tilde\cK^{-1}(K^2)]^{D-1}}{\tilde\cK'(k)|_{k=\tilde\cK^{-1}(K^2)}}\,,
\ee
where we assumed that we can invert $K(k)$ as $k=\tilde\cK^{-1}(K^2)$. Since a momentum volume of linear size $K$ is $\cV^{(K)}=\int\rmd K\, w(K)$, the Hausdorff dimension of the $K$-momentum space is
\be\label{dhk}
\dh^{(k)}=\frac{\rmd\ln\cV^{(K)}}{\rmd\ln K}=\frac{K w(K)}{\int\rmd K w(K)}\,.
\ee
On the other hand, the spectral dimension is
\be
\ds=\frac{\ell^2\int_0^{+\infty}\rmd k\,k^{D-1}\,\tilde\cK(k)\,\rme^{-\ell^2\tilde\cK(k)}}{\int_0^{+\infty}\rmd k\,k^{D-1}\,\rme^{-\ell^2\tilde\cK(k)}}=\frac{\ell^2\int_0^{+\infty}\rmd K\,w(K)K^2\,\rme^{-\ell^2K^2}}{\int\rmd K w(K)\,\rme^{-\ell^2K^2}}\,.
\ee
For simple dispersion relations, we know that $\ds=\dh^{(k)}$. For instance, taking the power law $\tilde\cK(k)=k^{2\g}$, we have $w(K)=K^{D/\g-1}/\g$, $\cV^{(K)}=K^{D/\g}/D$, and $\ds=D/\g=\dh^{(k)}$. Already for a binomial dispersion relation $\tilde\cK(k)=k^{2\g_1}+ a k^{2\g_2}$, one cannot get an exact result. Asymptotically, $\ds\simeq D/\g_{1,2}$ \cite{SVW2}, and clearly one also has $\dh^{(k)}\simeq D/\g_{1,2}$; transient regimes of $\ds$ and $\dh^{(k)}$ differ. Therefore, one should take eq.\ \Eq{dhk} as yet another definition of spacetime dimension.

\bigskip

\lolabel{02} \textbf{\emph{Are ``multiscale,'' ``multifractional,'' and ``multifractal'' synonyms?}}\addcontentsline{toc}{subsection}{\texorpdfstring{\loref{02}}{} Are ``multiscale,'' ``multifractional,'' and ``multifractal'' synonyms?}

No. Although there has been, in quantum gravity, a lot of confusion about ``fractal'' and ``multiscale'' geometries before the appearance of this proposal, and between ``multiscale'' and ``multifractional'' after that, now the terminology has been clarified \cite{trtls}. A geometry is \emph{multiscale} if the dimension of spacetime ($\dh$, $\ds$, and/or $\dw$) changes with the probed scale. By this, we mean that experiments performed at different energy or length scales are affected by different spacetime dimensionalities. In a multiscale geometry, at different length scales
\be\label{hier}
\ell_1>\ell_2>\ell_3>\dots\,,
\ee
one experiences different properties of the geometry. This is called \emph{dimensional flow}. In the infrared (IR, large scales $\ell>\ell_1$), the dimension of spacetime is known to be equal to the topological dimension $D$. In our case $D=4$, there are three spatial dimensions and one time dimension. The scales of the hierarchy \Eq{hier} are intrinsic to the geometry and appear in many (not necessarily all) physical observables. 

More precisely, a multiscale spacetime is such that dimensional flow occurs with three properties: [A1] at least two of the dimensions $\dh$, $\ds$, and $\dw$ vary; [A2] the flow is continuous from the IR down to an ultraviolet (UV) cutoff (possibly trivial, in the absence of any minimal length scale); [A3] the flow occurs locally, i.e., curvature effects are ignored (this is to prevent a false positive). [B] As a byproduct of A, a noninteger dimension ($\dh$, $\ds$, $\dw$, or all of them) is observed during dimensional flow, except at a finite number of points (e.g., the UV and the IR extrema).

On the other hand, \emph{multifractional} geometries are a special case of multiscale spacetimes. Their measure in position and momentum space and their Laplace--Beltrami operator are all factorizable in the coordinates:
\ba
\rmd^Dq(x) &:=& \rmd q^0(x^0)\,\rmd q^1(x^1)\cdots \rmd q^{D-1}(x^{D-1})\,,\label{facto}\\
\rmd^Dp(k) &:=& \rmd p^0(k^0)\,\rmd p^1(k^1)\cdots\rmd p^{D-1}(k^{D-1})\,,\label{rdp}\\
\cK_x &=& \sum_\mu\cK(x^\mu)\,,
\ea
in $D$ topological dimensions.

\emph{Weakly multifractal} spacetimes are multiscale spacetimes with the following property (inherited from fractal geometry, a standard branch of mathematics) in addition of A and B: [C] the relations
\be\label{proC}
\dw=2\frac{\dh}{\ds}\,,\qquad \ds\leq\dh
\ee
hold at all scales in dimensional flow. \emph{Strongly multifractal} geometries satisfy A, B, C, and [D] are nowhere differentiable in the sense of integer-order derivatives, at all scales except at a finite number of points (e.g., the UV and the IR extrema). For the traditional definition of \emph{fractal} set, which we will not use in this context of spacetime models, see \cite{trtls,Fal03} and references therein.

\bigskip

\lolabel{03} \textbf{\emph{How many multiscale, multifractional, and multifractal theories are there?}}\addcontentsline{toc}{subsection}{\texorpdfstring{\loref{03}}{} How many multiscale, multifractional, and multifractal theories are there?}

There are as many multiscale theories as the number of proposals in quantum gravity, plus some more. In fact, dimensional flow (mainly in $\ds$, but in some cases also in $\dh$) is a universal phenomenon \cite{tH93,Car09,fra1} found in all the main scenarios beyond general relativity: string theory \cite{CaMo1}, asymptotically-safe gravity ($\ds\simeq D/2$ in $D$ topological dimensions at the UV non-Gaussian fixed point; analytic results) \cite{LaR5,RSnax,CES}; CDT (for phase-C geometries, $\ds\simeq D/2$ in the UV \cite{AJL4,AJL5,BeH,SVW1} or, more recently, $\ds\simeq 3/2$ \cite{CoJu}; numerical results) and the related models of random combs \cite{DJW1,AtGW} and random multigraphs \cite{GWZ1,GWZ2}; causal sets \cite{EiMi}; noncommutative geometry \cite{Con06,CCM,ArAl1} and $\kappa$-Minkowski spacetime \cite{ACOS,CaRo1,Ben08,ArTr,AnHa1,AnHa2}; Stelle higher-order gravity ($\ds=2$ in the UV for any $D$ \cite{CMNa}); nonlocal quantum gravity ($\ds<1$ in the UV in $D=4$) \cite{Mod1}. 

In LQG, while there is no conclusive evidence of variations of the spectral dimension for individual quantum-geometry states based on given graphs or complexes \cite{COT2}, genuine dimensional flow has been encountered in nontrivial superpositions of spin-network states \cite{COT3}, as an effect of quantum discreteness of geometry. These states appear also in spin foams (where there were preliminary results \cite{CarM,MPM}) and GFT, so that both theories inherit the same feature. It must be said, however, that not all possible quantum states may correspond to multiscale geometries. 

Other examples, all based on analytic results, are Ho\v{r}ava--Lifshitz gravity ($\ds\simeq 2$ in the UV for any $D$) \cite{CES,SVW1,Hor3}, spacetimes with black holes \cite{CaG,Mur12,ArCa1}, fuzzy spacetimes \cite{MoN}, and multifractional spacetimes (variable model-dependent $\dh$ and $\ds$).

With the exception of noncommutative spacetimes, all these multiscale examples have factorizable measures in position and momentum space, either exactly or in certain effective limits (for instance, the low-energy limit in string field theory, or the continuum limit of discretized or discrete combinatorial approaches such as CDT, spin foams, and GFT). However, only multifractional geometries are characterized by factorizable Laplace--Beltrami operators (hence their name). There are one multifractional toy model and three theories in total, depending on the differential operators appearing in the action: the model $T_1$ with ordinary derivatives \cite{frc1,frc2,frc7,frc11} (a special case of the original nonfactorizable model $\tilde T_1$ of refs.\ \cite{fra1,fra2,fra3}) and the theories $T_v$, $T_q$, and $T_\g$ with, respectively, weighted derivatives \cite{frc3,frc4,frc6,frc7,frc9,frc11,frc13} (fixing the problems of $T_1$), $q$-derivatives \cite{frc2,frc7,frc9,frc11,frc12,qGW,frc14}, and fractional derivatives \cite{frc1,frc2,fra6,frc4}. We will explain their differences later. 

Finally, only a few of these theories have been explicitly checked to be weakly multifractal: asymptotic safety, certain multiscale states in LQG/spin foams/GFT, and the multifractional theory with $q$-derivatives. The multifractional theory with fractional derivatives is strongly multifractal. Noncommutative spacetimes where $\ds>\dh$ in the UV (as in most realizations of $\k$-Minkowski) and black-hole geometries described by a nonlocal effective field theory violate the inequality in \Eq{proC}, hence they are not multifractal. In the other cases, one should calculate the walk dimension $\dw$ to verify whether spacetime is multifractal or only multiscale. We should also mention some early studies of field theories on fractal sets \cite{Svo87,Ey89a,Ey89b}; by construction, these spacetimes are fractal but they are not multifractal (there is no change of spacetime dimensionality), hence they are not physical models.\footnote{A yet older attempt \cite{Sti77} defines a spacetime with fixed noninteger dimension but we do not know whether this can be considered a fractal.} On the other hand, Nottale's scale relativity \cite{Not93,Not97,Not08} is multiscale and presumably also multifractal. A proposal for ``fractal manifolds'' \cite{BA07} is multifractal but, like scale relativity, it is a principle rather than a physical theory, since the field dynamics is not defined systematically for matter sectors and gravity. Table \ref{tab2} summarizes the cases.
\begin{table}[ht]
\begin{center}
\begin{adjustwidth}{-2.1cm}{}
\footnotesize
\begin{tabular}{|l|cccccc|}\hline\hline
								& String theory 							 & Quantum gravities & $T_{1,v,q,\g}$ & $\tilde T_1$ & Early proposals & Scale relativity \\
								&															 &									 &								& \cite{fra1,fra2,fra3} & \cite{Svo87,Ey89a,Ey89b,Sti77} & \cite{Not93,Not97,Not08} \\ \hline
Multiscale 			& \ding{51} (low-energy limit) & \ding{51} (all)   & \ding{51}								 & \ding{51}								 & \ding{55}  & \ding{51}\\
Multifractional & \ding{55}										 & \ding{55} 		     & \ding{51}								 & \ding{55}								 & \ding{55}  & \ding{55}\\
Multifractal	  & ?														 & case dependent    & case dependent 					 & ?			 & \ding{55} (only fractal) & ? \\\hline
\end{tabular}
\caption{\label{tab2} Multiscale, multifractional, and (multi)fractal theories and models.}
\end{adjustwidth}
\end{center}
\end{table}

\bigskip

%%%%%%%%%%%%%%%%%%%%%%%%%%%%%%%%%%%%%%%%%%%%%%%%%%%%%%%%%%%%%%%%%%%%%%%%%%%%%
%%%%%%%%%%%%%%%%%%%%%%%%%%%%%%%%%%%%%%%%%%%%%%%%%%%%%%%%%%%%%%%%%%%%%%%%%%%%%

\section{Motivations}\label{moti}

\lolabel{04} \textbf{\emph{What are the motivations of multifractional theories?}}\addcontentsline{toc}{subsection}{\texorpdfstring{\loref{04}}{} What are the motivations of multifractional theories?}

There are at least four motivations to consider these theories. We call them the quantum-gravity-candidate argument, the flow-versus-finiteness argument, the uniqueness argument, and the phenomenology argument.
\begin{itemize}
\item[(i)] \emph{Quantum-gravity candidate.} Multifractional spacetimes were originally proposed as a class of theories where the renormalization properties of perturbative quantum field theory (QFT) could be improved, including in the gravity sector. The objective of obtaining a renormalizable quantum gravity was supported by a power-counting argument calculating the superficial degree of diverge of Feynman graphs for fields living on a multiscale geometry \cite{frc2}. Later on, it was shown that the theory $T_1$ with ordinary derivatives is only a toy model\footnote{Here and there in the text, we will make a small abuse of terminology and call $T_1$ a ``theory.''} due to the lack of a direct definition of a self-adjoint momentum operator \cite{frc5} (in other words, one has to prescribe an operator ordering in the field action \cite{frc7}) and to issues with microcausality \cite{frc2}. Also, explicit loop calculations and the general scaling of the Green function showed that renormalizability is not improved in the theories $T_v$ and $T_q$ with, respectively, weighted and $q$-derivatives \cite{frc9}. However, the theory $T_\g$ with fractional derivatives is likely to fulfill the original expectations (to see why, check question \lref{47}), but its study involves a number of technical challenges. Nevertheless, massive evidence has been collected that all multifractional models share very similar properties \cite{frc2,frc4,frc10,frc13,frc14}, especially $T_q$ and $T_\g$ (questions \lref{11} and \lref{34}). In preparation of dealing with the theory with fractional derivatives and to orient future research on the subject, it is important to understand in the simplest cases what type of phenomenology one has on a multiscale spacetime. In particular, $T_v$ and $T_q$ are simple enough to allow for a fully analytic treatment of the physical observables, while having all the features of multiscale geometries. Therefore, they are the ideal testing ground for these explorations. A better knowledge about the typical phenomenology occurring in multifractional spacetimes will be of great guidance for the study of the case with fractional derivatives.
\item[(ii)] \emph{Flow versus finiteness.} As soon as dimensional flow was recognized as a universal property of effective spacetimes emerging in quantum gravity \cite{tH93}, the possibility was considered that such property is related to the UV finiteness of a theory. This suspicion was mainly fueled by the fact that $\ds\simeq 2$ in the UV of many different models: having two effective dimensions would imply that two-point correlation functions (propagators, potentials, and so on) diverge logarithmically with the distance rather than as an inverse power law in the UV. Multifractional spacetimes are a class of theories where dimensional flow is under complete analytic control and where one can test the conjecture that multiscale geometries are related to UV finiteness. The counterexamples offered by the multifractional paradigm \cite{frc9}, regardless of the value of the spectral dimension in the UV, disproved this conjecture and reappraised the relative importance of dimensional flow with respect to UV finiteness. In parallel, the supposed universality of the magic number $\ds=2$ was later recognized as fictitious because based on a poor statistics; many models, supposedly UV finite, were in fact found where $\ds\neq 2$ at short scales, including some already considered in the past (such as CDT \cite{CoJu}).
\item[(iii)] \emph{Uniqueness.} Although renormalizability is a strongly model-dependent feature, it remains to understand why dimensional flow is so similar in so different and so many theories. A recent theorem explains why \cite{first}. Let dimensional flow of spacetime in the Hausdorff or spectral dimension $d=\dh,\ds$ be described by a continuous scale parameter $\ell$ (this is always the case, as stated in \loref{01}). Let also effective spacetime be noncompact, so that $d\simeq D$ in the IR and there are no undesired topology effects. As a further very general requirement, we also ask that dimensional flow is slow at large scales, meaning that the dimension $d$ forms a plateau in the IR (figure \ref{fig1}). Since the IR limit $\ell\to+\infty$ is asymptotic, this flatness of $d(\ell)$ in the IR is always guaranteed. IR flatness can be encoded perturbatively by requiring that $d\simeq d^{\rm IR}$ approximately at large scales. The accuracy of the approximation is governed by an order-by-order estimate of the logarithmic derivatives of $d$ with respect to the scale $\ell$, via the linear flow equation
\be\label{flow}
\sum_{j=0}^n c_j\frac{\rmd^j}{(\rmd\ln\ell)^j}[d^{(n)}(\ell)-d^{(n-1)}(\ell)]=0\,,\qquad d^{(0)}:=d^{\rm IR}\,,
\ee
where $c_j$ are constants. Then, given the three assumptions above (obeyed by \emph{all} known quantum gravities) and eq.\ \Eq{flow}, we can completely determine the profile $d(\ell)$ at large and mesoscopic scales once we also specify the symmetries of the measures in position and momentum space. The \emph{first flow-equation theorem} states that
\be\label{dhdsgen}
d(\ell)\simeq D+b\left(\frac{\ell_*}{\ell}\right)^c+\text{\rm (log oscillations)},
\ee
where $b$ and $c$ are constants fixed by the dynamics of the specific theory, $\ell_*$ is the largest characteristic scale of the geometry, and the omitted part is a combination of logarithmic oscillations in $\ell$. Using eqs.\ \Eq{dhdsgen} and \Eq{dh}, for $d=\dh$ (slow flow in position space) one can specify the scaling of spacetime volumes $\cV(\ell)$ with their linear size $\ell$, while for $d=\ds$ (slow flow in momentum space) one can derive the return probability $\cP(\ell)$ from \Eq{ds}. The proof of \Eq{dhdsgen} is independent of the dynamics of the theory and of the geometrical background, except for the requirement that dimensional flow exists [obviously, this implies that spacetime geometry is characterized by a hierarchy of fundamental scales \Eq{hier}]. The dynamics, and thus the details of the theory, determines the numerical value of the constants $b$ and $c$ and the identification of $\ell_*$ within the scale hierarchy of the theory. Many quantum-gravity examples are given in question \lref{45}.
\begin{figure}
\centering
\includegraphics[width=8.4cm]{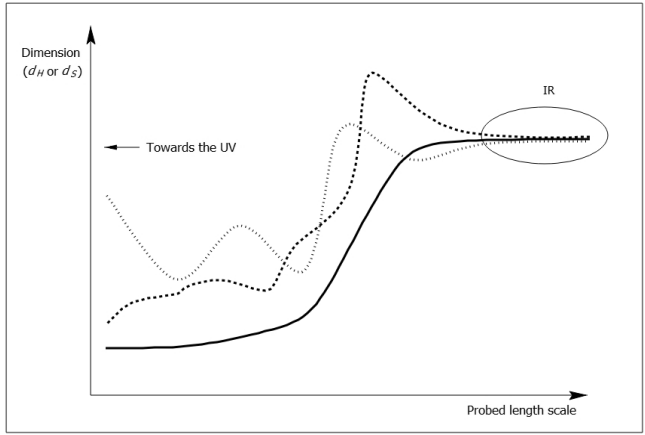}
\caption{\label{fig1} The central hypothesis of the theorems on dimensional flow described in the text.}
\end{figure}

Now, if the measures in position and momentum space are not Lorentz invariant but factorizable, and if the Laplace--Beltrami operator is also factorizable, we hit precisely the case of multifractional theories. Then, eq.\ \Eq{dhdsgen} ceases to be valid. In its stead, one has $D$ copies of it with $D=1$, one for each spacetime direction:
\be\label{dhdsgen2}
d(\ell)\simeq \sum_{\mu=0}^{D-1}d_\mu(\ell):=\sum_{\mu=0}^{D-1}\left[1+b_\mu\left(\frac{\ell_*^\mu}{\ell}\right)^{c_\mu}+\text{\rm (log oscillations)}\right],
\ee
where $b_\mu$ and $c_\mu$ are constant. This is the \emph{second flow-equation theorem}. Since a factorizable measure in position space can be written as eq.\ \Eq{facto} for $D$ independent profiles $q^\mu(x^\mu)$ (called geometric coordinates), in multifractional spacetimes volumes (of same linear size $\ell$ in all directions) are of the form $\cV(\ell)\sim\int_{\rm vol}\rmd^Dq(x)=\prod_\mu q^\mu(\ell)$. Plugging this expression into eq.\ \Eq{dh} and integrating using eq.\ \Eq{dhdsgen2}, we get an approximate $q^\mu(\ell)$ for each direction. The theorem determines the profiles $q^\mu(x^\mu)$ exactly. In this paragraph, we focus our attention on \emph{real} solutions to the flow equation \Eq{flow}, postponing the case of \emph{complex} solutions to question \lref{14}. At leading order in the perturbative expansion \Eq{flow} of $d(\ell)$ centered at the IR point, one has
\be\label{meamu}
q^\mu(x^\mu)\simeq x^\mu+\frac{\ell_*^\mu}{\a_\mu}{\rm sgn}(x^\mu)\left|\frac{x^\mu}{\ell_*^\mu}\right|^{\a_\mu} F_\om(x^\mu)\,,
\ee
where
\be\label{Fom}
F_\om(x^\mu)= 1+A_\mu\cos\left(\om_\mu\ln\left|\frac{x^\mu}{\ell_\infty^\mu}\right|\right)+B_\mu\sin\left(\om_\mu\ln\left|\frac{x^\mu}{\ell_\infty^\mu}\right|\right)\,,
\ee
all indices $\mu$ are inert (there is no Einstein summation convention), the first factor 1 in \Eq{Fom} is optional \cite{first}, $\ell_*^\mu$ and $\ell_\infty^\mu$ are $D+D$ length scales, and $\a_\mu$, $A_\mu$, $B_\mu$, and $\om_\mu$ are $D+D+D+D$ real constants. Going beyond leading order in the perturbative expansion of the dimension at the IR, one gets the even more general form, valid at \emph{all} scales,
\be\label{meag}
q^\mu(x^\mu) = x^\mu+\sum_{n=1}^{+\infty} \frac{\ell_n^\mu}{\a_{\mu,n}}{\rm sgn}(x^\mu)\left|\frac{x^\mu}{\ell_n^\mu}\right|^{\a_{\mu,n}} F_n(x^\mu)\,,
\ee
where $F_n(x^\mu)$ is $F_\om(x^\mu)$ with all real constants $\ell_\infty^\mu,\a_\mu,A_\mu,B_\mu,\om_\mu$ labeled by the sum index $n$. Equation \Eq{meag} describes the most general real-valued multifractional geometry along the direction $\mu$, characterized by an infinite hierarchy of scales $\{\ell_n^\mu,\ell_{\infty,n}^\mu\}$. Remarkably, exactly the same form of the geometric coordinates \Eq{meag} can be obtained in a totally independent way by asking \emph{a priori} that the measure \Eq{facto} on the continuum represent the integration measure on a multifractal \cite{frc2}. For each direction, one first considers a deterministic fractal set\footnote{\label{fu1}A deterministic fractal $\cF=\bigcup_i \cS_i(\cF)$ is the union of the image of some maps $\cS_i$ which take the set $\cF$ and produce smaller copies of it (possibly deformed, if the $\cS_i$ are affinities). Not all fractals are deterministic. Sets with similarity ratios randomized at each iteration are called random fractals. Cantor sets are popular examples of deterministic and random fractals. Let $\cS_1(x)=a_1 x+b_1$ and $\cS_2(x)=a_2 x+b_2$ be two \emph{similarity maps}, where $a_{1,2}$ (called similarity ratios) and $b_{1,2}$ (called shift parameters) are real constants and $x\in I$ is a point in the unit interval $I=[0,1]$. The image $\cS_i(A)$ of a subset $A\subset I$ is the set of all points $\cS_1(x)$ where $x\in A$. A Cantor set or Cantor dust $\cC$ is given by the union of the image of itself under the two similarity maps, $\cC=\cS_1(\cC)\cup \cS_2(\cC)$. For instance, the ternary (or middle-third) Cantor set $\cC_3$ has $a_1=1/3=a_2$, $b_1=2/3$, and $b_2=0$: $\cS_1(x)=\tfrac13 x+\tfrac23$, $\cS_2(x)=\tfrac13 x$. At the first iteration, the interval $[0,1]$ is rescaled by $1/3$ and duplicated in two copies: one copy (corresponding to the image of $\cS_2$) at the leftmost side of the unit interval and the other one (corresponding to $\cS_1$) at the rightmost side. In other words, one removes the middle third of the interval $I$. In the second iteration, each small copy of $I$ is again contracted by $1/3$ and duplicated, i.e., one removes the middle third of each copy thus producing four copies 9 times smaller than the original; and so on. Iterating infinitely many times, one obtains $\cC_3$, a dust of points sprinkling the line. The set is \emph{self-similar} inasmuch as, if we zoom in by a multiple of 3, we observe exactly the same structure. 

It is easy to determine the dimensionality of the Cantor set $\cC$. Since this dust does not cover the whole line, it has less than one dimension. Naively, one might expect that the dimension of $\cC$ is zero, since it is the collection of disconnected points (which are zero-dimensional). However, there are ``too many'' points of $\cC$ on $I$ and, as it turns out, the dimension of the set is a real number between 0 and 1. In particular, given $N$ similarity maps all with ratio $a$, the similarity dimension or capacity of the set is $\dc(\cC):=-\ln N/\ln a$, a formula valid for an exactly self-similar set made of $N$ copies of itself, each of size $a$. Note that $a= N^{-1/\dc}$: the smaller the size $a$, the smaller the copies at each iteration and the smaller the dimensionality of the set. In the case of the middle-third Cantor set, $N=2$ and $a=1/3$, so that $\dc=\ln2/\ln 3\approx 0.63$.} living on a line and obtains the typical (power law)$\times$(log oscillations) structure \cite{RYS,RLWQ,NLM}; summing over different scales, one obtains the multifractal profile \Eq{meag}. The independence of this derivation of the measure is important because it yields information not apparently available in the flow-equation theorems (see question \loref{07b}). In this review, we will not insist too much upon the beautiful formalism of fractal geometry implemented into multifractional spacetimes; a concise presentation can be found in refs.\ \cite[sections 3.2, 5.1, and 5.2]{frc2} and \cite{trtls}.

To summarize, the measure \Eq{meag} used in multifractional theories is the most general one when spacetime geometry is multiscale and factorizable \cite{first}. It also happens to coincide with the integration measure of a multifractal \cite{frc2}.\footnote{This is not inconsistent with what said in question \loref{02}. Even if the measure is multifractal, the geometry of spacetime may be nonmultifractal, depending on the symmetries enforced on the dynamics (i.e., type of derivatives).} Thus, multifractional theories are the most general factorizable framework wherein to study the phenomenon of dimensional flow. This can (and did) help to better understand the flow properties of other quantum gravities (even despite their nonfactorizability), either by recasting the dynamics of such theories as a multifractional effective model \cite{fra7,ACOS,CaRo1} or by employing the same mathematical tools endemic in multifractional theories \cite{CaMo1,CES,COT3,fra6}. The geometrical and physical reason beyond the existence of the flow-equation theorems and of a unique (in the sense of being described by the same multiparametric function) dimensional flow in all quantum gravities is the fact that the IR is reached as an asymptote where the dimension varies slowly. There is also a perhaps deeper physical reason, more delicate to track, that also sheds light into the flow-versus-finiteness issue. It consists in the fact that dimensional flow is the typical outcome of the combination of general relativity with quantum mechanics \cite{CaRo2a,CaRo2b}.
\item[(iv)] \emph{Phenomenology.} The search for experimental constraints on fractal spacetimes dates back to the 1980s \cite{ScM,ZS,MuS,CO}. Since early proposals of fractal spacetimes were quite difficult to handle \cite{Svo87,Ey89a,Ey89b,Sti77}, toy models of dimensional regularization were used and several bounds on the deviation $\e=D-4$ of the spacetime dimension from 4 were obtained. However, these models were not backed by any theoretical framework and they were not even multiscale. Multifractional theories are genuine realizations of multiscale geometries based on much more solid foundations, i.e., all the sectors one would possibly like to investigate are under theoretical control (classical and quantum mechanics, classical and quantum field theory, gravity, cosmology, and so on) and they give rise to well-defined physical predictions that can be (and actually have been) tested experimentally. Most notably, all the phenomenology extracted from multifractional scenarios comes directly from the full theory, with very few or no approximations. We will always use the term ``phenomenology'' in this sense, in contrast with its other use as a synonym of ``heuristic'' (i.e., inspired by a theory rather than derived from it rigorously) in some literature of quantum gravity. The questions left unanswered by the dimensional-regularization toy models can now receive proper attention; see section \ref{phen}. In the same section, we will see that multifractional theories make it possible to explore observable consequences of dimensional flow, which is not just a mathematical property.
\end{itemize}

\bigskip

\lolabel{05} \textbf{\emph{I understand that spacetimes endowed with a structure of weighted derivatives or $q$-derivatives are analyzed more in detail because they are simpler than the theory with fractional derivatives, which is most promising especially as far as renormalization is concerned. However, what is the physical reason why such extensions $T_v$ and $T_q$ should be of interest and relevance to particle-physics phenomenology? They are only distant relatives of a theory supposed to describe geometry (dimensional flow) and quantum gravity, with no connection to the Standard Model.}}\addcontentsline{toc}{subsection}{\texorpdfstring{\loref{05}}{} Why should multifractional theories with weighted and \texorpdfstring{$q$}{}-derivatives be of interest for particle-physics phenomenology?}

A first answer is given by the quantum-gravity-candidate argument of \loref{04}. All multifractional theories share similar phenomenology, as far as we can see. In the context of particle physics, it was shown that the scale hierarchy of $T_v$ is quite similar to the scale hierarchy of $T_q$, even if individual experiments may be sensitive to such scales in different ways [for instance, variations of the fine-structure constant in quantum electrodynamics (QED) are detectable only in the case with $q$-derivatives but not in the other] \cite{frc12,frc13}. In questions \loref{07b}, \lref{11}, and \lref{34}, we will find a striking similarity between $T_q$ and $T_\g$ when $\g=\a$, based on the dimensionality of the Laplace--Beltrami operator \cite{frc2}, on the form of the propagator in the UV, and on approximations of the integrodifferential calculi of the theories. Because of this approximate but crucial matching
\be
T_{\g=\a}\cong T_q\,,
\ee
we expect the phenomenology with $q$-derivatives to be very similar to that with fractional derivatives. Thus, it is useful to understand what type of experiments would be capable of constraining $T_\g$. 

Apart from this goal, it is important to recognize the impact of dimensional flow on physical observables. The quest for an observable imprint of quantum gravity is more feverish than ever and it is natural to look at possible effects of the most evident feature all these competitors have in common. The theories $T_v$ and $T_q$ are not mere toy models of a ``better'' theory: they represent autonomous realizations of physics on a geometry with dimensional flow. Even if their renormalizability is not better than in standard field theories, they display a full set of testable physical observables from particle and atomic physics to cosmology. Since the geometry described by the multifractional measures \Eq{meamu} and \Eq{meag} is the most general factorizable one if $\dh$ varies with the scale, the constraints on the scale hierarchy obtained in multifractional theories possibly have a much wider scope of validity, being somewhat prototypical of the whole class of multiscale theories; thus, including quantum gravities.

\bigskip

\lolabel{06} \textbf{\emph{These theories have been developed mainly by the author himself and hence their impact on the community at large might be limited. Will this line of research illuminate anything about quantum gravity?}}\addcontentsline{toc}{subsection}{\texorpdfstring{\loref{06}}{} Will this line of research illuminate anything about quantum gravity?}

Yes, mainly for the reason spelled out in \loref{05}. Multifractional theories did receive attention by the quantum-gravity community and have been actively studied not just by the author but also by researchers working in different fields such as quantum field theory \cite{frc6,frc7,frc9,frc12,frc13}, noncommutative spacetimes \cite{ACOS,CaRo1}, quantum cosmology and supergravity \cite{ACOS}, group field theory \cite{ACOS}, classical cosmology \cite{frc8,frc14}, and numerical relativity \cite{frc14}. As mentioned in \loref{04}, interest has not been limited to multifractional theories \emph{per se}, but extended to the possibility to use their machinery in different, quantum-gravity-related contexts \cite{CaMo1,CES,COT3}. However, despite the ongoing collaborative effort, the limited number of people involved is sometimes perceived as a signal that multifractional theories are not as interesting and useful as advertized.

There were two causes that led to this opinion. The first is the type of development that multifractional theories have undergone since the beginning \cite{fra1}. Many of their aspects have evolved slowly and heterogeneously from paper to paper and this has hindered a coherent exposition of the main ideas from the start. The present manifesto, with its overview and active integration of different elements, should help to clarify the context, advantages, and status of these theories. The second cause is that multifractional theories had to talk with a number of communities widely different from one another. On one hand, the original proposal was directed to the quantum-gravity sector, which is not at all annoyed by the breaking of Lorentz symmetries but is fragmented into, and busy with, a number of independent and very strong agend\ae\ based on elegant mathematical structures and convincing evidence (or proofs) of UV finiteness. Since there are hints that it is possible to quantize multifractional gravity but there is no proof yet, the present proposal is understandably regarded as unripe. On the other hand, the study of the multiscale Standard Models left gravity aside and was of more interest for the traditional QFT community, for which Lorentz invariance is a cornerstone and dimensional flow is an unnecessary concept. Consequently, the main motivation of the theories was lost (question \lref{30}).

The intrinsic difficulty in changing spacetime paradigm (a change of measure is relatively alien to ``usual'' quantum-gravity scenarios, with the exception of noncommutative spacetimes) and the lack of contact with observations have partially limited the reception of this proposal until now. However, the important conceptual clarifications and simplifications carried out in the last year (mainly in refs.\ \cite{frc13,trtls,first}) and the obtainment of the first observational constraints ever on the theory \cite{frc12,frc13,qGW,frc14} are already contributing to boost its visibility. It may also be relevant to recall that, contrary to popular quantum-gravity candidates, the case with $q$-derivatives is the first and only known example of a theory of exotic geometry that is efficiently constrained by gravitational waves alone \cite{qGW}. In this respect, as far as gravity waves are concerned, and until further notice, multifractional theories are proving themselves to be observationally as competitive as the usual quantum-gravity scenarios. This is the type of result one might like to find in the context of quantum gravity at the interface between theory and experiment. This and other results on phenomenology, together with the universality traits described in \loref{04}, make the multifractional paradigm not only a useful and general tool of comparison of different features in the landscape of quantum gravity, but also an independent theory that is legitimate to study separately. In this sense, it is not strictly subordinate to the problem of quantum gravity at large.

It is also relevant to recall that the idea underlying multifractional theories is not a prerogative of this framework. In other proposals \cite{Svo87,Ey89a,Ey89b,Sti77,Not93,Not97,Not08,BA07}, an \emph{Ansatz} for geometry and symmetries was made but no field-theory action thereon was given. The multifractional paradigm not only makes the ``fractal spacetimes'' idea systematic, but it also provides an explicit form for the dynamics (questions \lref{29} and \lref{37}). In particular, the ``fractal coordinates'' of scale relativity correspond to our binomial geometric coordinates but written as a power-law profile with a scale-dependent exponent, $q\sim x^{\a(\ell)}$ with $\a(\ell)=1+(\a-1)/[1+(\ell/\ell_*)^{\a-1}]$ \cite{frc2}.

%%%%%%%%%%%%%%%%%%%%%%%%%%%%%%%%%%%%%%%%%%%%%%%%%%%%%%%%%%%%%%%%%%%%%%%%%%%%%
%%%%%%%%%%%%%%%%%%%%%%%%%%%%%%%%%%%%%%%%%%%%%%%%%%%%%%%%%%%%%%%%%%%%%%%%%%%%%

\section{Geometry and symmetries}\label{geo}

\lolabel{07} \textbf{\emph{The choice of measure \Eq{facto} with eq.\ \Eq{meamu} and
\be\label{binom2}
\a_\mu=\a_0,\a\,,\qquad \ell_*^\mu=t_*,\ell_*\,,\qquad \ell_\infty^\mu=t_\infty,\ell_\infty\,,
\ee
so often used in multifractional models, is completely \emph{ad hoc}. On one hand, why should we limit our attention to factorizable measures \Eq{facto}? On the other hand, why should one choose the specific profile $q(x)$ in eq.\ \Eq{meamu}?}}\addcontentsline{toc}{subsection}{\texorpdfstring{\loref{07}}{} Why should we limit our attention to factorizable and binomial measures?}

A highly irregular geometry such as multidimensional fractals is generically described by a nonfactorizable measure $\vr(x^0,x^1,\dots,x^{D-1})$. There have been attempts to place a field theory on such geometries in the past \cite{Svo87,Ey89a,Ey89b} and even recently \cite{fra1,fra2,fra3} but, unfortunately, and regardless of their level of rigorousness, their range of applicability to physical situations was severely restricted. This was due to purely technical reasons, which include, for instance, the difficulty in finding a self-adjoint momentum operator and a self-adjoint Laplace--Beltrami operator compatible with the momentum transform. In order to make progress, factorizable measures $\rmd\vr(x^0,x^1,\dots,x^{D-1})=\prod_\mu\rmd q^\mu(x^\mu)$ [eq.\ \Eq{facto}] were considered starting from ref.\ \cite{frc1}. This choice has been successful in fully constructing a whole class of theories, in extracting observational constraints thereon, and in connecting efficiently with quantum-gravity frameworks. If we compare the 25-year stalling of nonfactorizable models with the 5-year advancement of factorizable models from theory to experiments, the practical justification of \Eq{facto} is evident. Also, from the point of view of the phenomenology of dimensional flow, there is nothing wrong with factorizable measures: they have exactly the same scaling properties of nonfactorizable measures, which is a necessary and sufficient condition to have the same change in dimensionality.

Of course, it may be that Nature, if multiscale, is not represented by factorizable geometries, in which case we have to look into other proposals. As discussed in ref.\ \cite{CaRo1}, the natural generalization of multifractional geometries to nonfactorizable measures are, arguably, noncommutative spacetimes, which overcome the problems associated with nonfactorizability with the introduction of a noncommutative product. The utility of factorizable multifractional theories is not exhausted even in that case because, although the mathematical and practical language describing noncommutative systems is different from the one employed in multiscale or fractal geometries, many contact points between these two frameworks are possible nonetheless \cite{ACOS,CaRo1}.

Once accepted the use of factorizable measures, according to the second flow-equation theorem the only possible choice is \Eq{meag}. We can walk the logical path \Eq{meag}$\to$\Eq{meamu}$\to$\Eq{binom2} as follows. Equation \Eq{meag} is an IR expansion with $D$ copies of an infinite number of free parameters (fractional exponents $\a_{n,\mu}$, frequencies $\om_{n,\mu}$, amplitudes, and the scales $\ell_n^\mu$ and $\ell_{\infty,n}^\mu$), which means that one can fit any wished profile when no dynamical input on the values of such parameters is given (it \emph{is} given in quantum gravities). The first step in reducing this ambiguity in multifractional theories comes from the scale hierarchy itself, which is divided in two sets. Omitting the index $\mu$ from now on, the first is the set of scales $\{\ell_n\}=\{\ell_1\geq\ell_2\geq\dots\}$ characterizing regimes where the dimension of spacetime takes different values (we will see which values in question \lref{13}); it is the scale hierarchy \emph{par excellence}, the one defining dimensional flow via the polynomials of \Eq{meag}. Superposed to that is the set of scales $\{\ell_{\infty,n}\}$, called harmonic structure in fractal geometry \cite{frc2}. The harmonic structure does not govern the main traits of dimensional flow but it modulates it with a superposition of $n$ patterns of logarithmic oscillations; such modulation affects even scales much larger than $\ell_\infty$, as cosmology shows \cite{frc11,frc14}. The scale hierarchies $\{\ell_n\}$ and $\{\ell_{\infty,n}\}$ are mutually independent but, from the derivation of eq.\ \Eq{meag}, it is easy to convince oneself that $\ell_n\geq \ell_{\infty,n}$ for each $n$ \cite{first}. Thus, the long-range modulation of the harmonic structure and the theoretical ``coupling'' $\ell_n\leftrightarrow\ell_{\infty,n}$ leads to the conclusion that the first multiscale effects we could observe in experiments would be at scales $\gtrsim\ell_*\equiv\ell_1$, possibly modulated by log oscillations with scale $\ell_\infty\equiv\ell_{\infty,1}$. In other words, eq.\ \Eq{meamu} is the approximation of \Eq{meag} at scales $\gtrsim\ell_*$. But this is already sufficient to extract all relevant phenomenology. Scales below $\ell_*$ are too small to be constrained by experiments, and $\ell_*$ acts as a sort of ``screen'' hiding the yet-unreachable microscopic structure of the measure at smaller scales. Whatever happens at smaller scales, no matter the number of transient regimes with different dimensionalities from $\ell_*$ down to Planck scales, from the point of view of a macroscopic observer the first transition to an anomalous geometry will occur near $\ell_*$. Experiments constrain just this scale, the end of the multiscale hierarchy. Thus, for all practical purposes there is no loss of generality in considering eq.\ \Eq{meamu} instead of the too formal \Eq{meag}. The further simplification from \Eq{meamu} to \Eq{binom2} is an isotropization of the scale hierarchies and dimensions to all spatial directions, while the time direction is left free to evolve independently. Full isotropization is achieved when $\a_0=\a$, but this is almost never needed in calculations. If one wishes to consider geometries which are multiscale only in the time or space directions, it is sufficient to set $\a_0\neq1,\a=1$ or $\a_0=1,\a\neq 1$, respectively. Having an isotropic spatial hierarchy (one scale $\ell_*^i=\ell_*$ for all directions) partially compensates for the restrictions of factorizability and makes observables easier to compute. One can even invoke this choice as a symmetry principle defining the theory, since there is no reason \emph{a priori} to have a strongly different dimensional flow along different spatial directions. One can consider this as part of a multiscale version of the principle of special relativity.

\bigskip

\lolabel{07b} \textbf{\emph{What is the parameter space of these theories?}}\addcontentsline{toc}{subsection}{\texorpdfstring{\loref{07b}}{} What is the parameter space?}

There are severe theoretical priors on $(\a_\mu,t_*,\ell_*,t_\infty,\ell_\infty,A,B,\om)$. 
\begin{itemize}
\item[--] The fractional exponents $\a_0$ and $\a$ are taken within the interval
\be\label{ranga}
0\leq\a_\mu\leq 1\,.
\ee
The lower bound $\a_\mu\geq 0$ guarantees that the UV Hausdorff dimension $\a_\mu$ of each direction in spacetime be non-negative, a minimal requirement if we want to be able to probe the geometry with conventional rules. The upper bound $\a_\mu\leq 1$ guarantees that the dimension in the UV is always smaller than the topological dimension $D$. Neither bound can be easily extended in the theories $T_1$, $T_v$, and $T_q$. The lower limit $\a_\mu\geq 0$ can be replaced by $\sum\a_\mu\geq 0$ (e.g., ref.\ \cite{frc9}); in general, not all $\a_\mu$ can be negative, lest $\dh\simeq\sum_\mu\a_\mu<0$ [see eq.\ \Eq{dhuv}]. However, this would lead to a negative-definite dimension either of time or of spatial slices, and it is not clear whether such a configuration would make sense physically. On the other hand, if we took the upper limit arbitrarily large, we would get a dimensionally larger UV geometry that has very few examples in quantum gravity; still, there exist a minority of cases where $\ds>D$ in the UV, as in $\k$-Minkowski spacetime \cite{ArAl1,ArTr} o near a black hole \cite{ArCa1}. However, multiscale corrections of physical observables are always of the form
\be
v_\mu(x^\mu):=\p_\mu q^\mu(x^\mu)=1+O(|x^\mu/\ell_*|^{\a_\mu-1}).
\ee
Therefore, an $\a_\mu>1$ always leads to a wrong IR limit, which is defined by the largest fractional charge $\a_{\mu,n}$ in eq.\ \Eq{meag}. By definition, this is equal to 1 (nonanomalous scaling of the coordinates). The special value 
\be\label{a12}
\a_\mu=\frac12
\ee
at the center of the interval \Eq{ranga} plays a unique role, not only because it is the average representative of this class of geometries (it is typical and instructive to compare experimental constraints with $\a_\mu\ll 1$, $\a_\mu=1/2$, and the standard geometry $\a_\mu=1$), but also because it signals a phase transition across a critical point in the theory \cite{frc2}. Take, for instance, a scalar field in flat multifractional Minkowski space:
\be\label{Sphi}
S_\phi=\int\rmd^Dq(x)\left[\frac12\phi\cK\phi-V(\phi)\right],
\ee
where the signature of the Minkowski metric is $\eta_{\mu\nu}=(-,+,\cdots,+)_{\mu\nu}$ and $\cK$ is the Laplace--Beltrami operator. The engineering (scaling) dimension of the scalar field is 
\be\label{phik}
[\phi]=\frac{\dh-[\cK]}{2}\,,
\ee
where $\dh$ is the scaling of the coordinate-dependent part of the measure $\rmd^Dq$.\footnote{Note that $[\rmd^Dq]=-D$ for the measure given by \Eq{facto}, \Eq{meamu}, and \Eq{binom2} (or in the general case \Eq{meag}) because all elements in the sum scale in the same way. However, what matters here is the scaling of the nonconstant terms of the measure, which is $-\a_\mu$ for each direction.} From eq.\ \Eq{dhuv2} ($\a_\mu=\a$ for all $\mu$), it follows that $\phi$ becomes dimensionless when $\a=[\cK]/D$. In the model $T_1$ and in the theory $T_v$, the Laplace--Beltrami operator is
\be\label{Ks}
T_1\,:\quad \cK=\B\,,\qquad T_v\,:\quad \cK=\cD_\mu\cD^\mu=\frac{1}{\sqrt{v}}\B\left(\sqrt{v}\,\cdot\,\right),\qquad \cD_\mu:=\frac{1}{\sqrt{v}}\p_\mu\left(\sqrt{v}\,\cdot\,\right),
\ee
where
\be\label{vfacto}
v=v(x):=v_0(x^0)\,v_1(x^1)\cdots v_{D-1}(x^{D-1})\geq 0\,.
\ee
Thus, $[\cK]=2$ at all scales and the critical value of the isotropic fractional exponent is $\a=2/D$. This is precisely $1/2$ in $D=4$ dimensions. Thus, in $T_1$ and $T_v$ the value \Eq{a12} is somewhat preferred because the critical point is interpreted (as in Ho\v{r}ava--Lifshitz gravity) as a UV fixed point.

In the theories $T_q$ and $T_\g$ on Minkowski spacetime, the Laplace--Beltrami operator is (Einstein's sum convention is used) \cite{frc2,frc4}
\be\label{Kss}
T_q\,:\quad \cK=\B_q=\eta^{\mu\nu}\frac{\p}{\p q^\mu}\frac{\p}{\p q^\nu}\,,\qquad T_\g\,:\quad \cK=\cK_\g\,,
\ee
where $\cK_\g$ is composed by the operators ${}_\infty\p^{2\g}$ and ${}_\infty\bar\p^{2\g}$, respectively, the Liouville and Weyl fractional derivative of order $2\g$ \cite{frc1,KST} (see question \lref{11} for the explicit expression). The varying part of the Laplacian scales as $[\B_q]\simeq 2\a$ and $[\cK_\g]\simeq 2\g$ (in the UV) for the isotropic choices $\a_\mu=\a$ and $\g_\mu=\g$, and the scalar field scales as $[\phi]=(D\a-2\g)/2$. For $\a=\g$, there is no UV critical point but the behaviour of $T_q$ and $T_\g$ is quite similar.

In the case with fractional derivatives $T_{\g=\a}$, the range \Eq{ranga} is further shrunk to $1/2\leq\a_\mu\leq 1$ by requiring multifractional spacetime to be normed (that is, distances obey the triangle inequality) \cite{frc1}.\footnote{There is no such restriction in $T_q$, which has a well-defined norm for any positive value of $\a$ \cite{frc11}.} Then, the value $\a_\mu=1/2$ is even more special being it the lowest possible in the theory. Equation \Eq{a12} is also supported independently by a recent connection with a heuristic estimate of quantum-gravity effects on measurement uncertainties \cite{CaRo2a,CaRo2b}. A parallel estimate, however, selects
\be\label{a13}
\a_\mu=\frac13
\ee
as an alternative preferred value \cite{CaRo2a,CaRo2b}. This lies in the region of parameter space where $T_{\g=\a}$ is not normed, but in questions \lref{27} and \lref{47} we will reconsider the restriction \Eq{ranga}. Last, the arguments of \cite{CaRo2a,CaRo2b} also fix the fractional exponents to the fully isotropic configuration
\be
\a_0=\a\,,
\ee
although in general we will not enforce this relation.
\item[--] There is no prior on $t_*$, $\ell_*$, $t_\infty$, and $\ell_\infty$, except that they are positive; there are also other free parameters $E_*$, $k_*$, $E_\infty$, and $k_\infty$ in the momentum-space measure. One can reduce the number of free parameters by relating time and length scales by a unit conversion. In a standard setting, one would make such conversion using Planck units. Here, the most fundamental scale of the system is the one appearing in the full measure with logarithmic oscillations, denoted above as $\ell_\infty$. For the time direction one has a scale $t_\infty$, while in the measure in momentum space the fundamental energy $E_\infty$ and momentum $p_\infty$ appear. Then, one may postulate that the scales $\ell_*\geq\ell_\infty$, $t_*\geq t_\infty$ and $E_*\leq E_\infty$ are related by
\be\label{formu}
E_*=\frac{t_\infty E_\infty}{t_*}\,,\qquad t_*=\frac{t_\infty \ell_*}{\ell_\infty}\,,
\ee
and so on with momenta. The origin of these formul\ae\ was left unexplained in \cite{frc12,frc13}, but we can understand them better by a simple observation \cite{CaRo2a,CaRo2b}. The origin of $\ell_\infty^\mu$ is a partition of the scales in fractional complex measures. As we will see in \lref{14}, the general real-valued leading-order solution of the flow equation has terms of the form $|x^\mu/\ell_*^\mu|^{\a+\rmi\om}\pm|x^\mu/\ell_*^\mu|^{\a-\rmi\om}$. Splitting $|x^\mu/\ell_*^\mu|^{\a\pm\rmi\om}=\la_{(\mu)}|x^\mu/\ell_*^\mu|^\a|x^\mu/\ell_\infty^\mu|^{\pm\rmi\om}$, where $\la_{(\mu)}=(\ell_\infty^\mu/\ell_*^\mu)^{\pm\rmi\om}$ is purely imaginary and $\ell_\infty^\mu$ is an arbitrary length, we get the log-oscillating measure \Eq{meamu}. If $\la_{(\mu)}=\la$ for all $\mu$ (same partition in all directions) and a space-isotropic hierarchy, we get $(t_*/t_\infty)^{\pm\rmi\om}=\la_{(0)}=\la_{(i)}=(\ell_*/\ell_\infty)^{\pm\rmi\om}$, hence the second equation in \Eq{formu}. On the other hand, the scales $k_*^\mu$ and $k_\infty^\mu$ in momentum space are conjugate to $\ell_*^\mu$ and $\ell_\infty^\mu$, in the sense that $k_*^\mu\propto 1/\ell_*^\mu$ and $k_\infty^\mu\propto1/\ell_\infty^\mu$ with the same proportionality coefficient. This is clear from dimensional arguments but it is made especially rigorous in $T_q$, where the momentum measure \Eq{rdp} is completely determined by asking that the geometric momentum coordinate $p^\mu(k^\mu)$ be conjugate to $q^\mu(x^\mu)$. For each direction, one has 
\be\label{mompk}
p^\mu(k^\mu)=\frac{1}{q^\mu(1/k^\mu)}\,,
\ee
where all scales $\ell_n^\mu$ in \Eq{meag} are replaced by energy-momentum scales $k_n^\mu$ \cite{frc11,frc14}. Therefore, in the case of a binomial space-isotropic measure we have $k_*^\mu\ell_*^\mu=k_\infty^\mu\ell_\infty^\mu$, which reduces to the first equation in \Eq{formu} for $\mu=0$.

Having understood eq.\ \Eq{formu}, one recalls that log-oscillating measures provide an elegant extension of noncommutative $\kappa$-Minkowski spacetime and explain why the Planck scale does not appear in the effective measure thereon \cite{ACOS} (see also question \lref{45}). In turn, this connection suggests that the fundamental scales in the log oscillations coincide with the Planck scales:
\be\label{infpl}
t_\infty=t_\Pl\,,\qquad \ell_\infty=\ell_\Pl\,,\qquad E_\infty=k_\infty=\ep=\mpl c^2\,.
\ee
In four dimensions, $t_\Pl =\sqrt{\hbar G/c^3}\approx 5.3912 \times 10^{-44}\,\text{s}$, $\ell_\Pl=\sqrt{\hbar G/c^5} \approx 1.6163 \times 10^{-35}\,\text{m}$, and $m_\Pl=\sqrt{\hbar c/G} \approx 1.2209\times 10^{19}\,\text{GeV}c^{-2}$. Remarkably, eq.\ \Eq{infpl} connects, via Newton constant, the prefixed multiscale structure of the measure with the otherwise independent dynamical part of the geometry. Also, it makes the log-oscillating part of multiscale measures ``intrinsically quantum'' in the sense that Planck constant $\hbar=h/(2\pi)$ appears in the geometry. An interesting follow-up of this concept will be seen in \lref{27}.

With eqs.\ \Eq{formu} and \Eq{infpl}, one reduces the number of free scales of the binomial measure \Eq{meamu} with \Eq{binom2} to one: $t_*$ or $\ell_*$ or $E_*$.
\item[--] The real amplitudes $A$ and $B$ can be set to be non-negative, since they multiply trigonometric functions. Also, they must be no greater than 1 in order to avoid negative distances \cite{trtls}. Therefore,
\be\label{ABpr}
0\leq A,B\leq 1\,.
\ee
\item[--] The frequency $\om$ stands out with respect to the other parameters because it takes a discrete set of values. As mentioned in \loref{04}, the measures \Eq{meag} and \Eq{meamu} can be derived from a pure calculation in fractal geometry. The geometry of the measure without log oscillations is a random fractal, namely, a fractal endowed with symmetries whose parameters are randomized each time they are applied over the set \cite{frc2,RLWQ}. On the other hand, the measure with logarithmic oscillations corresponds to a deterministic fractal where the symmetry parameters are fixed (see footnote \ref{fu1}). For the binomial measure \Eq{meamu} with \Eq{binom2}, $\a_0=\a$, and only one frequency $\om>0$, the underlying fractal $\cF=\otimes_\mu \cF_\mu$ is given, for each direction, by the union of $N$ copies of itself rescaled by a factor $\la_\om=\exp(-2\pi/\om)$ at each iteration. Since the capacity of $\cF_\mu$ is equal to the Hausdorff dimension and reads $\dc=-\ln N/\ln\la_\om=\dh=\a$, the number of copies is $N=\exp(-\a\ln\la_\om)=\exp(2\pi\a/\om)$. $N$ is a positive integer, so that $\om$ can only take the irrational values\footnote{Here we discover that, for consistency, we can have $\om_\mu=\om$ for all $\mu$ only if the measure \Eq{meamu} is isotropic, $\a_\mu=\a$ for all $\mu$. This piece of information has never been used in the literature but it does not affect observations much.}
\be\label{omN}
\om=\om_N:=\frac{2\pi\a}{\ln N}\,.
\ee
For $\a=1/2$ and $N=2,3,\ldots$, we have $\la_\om=1/N^2$ and $\om_2\approx 4.53> \om_3\approx 2.86>\dots$. The case $N=1$ is not a fractal [eq.\ \Eq{omN} is ill defined then], while for each $N$ one has a different fractal in the same class \cite{trtls}. Overall, the prior on $\om$ is
\be\label{ompr}
0<\om<\om_2=\frac{2\pi\a}{\ln 2}\,,
\ee
with irrational values picked in between.
\end{itemize}

\bigskip

\lolabel{08} \textbf{\emph{Are theories on multifractional spacetimes predictive and falsifiable? The reason of this concern is the presence of a largely arbitrary element, the measure profiles $q^\mu(x^\mu)$. Their choice is dictated only by mathematics (in particular, by multifractal geometry) and by very general properties of dimensional flow, but not by physics and physical observations or experiments. In most papers of the subject, the simplest form \Eq{meamu} with \Eq{binom2} of the measure is chosen, but still mathematically infinitely many other measures are possible which satisfy criteria of fractal geometry. The ambiguity in the selection of the measure is equivalent to having infinitely many parameters of the theory and this renders the theory nonpredictive. Nothing prevents one from using polynomial distributions or multiple logarithmic oscillations, such as in the measure \Eq{meag}. The criterion of subjective simplicity should never be used to substitute the requirement of physical predictability. Since the measure $q(x)$ is part of the definition of multifractional spacetimes, it cannot be verified and tested physically. Or, in other words, it can always be fine-tuned to correctly reproduce any phenomenological data. This means that these theories are not falsifiable.}}\addcontentsline{toc}{subsection}{\texorpdfstring{\loref{08}}{} Are multifractional theories predictive and falsifiable?}

We already answered to this in \loref{07}. Theories with the binomial measure \Eq{meamu} are representative for the derivation of phenomenological consequences of the whole class of theories on multifractional spacetimes. No matter what the detailed behaviour of the most general measure \Eq{meag} is, the physical consequences are universal and the theory is back-predictive. 

Furthermore, the ranges \Eq{ranga}, \Eq{ABpr}, and \Eq{ompr} of the values of the free parameters in \Eq{meamu} with \Eq{binom2} is so limited that it is extremely easy to falsify the theory and exclude large portions of the parameter space $(\a_0,\a,t_*,\ell_*,t_\infty,\ell_\infty,A,B,\om)$, as done by observations of gravitational waves \cite{qGW} and of the cosmic microwave background (CMB) \cite{frc14}.

\bigskip

\llabel{09} \textbf{\emph{What is the physical motivation of the choice of measure? I agree that, once the measure is chosen, the theory is fully predictive and experimental consequences can be derived. The problem, however, is how to predict such measure in the first place, based on physical considerations. If a measure $q(x)$ is fixed, then predictability and falsifiability are recovered, but then the new question is to physically motivate the choice of $q(x)$. I view its lack as the big drawback of this class of theories.}}\addcontentsline{toc}{subsection}{\lref{09} What is the physical motivation of the choice of measure?}

This type of remark, redundant with \loref{07} and \loref{08}, used to arise before the formulation of the flow-equation theorems \cite{first}. It is true that general theories of multifractional spacetimes with measure $q(x)$ unspecified lack predictability and falsifiability, but the same could be said about the general framework of ``quantum field theory'' with arbitrary interactions as opposed to the very concrete Standard Model. In our case, the measure $q(x)$ is given by \Eq{meag} or its approximation \Eq{meamu}, which is the general factorizable solution of eq.\ \Eq{flow}. Any other measure corresponds to different regimes of the general expression \Eq{meag}. The physical mechanism which determines the measure is precisely this flow equation (almost constant dimension in the IR) and it agrees completely with the arguments and calculations in fractal geometry invoked in early papers. We say ``physical'' rather than ``geometric'' because the geometry expressed by dimensional flow has a direct impact on physical observables.

\bigskip

\llabel{09b} \textbf{\emph{You said that the binomial measure captures all the main properties of a multifractal geometry. Can you illustrate that in a pedagogical way?}}\addcontentsline{toc}{subsection}{\texorpdfstring{\lref{09b}}{} Can you illustrate the multifractal properties of the binomial measure in a pedagogical way?}

\begin{figure}
\centering
\includegraphics[width=8.4cm]{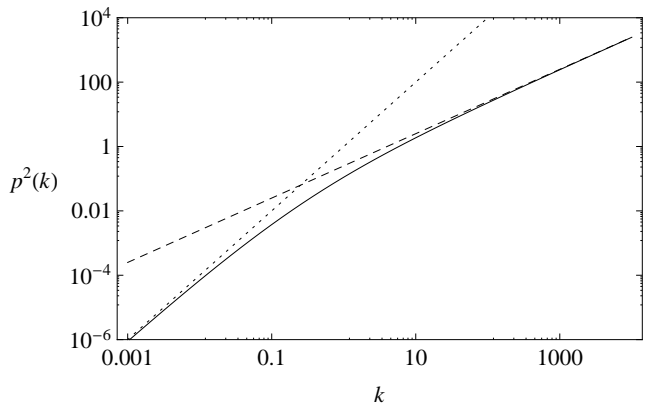}
\caption{\label{fig2} The binomial distribution \Eq{p2bin} of the Laplacian eigenvalues (solid curve) corresponding to a bifractal, compared with the ordinary distribution $k^2$ (usual Laplacian, standard geometry, dotted line) and with the fractional distribution $-|k|^{2\a}$ corresponding to a monofractal (dashed line). Here $k_*=1$ and $\a=1/2$.}
\end{figure}

Consider the theory $T_q$ with binomial measure \Eq{meamu} with $F_\om=1$. From eq.\ \Eq{mompk}, we get the measure in momentum space
\be\label{binomp}
p^\mu(k^\mu)=k^\mu\left[1+\frac{1}{\a_\mu}\left|\frac{k^\mu}{k_*^\mu}\right|^{1-\a_\mu}\right]^{-1}\,.
\ee
The eigenvalue equation of the Laplace--Beltrami operator $\B_q$ in eq.\ \Eq{Kss} is $\B_q\bE(k,x)=-p^2(k)\,\bE(k,x)$, where $\bE(k,x)=\exp[\rmi q_\mu(x_\mu) p^\mu(k^\mu)]$ and $p^2=p_\mu p^\mu$. In one dimension, this means that the spectrum of $-\p_q^2$ follows the distribution
\be\label{p2bin}
p^2(k)=k^2\left[1+\frac{1}{\a}\left|\frac{k}{k_*}\right|^{1-\a}\right]^{-2}.
\ee
(Including log oscillations, we would get the same spectrum but with a periodic modulation.) We can compare this distribution with the ordinary spectrum $k^2$ and with the distribution $|k|^{2\a}$ of a purely fractional measure (obtained by removing the factor 1 in eq.\ \Eq{p2bin} or by taking a fractional Laplacian \cite{frc4}). As one can appreciate from figure \ref{fig2}, the binomial profile \Eq{p2bin} interpolates between the fractional and the integer spectra.

The spectral distribution \Eq{p2bin} plotted in the figure is an idealization (but a faithful one) of what is found in actual experiments or observations involving multifractals, not only in physics, but also in fields of research as diverse as geology, ethology, and human cognition \cite{GiTM,Kel10}. Adding an extra power law to the binomial measure (i.e., considering a trinomial measure with two scales $\ell_1>\ell_2$), one would bend the right part of the solid curve in the figure towards a different asymptote. And so on.

\newpage

%\bigskip

\llabel{10} \textbf{\emph{Are multifractional theories Lorentz invariant?}}\addcontentsline{toc}{subsection}{\lref{10} Are multifractional theories Lorentz invariant?}

No, they are not because factorizable measures \Eq{facto} explicitly break rotation and boost invariance. They are not Poincaré invariant either, because they also break translations. An early nonfactorizable version $\tilde T_1$ of multifractal theories proposed a Lorentz-invariant measure, working on the assumption that keeping as many Lorentz symmetries as possible would lead to viable phenomenology \cite{fra1,fra2,fra3}. However, problems in finding an invertible Fourier transform associated with a self-adjoint momentum operator soon paved the way to the factorizable \emph{Ansatz} \Eq{facto}, as described in \loref{07}. As a consequence, the Poincaré symmetries
\be\label{poinor}
{x'}^\mu=\Lambda_\nu^{\ \mu}x^\nu+b^\mu
\ee
of standard field theory on Minkowski spacetime are not enjoyed by multifractional field theories.

\bigskip

\llabel{11} \textbf{\emph{Then, what are their local symmetries?}}\addcontentsline{toc}{subsection}{\lref{11} What are their local symmetries?} 

The symmetries of the dynamics depend on the structure of the action. Consider first the case without gravity (gravity will be included in question \lref{12}). All multifractional theories have the same measure $\rmd^Dq(x)$ invariant under the nonlinear \emph{$q$-Poincaré transformations} $\rmd^Dq(x)\to \rmd^Dq(x')$, where for each individual $q^\mu(x^\mu)$
\be\label{qlort} 
{q}^\mu({x'}^\mu)=\Lambda_\nu^{\ \mu}q^\nu(x^\nu)+a^\mu\,,
\ee
$\Lambda_\nu^{\ \mu}$ are the usual Lorentz matrices, and $a^\mu$ is a constant vector. Seen as a change on the geometric coordinates $q^\mu$, this looks like a standard Poincaré transformation. Seen as a transformation on the coordinates $x^\mu$, it is highly nonlinear and, in general, noninvertible. Looking at eq.\ \Eq{meag}, we cannot write $x^\mu(q^\mu)$ explicitly, unless we ignore log oscillations.

The $q$-Poincaré transformations \Eq{qlort} are a symmetry of the measure but, in general, not of the dynamics. Multifractional theories may still be invariant under other types of symmetries, which typically are a deformation of classical Poincar\'e and diffeomorphism symmetries. Before discussing that, it is useful to recall a few basic facts on symmetry algebras.

Ordinary Poincaré symmetries are defined in three mutually consistent manners: as coordinate transformations, as an algebra of operators on a vector space, and as an algebra of field operators. Meant as coordinate transformations, they are defined by eq.\ \Eq{poinor}. At the level of operators on a vector space, they are defined by an infinite-dimensional representation of operators $\hat P_\mu=\hat p$ and $\hat J=\hat \jmath$ satisfying the undeformed Poincar\'e algebra
\bs\label{poin}\ba 
&& [\hat P_\mu,\hat P_\nu]=0\,,\label{PP}\\
&& [\hat P_\mu,\hat J_{\nu\rho}]=\rmi(\eta_{\mu\rho}\hat P_\nu-\eta_{\mu\nu}\hat P_\rho)\,,\label{PJ}\\
&& [\hat J_{\mu\nu},\hat J_{\s\rho}]=\rmi (\eta_{\mu\rho}\hat J_{\nu\sigma}-\eta_{\nu\rho}\hat J_{\mu\sigma}+\eta_{\nu\s}\hat J_{\mu\rho}-\eta_{\mu\s}\hat J_{\nu\rho})\,.\nonumber\\ \label{JJ}
\ea\es
Ordinary time and spatial translations are generated by $\hat p_\mu:=-\rmi\p_\mu$, while ordinary rotations and boosts are generated by $\hat\jmath_{\mu\nu}:=x_\mu \hat p_\nu-x_\nu\hat p_\mu$. The mass and spin of a particle field can be defined by finding first a vector space where the operators $\hat P$ and $\hat J$ act upon, and then the eigenstates of $\hat P^2$ and $\hat W^2$ (where $\hat W^\mu=\e^{\mu\nu\rho\s}\hat P_\nu \hat J_{\rho\s}/2$ is the Pauli--Lubanski pseudovector). For a local relativistic theory, there is the further requirement that such vector space be invariant under representations of $\hat P$ and $\hat J$. At the level of field operators, ordinary Poincaré symmetries are encoded in some operators (without hat) $P_\mu=P_\mu[\phi^i]$ and $J_{\mu\nu}=J_{\mu\nu}[\phi^i]$ obeying the algebra \Eq{poin} where commutators $[\,\cdot\,,\,\cdot\,]$ are replaced by Poisson brackets $\{\,\cdot\,,\,\cdot\,\}$:
\bs\label{poin2}\ba 
&& \{P_\mu,\hat P_\nu\}=0\,,\label{PP2}\\
&& \{P_\mu,\hat J_{\nu\rho}\}=\rmi(\eta_{\mu\rho}P_\nu-\eta_{\mu\nu}P_\rho)\,,\label{PJ2}\\
&& \{J_{\mu\nu},\hat J_{\s\rho}\}=\rmi (\eta_{\mu\rho}J_{\nu\sigma}-\eta_{\nu\rho}J_{\mu\sigma}+\eta_{\nu\s}J_{\mu\rho}-\eta_{\mu\s}J_{\nu\rho})\,.\nonumber\\ \label{JJ2}
\ea\es

In quantum gravity (including noncommutative spacetimes) and in multifractional classical theories, quantum and/or multiscale effects (in quantum gravity, multiscale effects are quantum by definition) can deform the above algebra of generators $A_i=\hat p_\mu,\hat \jmath_{\mu\nu}$ in two ways. One is by deforming the generators $A_i\to A_i'$, which corresponds to a deformation of classical symmetries. For instance, one can have a momentum operator $A_i'=\hat P_\mu$ which generates a symmetry $x^\mu\to f(x^\mu)$ analogous to the usual spatial and time translations $x^\mu\to x^\mu+b^\mu$ generated by $\hat p_\mu$, such that $f(x_i)\simeq x_i+b_i$ when quantum corrections are negligible. In this case, one regards $\hat P^\mu$ as the generator of ``deformed translations.'' The other way in which an algebra is deformed is by a change in its structure. For instance, given an algebra $\{A_i,A_j\}=f_{ij}^kA_k$ in ordinary spacetime or in a classical gravitational theory, one might end up with an algebra $\{A_i',A_j'\}=F(A_k')$ in a multifractional or quantum theory, which can be written also in terms of the generators of the classical symmetries, $\{A_i,A_j\}=G(A_k)$, for some $G\neq F$. Depending on the specific multifractional theory, we can have no symmetry algebra at all (case $T_1$), a symmetry algebra with deformed operators and deformed structure (case $T_v$), or a symmetry algebra with deformed operators but undeformed structure (case $T_q$). Question \lref{46} retakes the topic of deformed algebras in the context of gravity.

Let $\phi^i$ be a generic family of matter fields (scalars, gauge vectors, bosons, and so on) and let $S[{\rm weight},{\rm derivatives},\phi^i]$ be the action of the theory with a specific choice of measure weight \Eq{vfacto} and of derivatives in kinetic terms.
\begin{itemize}
\item[--] In the model $T_1$ with ordinary derivatives, the Lagrangian is defined exactly as the usual one, for a scalar, for the Standard Model, and so on. As an example, for a scalar field the action
\be\label{S1}
S_1[v,\p,\phi^i]=\int\rmd^Dx\,v(x)\,\cL_1[\p,\phi^i]
\ee
is eq.\ \Eq{Sphi} with $\cK=\B$. Therefore, the Lagrangian $\cL_1$ is invariant under ordinary Lorentz transformations but the action $S_1$ is not. Since the profiles $q^\mu(x^\mu)$ are given \emph{a priori} by the flow equation (or by fractal geometry), the dynamics will not enjoy any symmetry at all. In other words, the structure of the geometric coordinates $q^\mu(x^\mu)$ is irreconcilable with that of the differential operators $\p_\mu$. Said in a more formal way, the operator $\hat p_\mu$ generating ordinary translations is not self-adjoint \cite{frc7} with respect to the natural inner product on the space of test functions defined on multifractional Minkowski space:
\be
(f_1,\hat p_\mu f_2):=\int_{-\infty}^{+\infty}\rmd^Dq(x)\,f_1(x)\,\hat p_\mu\,f_2(x)\neq (\hat p_\mu f_1,f_2)\,.
\ee
Consequently, the system is not translation invariant and ordinary momentum is not conserved. The proof for rotations and boosts is similar. This absence of symmetries is clearly a problem of this theory. Notice that one can define a self-adjoint momentum operator 
\be\label{hatPmu}
\hat P_\mu:=-\frac{\rmi}{2}\,\left[\p_\mu+\frac{1}{v}\p_\mu(v\,\cdot\,)\right]=-\rmi\left(\p_\mu+\frac{\p_\mu v}{2v}\right)\,,
\ee
but this is equivalent to the momentum operator in $T_v$. As a matter of fact, $T_v$ was born as the ``upgrade'' of $T_1$ and we should talk about three multifractional theories ($T_v$, $T_q$, and $T_\g$) rather than four. For this reason, we regard $T_1$ only as a phenomenological model, in the sense of being inspired by the multiscale principle without the pretension of being a rigorous theoretical construct. %Therefore, we will not attempt to fill many of the gaps of knowledge about $T_1$ (such as the details of the QFT therein).
\item[--] In the theory $T_v$ with weighted derivatives, field Lagrangians are defined by replacing standard operators $\p_\mu$ with the weighted derivatives defined in eq.\ \Eq{Ks}:
\be\label{Sv}
S_v[v,\cD,\phi^i]=\int\rmd^Dx\,v(x)\,\cL_v[\cD,\phi^i]\,.
\ee
The scalar-field example is eq.\ \Eq{Sphi} with $\cK=\cD_\mu\cD^\mu$. The action of the Standard Model of electroweak and strong interactions can be found in ref.\ \cite{frc13} and in question \lref{29}. Just like in the case with ordinary derivatives, the symmetry structure of the measure and of the operators $\cD_\mu$ is different and $S_v$ is not invariant under standard Poincaré symmetries. However, contrary to $T_1$, the theory $T_v$ is invariant under new symmetries encoded in deformed Poincaré transformations, but only in the absence of nonlinear interactions of third or higher order in at least one field. Let us explain. The weighted derivative defines the operator $\hat P_\mu:=-\rmi\cD_\mu$ [clearly equivalent to eq.\ \Eq{hatPmu}], which is self-adjoint:
\ba
(f_1,\hat P_\mu f_2)&:=&-\rmi\int_{-\infty}^{+\infty}\rmd^Dq\,f_1\,\cD_\mu\,f_2=-\rmi\int_{-\infty}^{+\infty}\rmd^Dx\,v\,f_1\,\cD_\mu\,f_2\nonumber\\
&=&-\rmi\int_{-\infty}^{+\infty}\rmd^Dx\,\sqrt{v}\,f_1\,\p_\mu(\sqrt{v}f_2)=\rmi\int_{-\infty}^{+\infty}\rmd^Dx\,\sqrt{v}\,\p_\mu(\sqrt{v}f_1)\,f_2\nonumber\\
&=&(\hat P_\mu f_1,f_2)\,.
\ea
This operator generates ``fractional'' translations rather than ordinary ones. The transformation law of fields can be worked out explicitly \cite{frc6,frc13}, but here suffice it to note that a field redefinition
\be\label{vpi}
\vp^i:=\sqrt{v}\,\phi^i
\ee
permits to write fractional expressions as ordinary ones, e.g., $\hat P_\mu\phi^i=v^{-1/2}\hat p_\mu\vp^i$. The same holds for the generators of rotations and boosts. Thus, it is possible for the theory to be invariant under \emph{weighted Poincaré transformations} (deformed translations, rotations, and boosts) generated by
\be\label{PJwei}
T_v\,:\qquad \hat P_\mu:=-\rmi\cD_\mu=\frac{1}{\sqrt{v}}\,\hat p_\mu\,\sqrt{v}\,,\qquad \hat J_{\nu\rho}:=x_\nu \hat P_\rho-x_\rho\hat P_\nu= \frac{1}{\sqrt{v}}\,\hat\jmath_{\nu\rho}\,\sqrt{v}\,,
\ee
which satisfy the undeformed algebra \Eq{poin} or its field-operator equivalent \Eq{poin2}, but only if the action has no third- or higher-order terms in one or more fields. If it does, then the algebraic structure \Eq{poin} and \Eq{poin2} is deformed. In the scalar-field example, it is easy to show that, given the Hamiltonian and spatial momentum (here $i=1,\dots,D-1$ are spatial directions)
\bs\label{hpo}\ba
H &:=& P^0 = \int\rmd^{D-1}{\bf x}\,v({\bf x})\,\left[\frac12(\cD_t\phi)^2+\frac12\cD_i\phi\cD^i\phi+V(\phi)\right]\,,\\
P^i &=&-\int\rmd^{D-1}{\bf x}\,v({\bf x})\,\cD_t\phi\cD^i\phi\,,
\ea\es
one has \cite{frc6}
\be\label{PH}
\{P^i,H\} =\int\rmd^{D-1}{\bf x}\,\p^i v({\bf x})\left[\frac12 \phi\,V_{,\phi}(\phi)-V(\phi)\right]\,,
\ee 
which vanishes only if $V\propto\phi^2$. Therefore, eq.\ \Eq{PP2} is violated and the Poincaré algebra is deformed not only in the form of the generators, but also in its structure. Similar violations occur in eqs.\ \Eq{PJ2} and \Eq{JJ2}.

It is important to distinguish between the symmetries of a generic field theory with weighted derivatives and the specific field theory describing natural phenomena. In the second case, the theory $T_v$ \emph{is} invariant under weighted Poincaré transformations. The ${\rm SU}(3)\otimes {\rm SU}(2)\otimes {\rm U}(1)$ Standard Model of electroweak and strong interactions has been constructed in ref.\ \cite{frc13} for the theory $T_v$. The only nonlinear terms arising are those of gauge derivatives and in the Higgs potential. The first type is of the form $\bar\psi A_\mu\g^\mu\psi$, linear in gauge vectors and quadratic in fermions (see \lref{29}), so that all spacetime dependence can be reabsorbed in field and couplings redefinitions. Since there is no $O[(\phi^i)^3]$ or higher-order term, the structure of the Poincaré algebra is undeformed (although the generators are). The Higgs potential does have third- and fourth-order terms (again, see \lref{29}), but their measure dependence is reabsorbed in the fields and in the couplings. The crucial point here, which solves the apparent contradiction with eq.\ \Eq{PH}, is that not only fields can be redefined, but also the physical couplings \cite{frc13}.
\item[--] In the theory $T_q$ with $q$-derivatives, the action is defined by taking the ordinary action (of a scalar field, of the Standard Model, of gravity, and so on) and replacing all coordinates $x^\mu$ therein with the geometric coordinates $q^\mu(x^\mu)$:
\be\label{Sq}
S_q[v,\p_q,\phi^i]=\int\rmd^Dx\,v(x)\,\cL_q[\p_{q(x)},\phi^i]=S[v,v^{-1}\p_x,\phi^i]\,.
\ee
Clearly, the theory is invariant under $q$-Poincaré transformations \Eq{qlort}. The symmetry algebra is undeformed, eq.\ \Eq{poin} with (obviously, no Einstein summation)
\be\label{PJq}
T_q\,:\qquad \hat P_\mu:=-\rmi\p_{q^\mu}=\frac{1}{v_\mu}\,\hat p_\mu\,,\qquad \hat J_{\nu\rho}:=x_\nu \hat P_\rho-x_\rho\hat P_\nu\,,
\ee
where $\p_{q^\mu}=\p/\p q^\mu(x^\mu)=[v_\mu(x^\mu)]^{-1}\p_\mu$. These operators are quite different from the $T_v$ case \Eq{PJwei} but, just like that, they describe deformed translations, rotations, and boosts.
\item[--] In the theory $T_\g$ with fractional derivatives, the action sports fractional derivatives (or differintegrals) ``$\p^\g$,'' for which there are many available definitions in the literature (see \cite{frc1} for a review and \cite{KST} for a textbook on the subject). For example, in refs.\ \cite{frc1,frc2} the left and right Caputo derivatives were preferred among other choices to define $T_{\g=\a}$, because of the possibility to define geometric coordinates such that $\p^\a_\mu q^\nu=\de_\mu^\nu$; later on, the Liouville and Weyl derivatives were chosen in the second definition of eq.\ \Eq{Kss}, since they are one the adjoint of the other \cite{frc4}. Omitting the $\mu$ index everywhere, along the $\mu$ direction the Liouville and Weyl derivatives are
\bs\label{liowey}\ba
{}_\infty\p^\a_x f(x) &:=& \frac{1}{\Gamma(1-\a)}\int_{-\infty}^{+\infty}\rmd x'\, \frac{\theta(x-x')}{(x-x')^\a}\p_{x'}f(x')\,,\qquad 0\leq \a<1\,,\\
{}_\infty\bar\p^\a_x f(x) &:=& \frac{1}{\Gamma(1-\a)}\int_{-\infty}^{+\infty}\rmd x'\, \frac{\theta(x'-x)}{(x'-x)^\a}\p_{x'}f(x')\,,\qquad 0\leq \a<1\,,
\ea\es
where $\theta$ is the Heaviside step function. %. Assuming $x>0$ for the sake of illustration (a condition that can be relaxed after some not too laborious work \cite{frc2,frc4}) and omitting the $\mu$ index everywhere, along the $\mu$ direction the left Caputo derivative is
%\be\label{capu}
%\p^\a_x f(x) := \frac{1}{\Gamma(1-\a)}\int_0^x \frac{\rmd x'}{(x-x')^\a}\p_{x'}f(x')\,,\qquad 0\leq \a<1\,.
%\ee
%Then, for $q=q_\a:=x^\a/\Gamma(\a+1)$ one has $\p^\a q=1$, a property of calculus bringing a tremendous number of advantages 
In particular, one can consider the combination
\ba
\tilde\cD^\a_x:=\frac12({}_\infty\p^\a_x+{}_\infty\bar\p^\a_x) &=&\frac{1}{2\Gamma(1-\a)}\int_{-\infty}^{+\infty}\rmd x'\,\left[\frac{\theta(x-x')}{(x-x')^\a}+\frac{\theta(x'-x)}{(x'-x)^\a}\right]\p_{x'}\nonumber\\
&=&\frac{1}{2\Gamma(1-\a)}\int_{-\infty}^{+\infty}\frac{\rmd x'}{|x-x'|^\a}\p_{x'}\,.\label{tidel}
\ea
Since the definition of ${}_\infty\p^\a=I^{1-\a}\p$ is inspired by the Cauchy formula for the $n$-repeated integration $I^n$, when $\a\to 1$ one obtains the ordinary derivative $\p_x$ in both the Liouville and Weyl case; this explains the prefactor $1/2$ in eq.\ \Eq{tidel}, $\tilde\cD^1_x=\p_x$. Caputo left and right derivatives are defined as in eq.\ \Eq{liowey} but with integration domains $(0,+\infty)$ and $(-\infty,0)$, respectively.

The theory $T_\g$ is invariant under $q$-Lorentz transformations [eq.\ \Eq{qlort} with $a^\mu=0$] but, contrary to the $T_q$ case, only up to boundary terms and only at individual plateaux in dimensional flow (i.e., in the toy-model limit of no-scale fractional geometries, pure power-law measure $q\propto |x|^\a$ \cite{frc1,frc2}). Therefore, $q$-Lorentz invariance is exact in $T_q$ (and extendable to $q$-Poincaré invariance) but approximate in $T_\g$, and the phenomenology of the two theories is thus expected to be more similar than between $T_\g$ and $T_v$ or $T_1$. To show this, one must first extend the definition of fractional derivatives to multiscale geometries. This can be done in two ways, which we will explore in greater detail in the future. One is to implement multiscaling ``externally'' with respect to the definition of fractional derivatives. In this case, one defines a superposition of fractional derivatives (indices $\mu$ inert, as usual)
\be\label{dissum}
\tilde\cD_\mu:=\sum_n g_{\mu,n}\tilde\cD^{\a_{\mu,n}}_\mu,
\ee
where the scale hierarchy appears in the coefficients $g_{\mu,n}(\ell_n^\mu)$. This definition of multifractional derivative is self-adjoint in the scalar product with integration $\rmd x^\mu$. To have it self-adjoint with integration measure $\rmd q^\mu=\rmd x^\mu v_\mu(x^\mu)$, it is sufficient to decorate eq.\ \Eq{dissum} with weight factors, so that the operator $\cK_\g$ in eq.\ \Eq{Kss} becomes
\be\label{Kssg}
\cK_\g=\frac{1}{\sqrt{v}}\tilde\cD_\mu\tilde\cD^\mu\left(\sqrt{v}\,\cdot\,\right)\,.
\ee
Expression \Eq{dissum} is similar to the so-called distributed-order fractional derivatives D \cite{Cap69,Cap95,BT1,BT2,CGS,LoH02,Koc1,Koc2}, where an integration over the parameter $\a$ is performed instead of the sum: ${\rm D}:=\int_0^1\rmd\a\,m(\a)\,\p^\a$, where $m(\a)$ is a distribution on the interval $[0,1]$. We do not know whether a continuous distribution would be more convenient that the discrete sum \Eq{dissum}. In either case, a global notion of $q$-Lorentz invariance does not exist, and one can only count on a forcefully approximated ``local'' version of the symmetries \Eq{qlort}. It may be that other ``fractional Poincaré'' symmetries are enforced, but we have not checked it yet.

The other possibility is to have an ``internal'' notion of multiscaling, that is, we can modify the definition of fractional derivatives so that to include the scale hierarchy within. To generalize eq.\ \Eq{tidel} to multiscale profiles such as \Eq{meag} or \Eq{meamu}, we define (index $\mu$ restored and not summed over)
\be\label{capuq}
{}_q\cD_\mu:=\int_{-\infty}^{+\infty}\frac{\rmd {x'}^\mu}{q^\mu(x^\mu-{x'}^\mu)}\frac{\p}{\p {x'}^\mu}\,.
\ee
This expression is strikingly similar to a not much known proposal for so-called variable-order fractional derivatives \cite{LoH02}. The main difference is that, for us, the distribution $q(x-x')$ is determined from the start by the second flow-equation theorem. At plateaux in dimensional flow, eq.\ \Eq{capuq} reduces to the mixed Liouville--Weyl derivative \Eq{tidel} and there is a manageable fractional calculus in all regions of interest, i.e., in the deep UV, in the IR, and in whatever intermediate region the dimension of spacetime is approximately constant (we expect dimension gradients to be difficult or even impossible to detect). Equation \Eq{capuq} should be further explored to clarify the role of boundary terms and discontinuities.\footnote{The junction of the left and right derivatives in eq.\ \Eq{tidel} masks a potentially tricky point at $x=0$ \cite{frc1}.} In particular, one will have to show that ${}_q\cD_\mu q^\mu\simeq 1$ at least at each plateau in dimensional flow. The constant coefficients inside $q^\mu$ should be chosen so that ${}_q\cD\to \p$ at large scales (roughly speaking, in the $q\to\mathbbm{1}$ limit) and ${}_q\cD\simeq\tilde\cD^{\a_n}$ at the $n$-th plateau. 

Weight factors can appear to the left and to the right of \Eq{capuq} to guarantee self-adjointness with respect to the measure $\rmd^D q(x)$ \cite{frc4}. An alternative to eq.\ \Eq{Kssg} is thus
\be\label{Kssg2}
\cK_\g=\frac{1}{\sqrt{v}}\,{}_q\cD_\mu\,{}_q\cD^\mu\left(\sqrt{v}\,\cdot\,\right),
\ee
and the general action of the theory $T_\g$ in the absence of gravity is
\be\label{Sa1}
S_\g[v,{}_q\cD,\phi^i]=\int\rmd^Dq(x)\,\cL_\g[{}_q\cD,\phi^i]\,,
\ee
which has no integer picture associated. A tentative proposal for the Lagrangian of a scalar field is, modulo weight factors, $\cL_\g[\phi]=-(1/2) {}_q\cD_\mu\phi{}_q\cD^\mu\phi-V(\phi)$. Fractional derivatives, either of fixed order as eq.\ \Eq{tidel} or multiscale as in eqs.\ \Eq{dissum} and \Eq{capuq}, have a technical complication which is one of the reasons why the dynamics of $T_\g$ has not been studied adequately so far: the Leibniz rule $\p^\g(fg)=(\p^\g f) g+f(\p^\g g)+\dots$ is rather messy in the ``\dots'' part and it complicates the equations of motion (after integrating by parts to calculate the field variations). Therefore, the kinetic term $-({}_q\cD\phi)^2/2$ is not equivalent to $\phi\cK_\g\phi/2$. This issue will be tackled in a separate publication.

Another, more formal way to get a multifractional derivative is via the differentials of the theory. In $T_\g$, the exterior derivative $\rmd$ can be replaced by a new definition $\bd$ which was proposed in ref.\ \cite{frc1} for a fractional measure $q\propto |x|^\a$. Instead of repeating that discussion, we extend it directly to multiscale geometries and define $\bd$ implicitly by
\be
\bd q^\mu(x^\mu)=q^\mu(\rmd x^\mu)\,,
\ee
so that $\bd q\sim\bd x+\bd x^\a+\ldots=\rmd x+(\rmd x)^\a+\dots$. The line element in geometric notation is
\be
\bd q(s)=\sqrt{g_{\mu\nu} \bd q^\mu(x^\mu)\otimes\bd q^\nu(x^\nu)}\,,
\ee
or, in fractional notation with ordinary differential,
\be
q(\rmd s)= \sqrt{g_{\mu\nu} q^\mu(\rmd x^\mu)\otimes q^\nu(\rmd x^\nu)}\,.
\ee
The most natural multifractional derivative in this formalism is
\be\label{capuq2}
\mathbbm{D}_\mu:=\frac{\bd}{\bd q^\mu}\,.
\ee
The $T_{\g=\a}\cong T_q$ approximation corresponds to $\bd\simeq\rmd$ and $\mathbbm{D}_\mu\simeq \rmd/\rmd q^\mu(x^\mu)$. 

In the case of eq.\ \Eq{capuq2}, there is no need to insert weight factors and the Laplace--Beltrami operator in flat space is
\be\label{Kssg3}
\cK_\g=\mathbbm{D}_\mu\mathbbm{D}^\mu\,.
\ee
The general action in the absence of gravity is
\be\label{Sa2}
S_\g[v,\mathbbm{D},\phi^i]=\int \rmd^Dq(x)\,\cL_\g[\mathbbm{D},\phi^i]\,,
\ee
where for a scalar field $\cL_\g[\phi]=-(1/2)\mathbbm{D}_\mu\phi\mathbbm{D}^\mu\phi-V(\phi)$ or $\cL_\g[\phi]=-(1/2)\phi\cK_\g\phi-V(\phi)$; operator ordering issues will have to be studied carefully. The integral $\int$ can be replaced by a multiscale ``geometric'' integral $\fint$ generalizing the fractional operator of \cite[eq.\ (3.17)]{frc1}, so that $\int \rmd^Dq(x)=\fint\bd^Dq(x)$ and one can completely recast the system in geometric notation.

Part of future work will also be to see if we can identify ${}_q\cD_\mu$ with eq.\ \Eq{capuq2} at the plateaux of dimensional flow, but we anticipate a positive answer provided ${}_q\cD\simeq\tilde\cD^{\a_n}$ at the $n$-th plateau. In fact, there one has $\bd q\simeq(\rmd x)^{\a_n}$, so that $\mathbbm{D}\simeq \bd/(\rmd x)^{\a_n}=\tilde\cD^{\a_n}$ is the fractional derivative \Eq{tidel} of $\a_n$-th order. Moreover, notice the invariance of definitions \Eq{capuq} and \Eq{capuq2} under translations,
\be\label{capuqtr}
{}_q\cD_{x-\bar x}={}_q\cD_x\,,\qquad \mathbbm{D}_{x-\bar x}=\mathbbm{D}_x\,,
\ee
for which the integration domain on the whole real line in \Eq{capuq} is crucial. Equations \Eq{capuq}--\Eq{capuqtr} are given here for the first time.

To summarize, with the derivative \Eq{capuq} the Lagrangian is invariant under $q$-Lorentz transformation, but clearly not under $q$-Poincaré \Eq{qlort}: ${}_q\cD_x$ is invariant under a translation in $x$ but not under a translation in $q$. On the other hand, a preliminary inspection seems to find that $\mathbbm{D}_x$ is $q$-Poincaré invariant.
\end{itemize}
Poincaré (in the absence of gravity) or local Lorentz symmetry (with gravity, in inertial frames) are restored at large scales and late times, where $q^\mu(x^\mu)\simeq x^\mu$ and the geometry measure becomes the standard Lebesgue measure on a smooth manifold. Whether a residual violation of Lorentz invariance is observable and what constraints on it are, will be the subject of section \ref{phen}.

Other local symmetries of multifractional theories are the gauge symmetries of QFT, which are deformations of the usual gauge invariance in ordinary Minkowski spacetime. These are discussed in ref.\ \cite{frc13}.

\bigskip

\llabel{12} \textbf{\emph{Is diffeomorphism invariance respected in multifractional theories?}}\addcontentsline{toc}{subsection}{\lref{12} Is diffeomorphism invariance respected in multifractional theories?}

No, except in $T_v$ in the absence of matter and in $T_q$. The reason is that the measure weight \Eq{vfacto} is not a scalar field but a fixed coordinate profile. Therefore, any coordinate transformation would change $v(x)$, which is not allowed by the flow equation \Eq{flow} if the measure is imposed to be factorizable. The lack of diffeomorphism invariance in most multifractional theories is not in contradiction with the fact that all of them are covariant. The reason is that covariance and diffeomorphism (in short, diffeo) invariance can be confused without damage in the absence of a nondynamical structure, while they are clearly separated concepts in the presence of such a structure (the measure, in the multifractional case).\footnote{To illustrate the point, we report the general discussion made in ref.\ \cite{frc11} and inspired by \cite{Giu06}. Let $\cM$ be a manifold endowed with some nondynamical structure $\Sigma$, and obeying the equations of motion $F[\phi^i,\Sigma]=0$. Covariance determines that, under a diffeomorphism $f$, the transformed fields $f\cdot \phi^i$ obey equations of motions with transformed nondynamical structure: $F[\phi^i,\Sigma]=0=F[f\cdot \phi^i,f\cdot\Sigma]$. On the other hand, diffeomorphism invariance limits the amount of nondynamical structure: it requires that the same equation of motion be satisfied by the fields and their transforms, $F[\phi^i,\Sigma]=0=F[f\cdot \phi^i,\Sigma]$ (active diffeomorphism), or, equivalently, that any solution $\phi^i$ of the equations of motion is also solution of a different set of equations parametrized by a transformed nondynamical structure, $F[\phi^i,\Sigma]=0=F[\phi^i,f\cdot \Sigma]$ (passive diffeomorphism).}

To answer in more detail, we have to turn gravity on and consider a curved embedding manifold (so far in this review, we have discussed only field theories on flat Minkowski spacetime). For instance, the multifractional action of gravity with a minimally coupled matter scalar field is of the form
\be
S=S[g]+\int\,\rmd^Dx\,v\,\sqrt{-g}\,\cL[\phi]\,,
\ee
where $S[g]$ is the action for the metric (which can be found in ref.\ \cite{frc11} and in question \lref{37} for $T_1$, $T_v$, and $T_q$), $g$ is the determinant of the metric, $\cL[\phi]$ is the Lagrangian in \Eq{Sphi} of the scalar and, everywhere in the total action, indices are contracted with the metric $g_{\mu\nu}$. In $T_1$ and $T_v$, there is no field or metric redefinition absorbing completely the dependence on the trivial measure. Even if one can do so in a Standard-Model matter sector, measure factors pop back in the gravitational action and in any non-Standard-Model matter sector with nonlinear interactions \cite{frc11}. Still, in the case of $T_v$ we can identify the matter sector as the responsible for violating diffeo invariance: in the absence of matter, the algebra of the canonical constraints of gravity is preserved (see \lref{46}) \cite{CaRo1}.

On the other hand, one immediately recognizes that the theory $T_q$ is diffeo invariant under active diffeomorphisms with respect to the geometric coordinates $q^\mu(x^\mu)$, but only in the absence in $S_q[g]$ of geometric Lagrangian terms made purely by the measure weight (question \lref{46}). The gravitational sector of the theory $T_\g$ has not been built yet, and presently we cannot comment on that. However, by analogy with the theory with $q$-derivatives and encouraged by the $T_{\g=\a}\cong T_q$ approximation, it should be possible to generalize the notion of diffeo invariance, at least approximately at the scales corresponding to plateaux in dimensional flow.

\bigskip

\llabel{13} \textbf{\emph{What is the dimension of multifractional spacetimes?}}\addcontentsline{toc}{subsection}{\lref{13} What is the dimension of multifractional spacetimes?}

It is not difficult to compute the dimensions $\dh$, $\ds$, and $\dw$ \cite{frc2,frc4,frc7}. The volume of a $D$-hypercube of size $\ell$ oriented along the Cartesian axes with a corner at $x^\mu=0$ is $\cV(\ell)\propto\prod_\mu q^\mu(\ell)$. Centering the hypercube elsewhere, with a corner at $x^\mu=\bar x^\mu$ would only bring the change $q^\mu(\ell)\to q^\mu(\ell-\bar x^\mu)- q^\mu(\bar x^\mu)$, which does not change the $\ell$ scaling of $\cV$. Using a $D$-ball instead of the cube would lead (up to immaterial centering effects) to $\cV(\ell)\propto\sqrt{\sum_\mu [q^\mu(\ell)]^2}$, again with no new impact on the overall scaling. Writing eq.\ \Eq{meag} evaluated at $x^\mu=\ell$ for all $\mu$ as $q^\mu(\ell)=\ell[1+\sum_n b_{\mu,n}(\ell/\ell_n^\mu)^{\a_{\mu,n}-1}F_n(\ell)]$, the Hausdorff dimension \Eq{dh} is (no index contraction, of course)
\be\label{dhmf}
\dh(\ell)=\sum_\mu\frac{1+\sum_n b_{\mu,n}(\ell/\ell_n^\mu)^{\a_{\mu,n}-1}[\a_{\mu,n}+F_n'(\ell)]}{1+\sum_n b_{\mu,n}(\ell/\ell_n^\mu)^{\a_{\mu,n}-1}F_n(\ell)}\,,
\ee
where $F'_n=\rmd F_n(\ell)/\rmd\ln\ell$. This result is independent of the dynamics and is therefore valid for all multifractional theories. It is easy to convince oneself that this expression has all the properties we would expect in dimensional flow. In the IR, $\dh\simeq D$, while at the $n$-th plateau $\dh\simeq \sum_\mu\a_{\mu,n}$. Taking only $n=1$ and one scale $\ell_1=\ell_*$ for all directions [binomial measure \Eq{meamu} with \Eq{binom2}], we have
\be\label{dhmf2}
\dh(\ell)=\sum_\mu\frac{1+b_\mu(\ell/\ell_*)^{\a_\mu-1}[\a_\mu+F_\om'(\ell)]}{1+b_\mu(\ell/\ell_*)^{\a_\mu-1}F_\om(\ell)}\,.
\ee
Near the IR, an expansion of \Eq{dhmf2} for $\ell/\ell_*\gg 1$ yields eq.\ \Eq{dhdsgen2} with $c_\mu=1-\a_\mu$. Thus, $\dh\simeq D$ at large spacetime scales. Near the UV ($\ell/\ell_*\ll 1$), 
\be\label{dhuv}
\dh\stackrel{\rm UV}{\simeq} \sum_\mu\frac{\a_\mu+F_\om'(\ell)}{F_\om(\ell)}\simeq \sum_\mu\a_\mu+\text{(log oscillations)}\,.
\ee
Ignoring logarithmic oscillations, the spacetime UV Hausdorff dimension is $\dh\simeq\sum_\mu\a_\mu$, as anticipated in \loref{07b}. For an isotropic measure,
\be\label{dhuv2}
\dh\stackrel{\rm UV}{\simeq} D\a\,.
\ee

The spectral dimension is calculated from the diffusion equation and the latter can be derived from the microscopic stochastic dynamics of the diffusing particle, governed by the Langevin equation \cite{frc7}. If the Laplacian $\bar\cK$ appearing in the diffusion equation is not self-adjoint (as it may happen in transport phenomena), then it does not necessarily coincide with the Laplace--Beltrami operator $\cK$ of theory. This is the case of the theories $T_1$ and $T_v$, whose diffusion equations are one the adjoint of the other. In both cases, one can show that
\be
T_1,T_v\,:\quad \ds=D\frac{\rmd\ln L^2(\s)}{\rmd\ln\s}\,,\qquad L^2(\s):=\int^\s\frac{\rmd\s'}{v(\s')}\,,
\ee
where $\s$ is the diffusion scale and, if it is anomalous, it is weighted by a distribution $v(\s)$. In the diffusion interpretation, there is no guiding principle telling us what $v(\s)$ should be, but assuming that it behaves like the multifractional measure weight of spacetime, we can take the profile $v(\s)=1+\sum_n \tilde b_n(\s/\s_n)^{\b_n-1}F_n(\s)$. At the $n$-th plateau of dimensional flow, $\ds\simeq D(2-\b_n)$, while for a binomial profile and $0<\b\equiv \b_1<1$ one obtains \cite{frc7}
\be
T_1,T_v\,:\quad \ds\stackrel{\rm IR}{\simeq} D\,,\qquad \ds \stackrel{\rm UV}{\simeq}D(2-\b)\,.
\ee
A fact gone unnoticed in previous works is that the QFT interpretation of the spectral dimension \cite{CMNa} does not have any of the ambiguities of the diffusion interpretation and fixes $\ds$ for this class of theories. Both $T_1$ and $T_v$ have standard propagator in position space and, for a massless scalar particle, $\tilde G(k^2)=-1/\tilde\cK(k)=-1/k^2$ \cite{frc2,frc6,frc13}. From the Schwinger representation \Eq{Sch} of this expression, one derives the running equation in momentum space $(\p_{L^2}+k^2)\tilde P(k^2,\ell)=0$. Seeing $L$ just as in integration parameter of the Schwinger representation, there is no reason to give it a nontrivial measure weight. Then, $\b=1$ and $\ds=D>\dh$ at all scales and eq.\ \Eq{proC} is violated: these geometries are not multifractal. Changing the initial condition of the solution of the diffusion equation, one can even produce dimensional flows from $0$ to $D$.

The diffusion equation for the theory $T_q$ is straightforward: $[\p_{L^2(\ell)}-\N_{q(x)}^2]P(x,x';\ell)=0$. Both in the diffusion and QFT interpretation, one considers the multiscale version of diffusion time or Schwinger parameter and a profile $L(\ell)$. In the QFT interpretation of the running equation, $L$ is a length or a time, whose inverse gives the spatial and temporal resolution of the measurement. In these geometries, $L$ is not the scale $\ell$ directly measured but it is related to that by a scale-dependent relation $L(\ell)=\ell[1+\sum_n b_n(\ell/\ell_n)^{\b_n-1}F_n(\ell)]$. If we chose $L$ to be on a specific space or time direction, we would have $\b_n=\a_{i,n}$ or $\b_n=\a_{0,n}$, and the spectral dimension at the $n$-th plateau would be insensitive to the geometry of the other directions. Therefore, it is more sensible to identify $\b_n$ with the average $n$-th fractional charge of the measure,
\be\label{avga}
\b_n=\frac{\sum_\mu\a_{\mu,n}}{D},
\ee
which corresponds to $\b_n=\a_n$ in the isotropic case. The spectral dimension in the theory $T_\g$ is more difficult to calculate than for $T_q$ \cite{frc4} but, at the end of the day, both cases agree:
\be
T_q,T_\g\,:\quad \ds=D\frac{1+\sum_n b_n(\ell/\ell_n)^{\b_n-1}[\b_n+F_n'(\ell)]}{1+\sum_n b_n(\ell/\ell_n)^{\b_n-1}F_n(\ell)}\,.
\ee
In the IR and ignoring log oscillations, $\ds\simeq D$, while at the $n$-th plateau $\ds\simeq D\b_n$. In the binomial case,
\be
T_q,T_\g:\quad \ds\stackrel{\rm IR}{\simeq} D\,,\qquad \ds \stackrel{\rm UV}{\simeq}D\b\,.
\ee
Taking eq.\ \Eq{avga},
\be\label{dsuv}
\ds=\dh
\ee
at all scales. This is the first case to our knowledge that agreement between the two interpretations of $\ds$ (diffusive or QFT) fixes a free parameter in one of them ($\b$ in the diffusion case).

Finally, the walk dimension \Eq{dw} of spacetime in $T_1$ and $T_v$ is $\dw=2D/\ds$ (confirming that this is not a multifractal), while in $T_q$ it is $\dw=2\dh/\ds$, independently of the use of \Eq{avga} \cite{frc7}. We have not calculated yet $\dw$ for $T_\g$.\footnote{In \cite{frc4}, the definition of $\dw$ was naively assumed to be eq.\ \Eq{proC} rather than eq.\ \Eq{dw}.}

\bigskip

\llabel{14} \textbf{\emph{Can the dimension of spacetime become complex or imaginary?}}\addcontentsline{toc}{subsection}{\lref{14} Can the dimension of spacetime become complex or imaginary?}

Yes it can, in multiscale setups such as quantum gravities \cite{COT2,first,ThSt}, in multifractional theories \cite{first}, and in fractal geometry \cite{LvF,Tep07,Akk1}. 

In multiscale theories (including quantum gravity at large), the flow-equation theorems establish that the most general iterative solution of eq.\ \Eq{flow} at infinite order is the dimension (Hausdorff and/or spectral) $d(\ell):=\lim_{n\to+\infty}d^{(n)}$, where \cite{first}
\be\label{soln}
d^{(n)}(\ell)-d^{(n-1)}(\ell)=\sum_{i=0}^{n-1} b_{i,n}\,\ell^{\a_{i,n}+\rmi\om_{i,n}}\,,\qquad \sum_{j=0}^n c_j (\a_{i,n}+\rmi\om_{i,n})^j=0\,.
\ee
The complex exponents $\a_{i,n}+\rmi\om_{i,n}$ satisfy a characteristic equation for all $i$. All quantum gravities have dimensional flow but, formally, all dimensional flows follow the same universal profile, which can vary from case to case depending on how the dynamics determines the free parameters $b_{i,n}$, $\a_{i,n}$, $\om_{i,n}$, and $c_j$  within. Some quantum gravities may just have real-valued dimensions either because $\om_{i,n}=0$ for all $i$ and $n$ or because conjugate powers $\pm\rmi\om_{i,n}$ combine to give the log oscillations we discussed so far. Other quantum gravities, however, can display complex dimensionalities
\be
d(\ell)\in\mathbbm{C}
\ee
because conjugate powers do not combine. The question now is whether this feature is only an abstract mathematical property of the solution \Eq{soln} or is realized in concrete scenarios. There is evidence that such is indeed the case in spin foams \cite{COT2,ThSt}. In contrast with kinematical states, spin-foam sums of dynamical states generically contain \emph{degenerate geometries} (i.e., some component of the tetrad $e_\mu^I$ vanish identically), where the volume operator is not densely defined and has 0 as an eigenvalue. Preliminary calculations of the spectral dimension on small combinatorial complexes, using the graviton propagator in $(2+1)$-dimensional spin foams, show that the heat kernel $P$ acquires an imaginary part, from which it stems that also the return probability $\cP$ and the spectral dimension $\ds$ are complex-valued. These results were reported in ref.\ \cite{COT2} without giving the details; work in progress \cite{ThSt} on a recent model of spin foams on a hypercubic lattice \cite{BaSt} finds similar results.

The general solution \Eq{soln} of the flow equation affects also multifractional theories; therefore, they too can have complex dimensions. However, since the beginning \cite{fra4} and to date, attention has been limited to real-valued measures (i.e., with log oscillations rather than imaginary powers). As a further guarantee of avoiding ``unphysical'' situations with negative dimensions due to large oscillation amplitudes, the spacetime dimensions $\dh$ and $\ds$ have also been defined to be calculated after averaging out log oscillations, which is easily done by replacing $\cV$ and $\cP$ in eqs.\ \Eq{dh} and \Eq{ds} with their log average \cite{frc2}. These conditions are sufficient to have $\dh,\ds\geq 0$ but, after a few years of investigation, they might turn out to be too restrictive inasmuch as they exclude geometries that \emph{are} physical despite their highly unconventional features. Abandoning the averaging procedure (as done here) is not particularly dangerous: in practice, and in all known examples, log oscillations have a very small amplitude \cite{frc14,Akk1} and they reduce to tiny ripples around the average. Relaxing also the reality condition, we get access to complex dimensions \Eq{soln} and have to face the task of interpreting the ensuing spacetimes. 

The spin-foam results mentioned above could shed some light on this interesting subject and hint to an association between complex dimensions and degenerate geometries, for which the metric $g_{\mu\nu}=\eta_{IJ}e_\mu^I e_\nu^J$ has some ill-defined components. We argue here that fractal geometry supports this view and, therefore, that we might be on the right track. Given the Laplacian $\cK$ on a deterministic fractal, one can compute the Mellin--Laplace transform of the associated heat kernel $P$, which is a function $\zeta_\cK(s)$ of the Laplace momentum $s$ called spectral zeta function (usually proportional to the Riemann $\zeta$ function). The spectral zeta function is given by
\be\label{zetak}
\zeta_\cK(s)=\sum_j\la_j^{-s}\,,
\ee
where $\la_j$ are the nonzero eigenvalues of $\cK$. The poles of $\zeta_\cK(s)$ are complex-valued and of the form
\be\label{sm}
s_m =\frac{1}{2}(\ds+ \rmi \dcs)\,,\qquad m\in\mathbbm{Z}\,,
\ee
where $\dcs\propto 4\pi m$ are called \emph{complex spectral dimensions} \cite{LvF,Tep07} and accompany the usual spectral dimension, which is the real-valued pole of $\zeta_\cK(s)$. The complex poles \Eq{sm} are a typical feature of fractals (even of popular examples such as the Sierpi\'nski gasket, the Julia sets, diamond fractals, and the Cantor string, the complement of the middle-third Cantor set \cite{Tep07,Akk1}; see also \cite{Akk2,Akk12}), and their ``physical'' origin can be understood from eq.\ \Eq{zetak}. The infinite number of poles $m$ is due to the presence of an exponentially large degeneracy of some special eigenvalues of the Laplacian called iterated (in contrast, in ordinary manifolds this degeneracy factor is power-law) \cite{Akk1}. In turn, nonmetric geometries or labels on combinatorial graphs have spectral features that could easily lead to exponential degeneracies in the Laplacian eigenvalues, and hence could acquire complex dimensions. This was briefly commented upon in ref.\ \cite{Akk1} and agrees intriguingly with what found later in ref.\ \cite{COT2}. The relation between metric degeneracy and Laplacian eigenvalue degeneracy has not been clarified to date, but these few fragments we collected here are suggestive of a coherent picture awaiting further study.

\bigskip

\llabel{15} \textbf{\emph{Do multifractional theories really have dimensional flow? Regardless of the choice of derivatives, the measure \Eq{facto} is mathematically equivalent to the standard Lebesgue measure $\rmd^Dx$, where one uses the symbol ``$q$'' instead of ``$x$.'' If we compute the volume of a hypercube or of a $D$-ball, we find
\be\nonumber
\cV\sim\int_0^\ell\rmd^D q(x)=\int_0^L\rmd^D q=L^D,
\ee
where $L=q(\ell)$ (here we are ignoring $\mu$ indices for simplicity) is the edge size of the hypercube or the radius of the ball. Then, the Hausdorff dimension coincides with the topological dimension:
\be\label{dhfake}
\dh(L)=\frac{\p\ln\cV}{\p\ln L}=D\,.
\ee
One could make a similar calculation for the spectral dimension and show that $\ds=D$.}}\addcontentsline{toc}{subsection}{\lref{15} Do multifractional theories really have dimensional flow?}

The above calculation is mathematically correct but it neglects the physics. The step $x\to q(x)$ is not a coordinate transformation in multifractional theories, which break Lorentz invariance (see question \lref{22}). An absolutely indispensable ingredient of the multifractional recipe is the establishing of measurement units or, in other words, of a coordinate frame where all physical measurements must be carried out. This step is necessary because the profiles $q^\mu(x^\mu)$ are noninvariant under coordinate transformations, and one must fix the frame where the form \Eq{meag} is valid. By definition from the onset, the coordinates $x^\mu$ have the scaling dimension of lengths and time,
\be\label{x1}
[x^\mu]=-1\,,
\ee
and are called fractional coordinates. The frame $\{x^\mu\}$ is called \emph{fractional frame} or \emph{picture}. The geometric coordinates $q^\mu$, which define the \emph{integer frame} or \emph{picture} in the theory with $q$-derivatives, also have the dimension of lengths and time, $[q^\mu]=-1$ exactly, but their $x$-dependent part does not. At the $n$-th plateau in dimensional flow, i.e., at distances or times $\sim\ell_n^\mu$, this varying part scales as
\be\label{xa}
[|q^\mu|]\stackrel{x\sim\ell_n}{\sim} [|x^\mu|^{\a_{\mu,n}}]=-\a_{\mu,n}\neq -1\,.
\ee
This is what is meant in the literature by \emph{anomalous scaling}. 

The physical meaning of eqs.\ \Eq{x1} and \Eq{xa} is that, in the fractional picture constituted by the fractional coordinates $x^\mu$, measurements are taken by clocks and rods that do not change with the probed scale [there is no scale dependence in \Eq{x1}], while in the integer picture made of the geometric coordinates $q^\mu$ measurements are taken by clocks and rods that adapt with the probed scale (there is a scale dependence in \Eq{xa}). By definition, physical measurements in multifractional theories are performed in the fractional picture: clocks and rods are nonadaptive, rigid, not multiscale.\footnote{In asymptotic safety, precisely the opposite holds and physical rods are adaptive \cite{fra7}. We will comment on this in question \lref{45}.} The reason beyond this choice instead of its complementary is simple. Measurement apparatus created by humans are local objects with definite size probing length, time, or energy scales in a definite range. Rods measuring the length of a goldfish are the same rods measuring a whale, only shorter. When we probe lengths at very different scales, such as of goldfish or atomic or elementary-particle size, we do not have one general ``rod'' marking centimeters and Compton lengths: we have to construct new ``rods'' for each scale, based on different principles.
%If there were an alien species taking different sizes spanning several orders of magnitude (a biological impossible, of course), they could measure the scaling of balls and hypercube with their linear size with rods taking the same range as the species and they

Having thus established nonadaptive rods (i.e., the fractional picture) as the measuring tool of physics, it is clear that the radius of the ball we measure is $\ell$, not $L$, so that its volume scales as $\ell^{\dh}$, not as $L^{\dh}$. Consequently, the Hausdorff dimension is \Eq{dhmf}, not \Eq{dhfake}. A similar reasoning holds for $\ds$.

\bigskip

\llabel{16} \textbf{\emph{Is prescribing measurement units in this way scientific? We all know that any theory of physics is based upon some principles or axioms, but we could obtain everything just by changing well-established axioms or by replacing them by something else, as you do in multifractional theories.}}\addcontentsline{toc}{subsection}{\lref{16} Is prescribing measurement units in this way scientific?}

And as done in scalar-tensor theories \cite{Dic62,FaNa2} or in varying-speed-of-light (VSL) models \cite{Mag00,Mag03}. The selection of special frames where physical observables are measured is not a novelty. There is nothing wrong in modifying well-established axioms, as long as the resulting theory is motivated from above, internally consistent, and testable by experiments.

In scalar-tensor theories, the change from the Jordan to the Einstein frame corresponds to a change of measurement units. In VSL theories, we are dealing with units adapted with the scales in the dynamics and, in particular, chosen such that the speed of light $c(x)$ varies in space and time. Time and space units are redefined so that the differentials scale as $\rmd t\to [f(x)]^a \rmd t$, $\rmd x^i\to [f(x)]^b \rmd x^i$, where $f$ is a function, $a$ and $b$ are constants, and local Lorentz invariance of the line element requires $c(x)\propto [f(x)]^{b-a}$. We recognize here a particular form of anisotropic multiscaling (one that distinguishes between space and time variables). In particular, when $b=0$ one formally reabsorbs $c$ in the coordinate $x^0=\int\rmd t\, c(t)$, which scales as a length. With this coordinate, all equations can be made formally identical to the usual ones provided some conditions are met. Models where the electric charge $e$ or the speed of light $c$ varies can be recast in new units such that, respectively, the electric charge and the speed of light become constant, but in both cases the dynamics can become substantially more complicated. This criterion of simplicity is not the only one which attaches one label or the other (varying-$e$ or varying-$c$) to these models: experiments are able to distinguish between them. The change of units at the base of scalar-tensor theories, VSL models, and varying-electric-charge models all map in one way or another \cite{frc8} to the Manichaean notion of ``adapting'' versus ``nonadapting'' rods in multifractional models \cite{fra7,trtls}. Furthermore, the multifractional paradigm can be discriminated from scalar-tensor, VSL, and other changing-unit proposals both by experiments and by their theoretical structure. Despite the striking similarity of the Einstein equations of scalar-tensor theories \cite{frc11}, the measure weight is not a Lorentz scalar and it heavily affects the gravitational dynamics (for instance in cosmology) in a way irreproducible by scalar-tensor models. The presence of a nontrivial measure consistently affects the definition of functional variations, Poisson brackets and Dirac distribution, in turn leading to a deformation of the Poincar\'e symmetries (see \lref{11}) not realized in varying-$e$ and varying-$c$ scenarios. In question \lref{48}, we will see an example of how one can measure departure from a standard space in a multiscale geometry.

\bigskip

\llabel{17} \textbf{\emph{Is the volume density $\sqrt{-g}$ from the metric implemented consistently? Therein, I do not see any change of anomalous geometry with the scale.}}\addcontentsline{toc}{subsection}{\lref{17} Is the volume density \texorpdfstring{$\sqrt{-g}$}{} from the metric implemented consistently?}

This somewhat vague question arises because in the majority of papers gravity is ignored and the measure is $\rmd^Dq(x)$ (with trivial volume density factor $\sqrt{-\eta}=1$), while when gravity is triggered the volume measure is $\rmd^Dq(x)\,\sqrt{-g}$ \cite{frc11}. This creates confusion because no show of dimensional flow seems to emanate from the volume density. The point is that the calculus structure and the metric structure are totally independent at the level of the action. On one hand, there is the calculus structure embodied by the integral measure $\rmd^Dq(x)$ and the choice of derivatives. On the other hand, there is the metric structure expressed by the volume density $\sqrt{-g}$, curvature terms, and covariant derivatives. When the calculus structure reduces to the usual one and $q^\mu\simeq x^\mu$, then standard general relativity is recovered (by construction). This limit is independent of the curvature of spacetime, so that to preserve covariance and diffeo invariance in the IR the factor $\sqrt{-g}$ must be there.

The mutual independence of the integrodifferential and the metric structures does not imply that they do not talk to each other. The multiscaling of the geometric coordinates $q^\mu(x^\mu)$ strongly affects the dynamics and, hence, the solutions to the Einstein equations. In particular, the background metric $g_{\mu\nu}(x)$ solving the dynamical equations is multiscale \cite{frc11,frc14}.

\bigskip

\llabel{18} \textbf{\emph{Is geometry discrete at the smallest scales?}}\addcontentsline{toc}{subsection}{\lref{18} Is geometry discrete at the smallest scales?}

Yes, it is. Take for simplicity the binomial measure \Eq{meamu} in one dimension: $q(x) = x+(\ell_*/\a){\rm sgn}(x)|{x}/{\ell_*}|^\a F_\om(x)$, where $F_\om(x)= 1+A\cos(\om\ln|{x}/{\ell_\infty}|)+B\sin(\om\ln|{x}/{\ell_\infty}|)$. The distribution $F_\om$ is invariant under the \emph{discrete scale invariance} (DSI)
\be\label{dsi}
x\,\to\, \la_\om^n x\,,\qquad \la_\om:=\exp\left(-\frac{2\pi}{\om}\right)\,,\qquad n\in\mathbbm{Z}\,.
\ee
This symmetry, often found in chaotic systems \cite{Sor98,JoS,JSH}, is a dilation transformation under integer powers of a prefixed scaling ratio $\la_\om$. Although $F_\om(\la_\om x)=F_\om(x)$, the measure $q(x)$ is not invariant (up to an overall constant factor), since
\be
q(\la_\om x)= \la_\om^\a q(x)+(\la_\om -\la_\om^\a)x\,.
\ee
The last term never vanishes. However, at scales $\lesssim \ell_*$ the overall scaling is determined by $\a$ and the dominant piece of the measure is DSInvariant. In the IR, the usual dilation symmetry $x\to \la x$ with arbitrary $\la$ is recovered, while a natural discrete-to-continuum transition happens at intermediate scales (for a detailed description, see \cite{fra4,frc2}). At one extremum of this transition, UV spacetime is effectively discrete and described by a lattice of size $\la_\om \ell_\infty$, even if the full integration measure is defined on a continuum.\footnote{Discreteness of a geometry can be encoded either in continuum models $\int_{\rm lattice}\rmd x\,v(x)\,\cL(x)$ with discrete integration domain (integrals in a continuous embedding weighted by measures with discrete support), or by a setting with discrete calculus, $\sum_n \cL(x_1,\dots,x_n)$. Multifractional theories adopt the first option, while CDT (as an artifact), GFT, LQG, and spin foams realize the second.}

\bigskip

\llabel{19} \textbf{\emph{Is $D=4$ assumed or predicted?}}\addcontentsline{toc}{subsection}{\lref{19} Is \texorpdfstring{$D=4$}{} assumed or predicted?}

In general, it is assumed, just like in any other theory except string theory. However, in question \loref{07b} we mentioned that there is a phase transition in the theories $T_1$ and $T_v$ for the special value $\a=2/D$ of the fractional exponent in the measure, so that $\dh\simeq D\a=2$ in the UV. Only in $D=4$ does this exponent $\a=1/2$ lie at the middle of the allowed interval \Eq{ranga}. Intriguingly, the value of the Hausdorff dimension in the UV ($\dh\simeq 2$) and in the IR ($\dh\simeq 4$) are mutually related rather than being independent as in many multiscale quantum gravities. Thus, $D=4$ is special among any other possibility, but only in $T_1$ and $T_v$ and only in relation with the UV value: in this sense, the above argument is circular and does not allow to make separate claims about the uniqueness of the UV and the IR dimension separately. In $T_v$, however, there is an independent argument selecting $D=4$ as the only case where the gravitational action simplifies (see question \lref{37}) and the metric $g_{\mu\nu}$ has the natural structure of a bilinear field with measure weight $-1$ \cite{frc11}. 

In the theory $T_q$, there is no phase transition relating the UV and the IR dimensions. In the theory $T_\g$, there is a stronger argument to select $\a=1/2$ (it is the lowest possible value to have a normed space), but it is not related to the IR dimension. In these cases, we are not aware of any robust argument to select $D=4$.

%%%%%%%%%%%%%%%%%%%%%%%%%%%%%%%%%%%%%%%%%%%%%%%%%%%%%%%%%%%%%%%%%%%%%%%%%%%%%
%%%%%%%%%%%%%%%%%%%%%%%%%%%%%%%%%%%%%%%%%%%%%%%%%%%%%%%%%%%%%%%%%%%%%%%%%%%%%

\section{Frames and physics}\label{frafi}

\llabel{20} \textbf{\emph{The theory with weighted derivatives is trivial. Consider for instance the scalar-field action \Eq{Sphi} with polynomial interactions:
\be\label{wS}
S_\phi=-\int\rmd^Dx\,v\left(\frac12\cD_\mu\phi\cD^\mu\phi+\sum_{n=2}^N\frac{\s_n}{n}\phi^n\right).
\ee
After the field redefinition \Eq{vpi}, $\vp=\sqrt{v}\,\phi$, the action becomes
\be\label{minS}
S_\phi=-\int\rmd^Dx\left(\frac12\p_\mu\vp\p^\mu\vp+\sum_n\frac{\tilde\s_n}{n}\vp^n\right),\qquad \tilde\s_n=\s_nv^{1-\frac{n}{2}}\,.
\ee
If we also assume that, originally, the $\s_n$ were spacetime dependent and such that $\s_n(x)\propto [v(x)]^{n/2-1}$ for all $n$, then the couplings $\tilde\s_n$ are constant (the mass $\s_2=m^2$ is constant also in the fractional picture) and eq.\ \Eq{minS} is the usual action in standard Minkowski spacetime. In \cite{frc6,frc9}, the $\s_n$ were assumed to be constant, but in the case of the Standard Model \cite{frc8,frc13} all the effective couplings $\tilde \la_i$ after the field transformation \Eq{vpi} were found to be constant. Nevertheless, it was concluded that the theory was nontrivial. I do not see how, since the actions \Eq{wS} and \Eq{minS} are equivalent.}}\addcontentsline{toc}{subsection}{\lref{20} Is the theory with weighted derivatives trivial? (i)}

As in the case of $T_q$, $T_v$ can be written in two different ways or frames (question \lref{15}). The one defining the theory, and where physical measurements have to take place, is called the fractional picture or fractional frame and corresponds to eq.\ \Eq{wS} and to the general action on flat space \Eq{Sv}. After the field redefinition \Eq{vpi}, the theory is simplified and takes exactly the form of a field theory on ordinary Minkowski space, provided all couplings $\la$ in the fractional picture have a spacetime dependence such that the couplings $\tilde\la$ in the integer frame or integer picture are constant.\footnote{Note that the integer picture in the theory $T_q$ is defined differently and does not involve field transformations (see \lref{15}).} In general and in the absence of gravity, the two frames are related by
\be\label{SvSi}
S_v[v,\cD,\phi^i,\la_i]=S_1[1,\p,\vp^i,\tilde\la_i]\,,\qquad \text{$v(x)$ in the left-hand side fixed}.
\ee
The claim in the question is that the right-hand side is the standard action of a QFT on Minkowski spacetime, hence the theory is trivial. However, there are three elements that should be taken into account: a general remark about the physical frame, information from non-QFT physics, and inclusion of gravity.

The general fact is that, by definition, physical observables must be evaluated in the fractional picture, which is the frame where physical measurements take place. This was stated in \lref{15}. The integer picture is a frame where the theory is simplified in such a way that all calculations in QFT can be carried out easily, that is, with standard perturbative QFT techniques. These techniques are not applicable in the fractional picture: the field theory described by \Eq{wS} or \Eq{Sv} has spacetime-dependent kinetic terms and couplings, which make Feynman rules difficult or practically impossible \cite{frc9,frc13}. The QFT in the integer picture is the usual one and we can calculate effective observables easily. However, the effective observables in the integer picture must be converted into the physical observables in the fractional picture, which are those to be compared with experiments. Therefore, the integer picture is only a convenient way to recast the theory and make calculations, but it is not physically equivalent to the fractional picture. Several observables have been computed and constrained experimentally which illustrate the point \cite{frc8,frc13,frc14}. A similar situation happens in scalar-tensor theories, although in that case the frame dilemma is shifted to the quantum level (see question \lref{26}).

Another general argument \cite{frc13} is that QFT is only part of the whole story. The QFT couplings in the theory $T_v$ are constant in the integer picture not only for necessity (masses are constant to allow for a manageable quantum perturbative treatment), but also as a requirement of gauge invariance \cite{frc13}. Such restrictions do not exist in the realm of statistical and particle mechanics. Examples are the random motion of a molecule \cite{frc7}, the dynamics of a relativistic particle \cite{frc10}, and the black-body radiation spectrum \cite{frc14}, all processes with a characteristic energy much smaller than that in the center of mass of subatomic scattering events. On one hand, the form of the couplings in QFT is constrained by the way we are able to deal with interacting quantum fields. On the other hand, statistical and particle mechanics are intrinsically nonlinear, either through the stochastic interaction of a degree of freedom with the environment (as in the multifractional Brownian motion of a particle \cite{frc7}), or by definition of the action (as for the relativistic particle \cite{frc10}), or via the collective description of microscopic degrees of freedom (as in the frequency distribution of a thermal bath of photons \cite{frc14}). These systems yield nontrivial predictions because they are not subject to requirements as severe as those we imposed on a quantum field theory. Therefore, one should not identify the theory $T_v$ with QFT alone, just like standard QFT cannot describe all possible systems of physics.

A third consideration to make is about gravity. On a curved background, the equivalence of frames after field and metric redefinitions is broken. In the integer picture, the theory $T_v$ is not general relativity with minimally coupled matter, and one can never trivialize the theory to the ordinary one as in the flat case \cite{frc11}. The gravitational dynamics of the theory with weighted derivatives was studied in ref.\ \cite{frc11}. The metric is not covariantly conserved and the geometry corresponds to a Weyl-integrable spacetime. The total action reads
\be
S_v[g,\phi^i] =\frac{1}{2\kappa^2}\int\rmd^Dx\,v\,\sqrt{-g}\left[{\cal R}-\om\cD_\mu v\cD^\mu v-U(v)\right]+S_v[\phi^i]\,,\label{eha}
\ee
where ${\cal R}$ is the Ricci scalar constructed with weighted derivatives of different weight \cite{frc11} (see question \lref{37}), $\om$ and $U$ are functions of the weight $v$, and in $S_v[\phi^i]$ the metric is minimally coupled. Absorbing weight factors into the matter fields $\phi^i$ with the picture change \Eq{vpi} requires a redefinition of the metric $g_{\mu\nu}\to\tilde{g}_{\mu\nu}$. Indeed, one can go to the integer picture (Einstein frame) where the gravitational action is $\propto \int\rmd^Dx\,\sqrt{-\tilde g}\,\tilde R$ but not without introducing nontrivial measure-dependent terms. These terms affect the cosmic evolution. Thus, a change of picture does not lead to standard general relativity plus matter and the dynamics is different from (and much more constrained than) that of scalar-tensor scenarios in both frames. In general, one should be careful about the issue of the physical inequivalence between the fractional and the integer picture. As for scalar-tensor models, from a simple visual inspection of the actions one cannot conclude that the Jordan and Einstein frames define different physics. What matters are the physical observables. The homogeneous classical cosmology of multifractional theories is physically distinguishable from the usual one even in the integer picture (Einstein frame), since $\tilde\om\neq 0\neq \tilde U$.

\bigskip

\llabel{21} \textbf{\emph{In the so-called fractional picture, the theory with weighted derivatives appears to violate Poincaré invariance explicitly, as also stated in \lref{10} and \lref{11}. But if Poincaré violations can be eliminated by redefining the fields (in the so-called integer picture), then where is the new physics? Fields are auxiliary concepts and redefining them should not change the physical content (for instance, the S-matrix) of the theory.}}\addcontentsline{toc}{subsection}{\lref{21} Is the theory with weighted derivatives trivial? (ii)}

The fractional and integer pictures are not related only by the field redefinition \Eq{vpi} (together with redefinitions of couplings \cite{frc13}). When an observable is computed (for convenience) in the integer picture, it must be mapped back to the measurement units of the fractional pictures, which is the frame where physical measurements take place with nonadaptive clocks, rods, and particle detectors. For instance, the observed electron charge $\tilde e=e_0$ is constant in the integer picture, but it is a time-dependent quantity $Q(t)$ in the fractional picture \cite{frc8}.\footnote{This property tells the electric charge apart from all other gauge couplings of the Standard Model \cite{frc13}. See question \lref{31}.} This means than all the phenomenology associated with the fine-structure constant will be standard in the integer picture but time-dependent in the fractional picture. What we constrain by observations is the second.

In general, the symmetries enjoyed in the integer frame (such as Poincaré invariance) can be violated in the physical frame, and observables are affected consequently.
We postpone to question \lref{26} a discussion on the S-matrix.

\bigskip

\llabel{22} \textbf{\emph{The theory with $q$-derivatives is trivial. Consider for instance the scalar-field action \Eq{Sphi} with a mass term and a higher-order interaction:
\be\label{qS}
S_\phi=-\int\rmd^Dx\,v\left[\frac12\eta^{\mu\nu}\frac{\p\phi}{\p q^\mu(x^\mu)}\frac{\p\phi}{\p q^\nu(x^\nu)}+\sum_{n=2}^N\frac{\s_n}{n}\phi^n\right].
\ee
Here there is no field redefinition available but one can consider
\be\label{xq}
x^\mu\to q^\mu(x^\mu)
\ee
simply as a change of coordinates. Since the physics should be invariant under such coordinate transformations, then the theory is equivalent to the usual one.}}\addcontentsline{toc}{subsection}{\lref{22} Is the theory with \texorpdfstring{$q$}{}-derivatives trivial? (i)}

In general, the mapping between the fractional and the integer picture is
\be\label{fipq}
S_q[v,v^{-1}\p_x,\phi^i,\la_i]=S_q[1,\p_q,\phi^i,\la_i]\,,\qquad \text{$v(x)$ in the left-hand side fixed}.
\ee
The fractional picture is the frame where the $x$-dependence of the composite coordinates $q(x)$ is manifest [left-hand side of \Eq{fipq}], while the integer picture is the frame described by the geometric coordinates $q$ [right-hand side of \Eq{fipq}]. Contrary to the mapping \Eq{SvSi} for the theory $T_v$, there is no redefinition of the couplings. As in the theory $T_v$, the difference between the fractional and the integer picture is in the way geometry is perceived by the dynamical degrees of freedom: as standard Minkowski spacetime in the integer picture, as an anomalous geometry with a fixed integrodifferential structure in the fractional picture. The presence of this predetermined structure does affect the physics because it prescribes the existence of a preferred frame where physical observables should be compared with experiments. As we already said, by definition of the theory, this frame is the fractional picture. This is an important conceptual novelty with respect to theories with an ordinary integrodifferential structure: a choice of frame is a mandatory step in the definition of multifractional spacetimes. 

In the case with $q$-derivatives, time intervals, lengths and energies are physically measured in the fractional picture where coordinate transformations are described by the nonlinear law \Eq{qlort}. We stress that eq.\ \Eq{xq} is \emph{not} a coordinate transformation. It governs the formal passage between the fractional picture described by the composite coordinates $q^\mu(x^\mu)$ and the integer picture described by coordinates $q^\mu$. The integer picture is a convenient frame for calculations, but it is no more than that, since eq.\ \Eq{xq} is not even invertible except in the simple case of a binomial measure without oscillations.

To illustrate in what sense the integer frame is ``convenient,'' we write down eq.\ \Eq{qS} in $D=1+1$ dimensions:
\ba
S_\phi&=&\int\rmd^2 q\,\left\{\frac12[\p_{q_0(t)}\phi]^2-\frac12[\p_{q_1(x)}\phi]^2-\sum_n\frac{\s_n}{n}\phi^n\right\}\nonumber\\
&=&\int\rmd^2 x\,\left[\vphantom{\sum_n}\frac{v_1(x)}{2v_0(t)}\dot\phi^2-\frac{v_0(t)}{2v_1(x)}(\p_x\phi)^2-\sum_n\frac{v_0(t)v_1(x)\s_n}{n}\phi^n\right].
\ea
Since we do not know how to define a quantum field theory with varying couplings and nonhomogeneous kinetic terms, it is necessary to perform all calculations in geometric coordinates. Therefore, we transform to the integer picture  via \Eq{xq} where the theory looks trivial and one can borrow all the known calculations in standard QFT. Any ``time'' or ``spatial'' interval or ``energy'' predicted in the integer picture are not a physical time or spatial interval or energy, since they are measured with $q$-clocks, $q$-rods, or $q$-detectors. The results must be reconverted to the fractional picture in order to interpret them correctly. QFT examples of this inequivalence of observables are the muon decay rate, the Lamb shift, and the variation of the fine-structure constant \cite{frc12,frc13}, while cosmological and astrophysical examples are given in refs.\ \cite{qGW,frc14}.

\bigskip

\llabel{23} \textbf{\emph{I am still not convinced, so let me rephrase my criticism. The theory $T_q$ tries to incorporate the effects of new fundamental energy, time and length scales at a microscopic scales while getting the standard physics at mesoscopic distances. This is done through a particular replacement of coordinates. As it is, it is unclear what this replacement actually is. I see two possibilities, it is either a change in the description or a change in the physical behavior. I will argue against any of these possibilities. Let us first assume that the replacement \Eq{xq} is a change of the description. This corresponds to a coordinate change but, as we know, the theory of relativity is built in such a way that a change of coordinates does not change the physics. The new effects claimed to be found are spurious and unphysical because the coordinate change is ill defined, since it is not invertible in general. To avoid the problems associated with invertibility, one would need to focus on a single chart where the $q^\mu(x^\mu)$ were invertible, but this restriction is not considered. In fact, this omission is the root of the DSI of the function $F_\om$, and ultimately of the supposed ``fractal'' nature of the theory. Therefore, this is not a valid mechanism to introduce new scales.}}\addcontentsline{toc}{subsection}{\lref{23} Is the theory with \texorpdfstring{$q$}{}-derivatives trivial? (ii)}

The theory of general relativity is built in such a way that a change of coordinates does not change the physics, but multifractional theories are not. It is a mistake to impose the principles of Einstein gravity to a multiscale geometry. %In fact, violation of diffeo invariance is the price to pay in multifractional theories in order to have a multiscale geometry, which does not mean that diffeo-invariant multiscale theories do not exist (asymptotic safety is an example).
 Noninvertibility, which is a consequence of the flow-equation theorems having nothing to do with ill-defined coordinate changes, is rather one of the reasons why eq.\ \Eq{xq} cannot be regarded as a coordinate transformation; the other reason is that different frames correspond to different measurement units and one must make a choice (see \lref{15} and \lref{22}). Making a frame/unit choice is not particularly exotic, as recalled in \lref{16}.

\bigskip

\llabel{24} \textbf{\emph{The change of coordinates \Eq{xq} is badly implemented inasmuch as the volume form is not corrected with the square root of the metric determinant $\sqrt{-g}$, nor is the inverse metric corrected in the kinetic term of the scalar field. Again, if these issues were considered, no new physics would arise.}}\addcontentsline{toc}{subsection}{\lref{24} Is the theory with \texorpdfstring{$q$}{}-derivatives trivial? (iii)}

We just argued against the interpretation of eq.\ \Eq{xq} as a change of coordinates. Letting aside this abuse of terminology, the volume density $\sqrt{-g}$ does appear in the theory as soon as gravity is switched on, and derivatives are made covariant accordingly \cite{frc11}. New physics does arise in that case \cite{qGW,frc14}, simply because the dichotomy between fractional and integer frame persists also when the embedding manifold is curved. We can even say more: the theory in the integer frame is invariant under a change of geometric coordinates $q^\mu\to {q'}^\mu$ \cite{frc11}, as stated in \lref{12}. This is not a symmetry of physical observables, since it is broken in the fractional picture where the form of the geometric coordinates is given by the second flow-equation theorem.

\bigskip

\llabel{25} \textbf{\emph{Let me give you a third argument against the interpretation of eq.\ \Eq{xq} as a change of description. In order to get a dispersion relation for a particle, the physical meaning for the $x$ coordinates should be specified. Interpreting them as the position of a particle (i.e., its worldline in an arbitrary parametrization), one immediately notes that the profile $q^0(t)$ must be monotonic in time, something that is not fulfilled by eqs.\ \Eq{meamu} or \Eq{meag}. Hence, in terms of the composite coordinates $q$ particles do not follow proper worldlines. This should be enough to understand that no new physics can be obtained in the $q$-theory, outside a single chart.}}\addcontentsline{toc}{subsection}{\lref{25} Is the theory with \texorpdfstring{$q$}{}-derivatives trivial? (iv)}

The profile $q^0(t)$ is not monotonic due to log oscillations, but this does not mean that time $t$ for the particle goes back and forth. Again, here one is confusing geometric coordinates with physical ones. Moreover, worldlines in a multiscale spacetimes are certainly not expected to behave as usual and, in fact, they do not, as was shown in the theory $T_v$ for a nonrelativistic and a relativistic particle \cite{frc5,frc10}. The case of the point particle in $T_q$ is straightforward; in this theory, the physical inequivalence of the fractional and integer pictures is further shown by the fact that dispersion relations are modified (question \lref{55}).

\bigskip

\llabel{26} \textbf{\emph{Even granting that the measure \Eq{facto} with \Eq{meag} comes from some different paradigm we are not accustomed to in general relativity, it breaks Poincaré invariance and, as any theory with Lorentz violation, fixes a preferred frame. While in general relativity frames are equivalent at least at the classical level, here one must make a frame choice. With what criteria? What exactly is the preferred frame in physical terms?}}\addcontentsline{toc}{subsection}{\lref{26} What are the criteria to choose the physical frame?}

We already answered in \lref{15}, \lref{16}, \lref{20}--\lref{22}. Here we make a couple of remarks on the similar problem of choice between the Einstein and the Jordan frame in scalar-tensor theories. After several years of debate, it has by now become accepted that the two frames are physically equivalent both classically \cite{DeS10,ChY} and at the quantum level to first order in perturbation theory (both in a QFT and a cosmological sense), but they differ in a nonlinear quantum regime \cite{Cho97,HuNi,ArPe,NiPi}. At that point, a choice of frame is necessary according to some criterion. For instance, one might regard the Jordan frame as the fundamental one because it is the frame where matter follows the geodesics. A choice of frame is a choice of measurement units \cite{Dic62}. In the case of the VSL models mentioned in \lref{16}, the criterion for the choice of units is simplicity of the dynamics. In the case of multifractional theories, it is to have nonanomalous clocks and rods at all scales in a multiscale spacetime (see \lref{15}).

A small caveat about quantum inequivalence of frames will conclude the discussion. Let us recall an argument by Duff against having quantum fields on a classical gravitational background \cite{Duf80}. Consider an ordinary (nonmultiscale) spacetime and an action $S[g,\phi^i]$ dependent on the metric and on some matter fields. Consider also a suitably regular field redefinition $\bar g_{\mu\nu}=\bar g_{\mu\nu} (g_{\mu\nu},\phi^i)$, $\bar\phi^i=\bar\phi(g_{\mu\nu},\phi^i)$, so that the actions $\bar S[\bar g,\bar\phi^i]=S[g,\phi^i]$ describe the same physics at the classical level. At the quantum level, if all fields (including $g_{\mu\nu}$) are quantized, then the two theories are equivalent on shell order by order in perturbation theory, although they differ as far as individual Feynman diagrams and off-shell S-matrix elements are concerned. This is because the on-shell S-matrix is invariant under field redefinitions. However, if gravity is purely classical only matter fields are quantized and the two theories are physically inequivalent. The intuitive reason is that internal graviton lines, which are essential to maintain the on-shell equivalence, are now absent. An example is the minimally-coupled massless scalar field theory
\be\label{ion1}
S[g,\phi]=\int\rmd^4x\sqrt{-g}\left(\frac{R}{2\k^2}-\frac12g^{\mu\nu}\p_\mu\phi\p_\nu\phi\right)\,.
\ee
At the one-loop level, UV divergences are removed if one adds a certain counterterm $\Delta S$ \cite{HoV2}. The classical theory \Eq{ion1} is equivalent to the nonminimal action
\be\label{ion2}
\bar S[\bar g,\bar\phi]=\int\rmd^4x\sqrt{-\bar g}\left[\bar R\left(\frac{1}{2\k^2}-\frac{\bar\phi^2}{12}\right)-\frac12\bar g^{\mu\nu}\p_\mu\bar\phi\p_\nu\bar\phi\right]
\ee
via a conformal transformation. One-loop finiteness of this theory requires a counterterm $\Delta\bar S$. When graviton internal lines are taken into account, on shell we have $\Delta S=\Delta\bar S$. However, when only the scalar field is quantized one finds that $\Delta S\neq\Delta\bar S$ \cite{Duf80}. Therefore, the same classical theory could be written in infinitely many different ways and one would have to invoke a criterion selecting one frame among all the others. This may be problematic, but the existence of such a criterion is not altogether unreasonable: for instance, one could impose positivity of energy and choose the Einstein frame $\bar g_{\mu\nu}$ as the frame where the fundamental theory is defined \cite{FaGN}.

Duff's example illustrates why two classically equivalent frames can differ at the quantum level and a frame choice must be made. In multifractional theories, the situation is different because in the fractional frame we do not know how to deal with the quantum theory \cite{frc9,frc13} (see \lref{20} and \lref{22}). In the multifractional case, the choice where to do QFT is somewhat mandatory: we move to the integer frame to do all intermediate QFT calculations before getting physical observables. The latter are obtained in the end in the fractional frame, which was selected as preferred already at the classical level. This marks a difference with respect to the scalar-tensor case, where the frame choice dictated by some principle is necessary only at the quantum level.

\bigskip

\llabel{27} \textbf{\emph{Even accepting that it is part of the definition of these theories to establish a frame choice, what is the meaning of the point $x=0$ in eqs.\ \Eq{meamu} and \Eq{meag}? If we write the measure in one direction as $\rmd q(x)=\rmd x\,v(x)$, then the measure weight $v(x)\sim 1+|x/\ell_*|^{\a-1}$ is singular at $x=0$ because $\a<1$. So where are we with respect to this singularity? What are the physical consequences of having this uniquely special spacetime point?}}\addcontentsline{toc}{subsection}{\lref{27} What is the meaning of the singularity in the measure?}

This question hits one of the most peculiar aspects of multifractional theories, known as the \emph{presentation problem}. Let us explain it in detail for the theory with $q$-derivatives, following \cite{trtls} up to some point but greatly improving on the interpretation and on the physics thanks to the second flow-equation theorem. The model $T_1$ and the theory $T_v$ face a similar issue, while the theory $T_\g$ is a separate case. 

Suppose we wish to measure the distance $\De x$ of two points A and B on a sheet of paper. If the paper is charted by a Cartesian system, then the distance is given by the two-dimensional Euclidean norm $\De x:=\sqrt{|x_{\rm B}^1-x_{\rm A}^1|^2+|x_{\rm B}^2-x_{\rm A}^2|^2}$. Then we make a coordinate transformation $x^i\to {x'}^i$ such that $\De x=F({x_{\rm A}'}^i,{x_{\rm B}'}^i)$ is a function of the new coordinates. For instance, going to polar coordinates $\{x^1,x^2\}\to \{\vr,\theta\}$ conveniently centered at $x_A$, one has $\De x=r$. The observed value of the distance is insensitive to the coordinates we choose to represent $\De x$ with. In the theory $T_q$, we can try to do the same in the fractional picture, which is one of the coordinate frames $\{x^1,x^2\}$ where the distance $\De x$ is calculated. However, to each of these fractional frames we must associate an integer frame described by geometric coordinates. Thus, the Cartesian fractional frame $\{x^1,x^2\}$ is mapped into the integer frame $\{q^1(x^1),q^2(x^2)\}$ and, after inverting to $x^i=x^i(q^i)$ (assuming it possible, which is not always the case) the Euclidean norm $\De x$ is mapped into some complicated expression $\De x(q_{\rm A}^i,q_{\rm B}^i)$ differing from the geometric Euclidean norm $\De q:=\sqrt{\sum_{i=1}^2|q_{\rm B}^i-q_{\rm A}^i|^2}$ by correction terms $\cX$ and $\cT$ we will calculate below. If we redo the mapping to geometric coordinates starting from polar fractional coordinates, we get another integer frame $\{q_r(r),q_\theta(\theta)\}$, where the relations between $q_r$ and the $q^i$ are $q^1=q_r\cos q_\theta$ and $q^2=q_r\sin q_\theta$. Thus, on which chart is eq.\ \Eq{meamu} or \Eq{meag} represented? In the example of the paper sheet, is eq.\ \Eq{meamu} the form of $q$ in the integer frame $\{q^1(x^1),q^2(x^2)\}$ based on Cartesian coordinates $\{x^1,x^2\}$ or the form of $q$ in the integer frame $\{q^1(r),q^2(\theta)\}$ based on polar coordinates $\{r,\theta\}$ [so that $q^1(r)=r+(\ell_*/\a)(r/\ell_*)^{\a}$], or something else? Ordinary Poincar\'e invariance is violated by factorizable measures \Eq{facto}. A change of presentation such as a translation, a rotation of the coordinates or an ordinary Lorentz transformation modifies the size of the multiscale corrections to the measure. One realizes that different choices of the fractional frame lead to different theories in the integer frame. Clearly, $q^1(r)\neq \sqrt{[q^1(x^1)]^2+[q^2(x^2)]^2}$ due to the nonlinear terms in the geometric coordinates.

In factorizable measures \Eq{facto}, coordinates never mix together due to the absence of rotation and boost invariance. The only transformations preserving this structure are translations, which encode the ambiguity of presentation:
\be\label{prese}
q^\mu(x^\mu)\to \bar q^\mu(x^\mu)=q^\mu(x^\mu-\bar x^\mu)\,.
\ee
Given an interval $\De x^\mu=|x_{\rm B}^\mu-x_{\rm A}^\mu|$ between two points A and B lying on the $\mu$-th direction, its geometric analogue $\De \bar q^\mu = |\bar q(x_{\rm B}^\mu)-\bar q(x_{\rm A}^\mu)|$ for a binomial measure is
\be\label{Dex1}
\De \bar q^\mu = \De x^\mu|1\pm\cX^\mu|\,,
\ee
where
\be\nonumber
\cX^\mu:=\pm\frac{1}{\a_\mu}\frac{\ell_*^\mu}{\De x^\mu}\left[\left|\frac{x_{\rm B}^\mu-\bar x^\mu}{\ell_*^\mu}\right|^{\a_\mu} F_\om(x_{\rm B}^\mu-\bar x^\mu)-\left|\frac{x_{\rm A}^\mu-\bar x^\mu}{\ell_*^\mu}\right|^{\a_\mu}F_\om(x_{\rm A}^\mu-\bar x^\mu)\right].
\ee
We define four different presentations characterized by special values of $\bar x^\mu$: \emph{null presentation} $\bar x^\mu=0$, \emph{initial-point presentation} $\bar x^\mu=x_{\rm A}^\mu$, \emph{final-point presentation} $\bar x^\mu=x_{\rm B}^\mu$, and \emph{symmetrized presentation} $\bar x^\mu=(x_{\rm B}^\mu+x_{\rm A}^\mu)/2$. At a first sight, one might want to discard all but the null presentation, which is the only one where the $\bar x^\mu$ do not depend on the ``beginning'' or ``end'' of the experiment (the measure of the geometry should be the same for all experiments). However, we now show that the most natural choice is quite the contrary, the initial- and final-point presentations! The origin of the multiscale measure of the theory has been recently clarified by the second flow-equation theorem and it reveals an important omission in eqs.\ \Eq{meamu} and \Eq{meag}, which we correct here for the first time. There, we interpreted $q^\mu(\ell^\mu)$ as the integral (indices $\mu$ inert)
\be\label{nullp}
q^\mu(\ell^\mu)=\int_0^{q^\mu(\ell^\mu)}\rmd q^\mu(x^\mu)\stackrel{?}{=}\int_0^{\ell^\mu}\rmd x^\mu\,v_\mu(x^\mu)\qquad \forall\,\mu\,,
\ee
but the following alternative is equally valid and based on the fact that the scales $\ell^\mu=|x_{\rm B}^\mu-x_{\rm A}^\mu|$ are distances:
\be\label{infip}
q^\mu(\ell^\mu)=\int_{x_{\rm A}^\mu}^{x_{\rm B}^\mu}\rmd x^\mu\,v_\mu(x^\mu-\bar x^\mu)\qquad \forall\,\mu\,,\qquad \bar x^\mu=x_{\rm A}^\mu, x_{\rm B}^\mu\,.
\ee
Equation \Eq{nullp} corresponds to the null presentation or, in other words, the null presentation is the choice of integration interval $[0,\ell^\mu]$. With posterior wisdom, it is almost obvious that this choice is not particularly happy. On one hand, it takes both coordinate extrema $x_{\rm A}^\mu$ and $x_{\rm B}^\mu$ on the upper limit of the integral, which should already sound an alarm bell because it fixes the edge origin. On the other hand, it eventually leads to corrections $\cX^\mu(x_{\rm A},x_{\rm B})$ that depend on the initial and final coordinate separately. The symmetrized presentation relies on an even more unnatural choice of integration domain and it leads to a trivial theory with $\cX^\mu=0$. In contrast, eq.\ \Eq{infip} is valid both in the initial-point presentation [its right-hand side is $q^\mu(\ell^\mu)-q^\mu(0)=q^\mu(\ell^\mu)$, since $q^\mu(0)=0$] and in the final-point presentation [the right-hand side is $q^\mu(0)-q^\mu(-\ell^\mu)=q^\mu(\ell^\mu)$, since the $q^\mu$ are odd in their argument]. The integration domain is now the natural one $[x_{\rm A}^\mu,x_{\rm B}^\mu]$ and the corrections $\cX^\mu(x_{\rm B}^\mu-x_{\rm A}^\mu)$ now depend only on the spatial distance or time interval, but not on the initial and final coordinates separately. The desirability of this feature for physical predictions is evident and it was implicitly used in all phenomenology-oriented papers \cite{frc8,frc12,frc13,qGW,frc14}.\footnote{Although incorrectly associated with a measure in null presentation.
In particle-physics experiments, one regards the point $\bar t$ as the beginning of the observation or the moment when a certain collision occurs or a certain particle is created, while $t_*$ is the time, measured from $\bar t$, before which multiscale effects are important \cite{frc13}. In cosmology, $\bar t$ is the discriminator between ``early'' times $\Delta t=t-\bar t\lesssim t_*$ and ``late'' times $\Delta t\gg t_*$; $\Delta t$ represents the moment when a cosmological phenomenon takes place with respect to some special instant $\bar t$ in the history of the universe, which may be the big bang \cite{frc8,frc11,frc14}. And so on.} Therefore, we supersede the discussion of \cite{trtls} and rule out the null and symmetrized presentations from the game, leaving only the initial- and final-point presentations. The correction in eq.~\Eq{Dex1} reads
\be\label{XT}
\cX^\mu=\frac{1}{\a_\mu}\left|\frac{\ell_*^\mu}{\ell^\mu}\right|^{1-\a_\mu}F_\om(\ell^\mu).
\ee
The sign in eq.\ \Eq{Dex1} depends on the choice between initial-point presentation ($+$) and final-point presentation ($-$). Thus, the presentation (i.e., the value of $\bar x^\mu$) affects the output of physical measurements via the sign in front of multiscale corrections. But how can we reconcile the initial- and final-point presentations with the requirement that the constant $\bar x^\mu$, fixed in the measure of geometry, be the same for all observers? In Minkowski spacetime for the theories $T_1$, $T_v$, and $T_q$, we cannot because the weights $v(x-\bar x)$ appearing in the derivatives in the equations of motion break translation invariance. On a curved background, however, the chart where the measure $q(x-\bar x)$ is defined is the local inertial frame of an observer, and the multiscale version of such frames exists for $T_v$ and $T_q$ (we do not know about $T_1$ and $T_\g$, but they probably exist at least for $T_\g$). Thus, a local observer is at liberty to choose $\bar x$ in such a way that it coincides with $x_{\rm A}$ or $x_{\rm B}$. We will stress on this point also in question \lref{28}. In the theory with multifractional derivatives \Eq{capuq} or \Eq{capuq2}, the problem is solved [if eq.\ \Eq{capuq} satisfies a set of requirements yet to be checked] without invoking gravity, already in the case of a Minkowski embedding: the derivatives appearing in the equations of motion are translation invariant [eq.\ \Eq{capuqtr}], independently of the choice of $q(x-x')$.

This is the presentation problem. We differentiate between two possible views of it. One, which we dub ``deterministic,'' has been advocated consistently from \cite{frc1} until the appearance of \cite{CaRo2a,CaRo2b}. The other, which we call ``stochastic,'' has been proposed in ref.\ \cite{CaRo2a,CaRo2b}.  Although the deterministic view works, the stochastic view may work even better because it solves the presentation problem not by the brute force of Aristotelian logic (either one presentation or the other, \emph{tertium non datur}), but by accepting \emph{both} presentations at the same time.
\begin{itemize}
\item[--] \emph{Deterministic view.} The tenet of this view is that a change of presentation changes the theory, i.e., the sign and magnitude of the corrections $\cX^\mu$. Due to the smallness of these corrections, all qualitative features are unaffected \cite{trtls}. It is well known that inequivalent presentations leave the anomalous scaling of the measure and the dimension of spacetime untouched \cite{frc1,frc2}, basically for the same scaling argument by which the volume of a hypercube or of a $D$-ball scale in the same way (question \loref{13}). Therefore, multifractional scenarios are robust across different presentations, including those that we disfavoured above. Picking a presentation  corresponds to defining the theory and allows us to make predictions which will change in another presentation (i.e., another theory connected to the first by a one-parameter transformation), but not by much. 
\item[--] \emph{Stochastic view.} Instead of making a choice between two inequivalent but equally valid theories, we can can try to have both theories coexist. Since it is impossible to choose between the initial- and final-point presentation without an external input, we conceive a ``macrotheory'' with an \emph{intrinsic uncertainty} in the presentation, so that the term $\pm\chi^\mu$ in eq.\ \Eq{Dex1} is interpreted as an irresoluble uncertainty in distance and time measurements \cite{CaRo2a,CaRo2b}. The mechanism to do so is not quantum mechanics but a stochastic reinterpretation of the coordinates of multifractional spacetimes \cite{trtls,CaRo2a,CaRo2b}. The stochastic view could be realized in two ways, which are still under study. One is by using log oscillations as a direct source of fluctuations, mimicking a stochastic effect when they average to zero \cite{CaRo2a,CaRo2b}; this possibility applies exactly to all multifractional theories. The other way, which we will consider here, goes through the integration and differential structure as a whole \cite{trtls}, in which case this view is naturally implemented in $T_\g$ (where the restriction to having a normed space is naturally lifted \cite{CaRo2a,CaRo2b}), while it is ``superposed'' to the structure of $T_1$, $T_v$, and $T_q$. Since the theory closer to $T_\g$ is $T_q$, we can apply this view successfully only in these two cases, the second being an approximation. For $T_1$ and $T_v$, we have to adopt the usual deterministic view, so that the presentation problem persists in the absence of gravity and is reduced to two presentations (determining the maximal uncertainty) in its presence.
\end{itemize}
Whenever we can choose either the initial- or the final-point presentation, in both views there is no reference to any special point in the coordinate chart $\{x^\mu\}$ defining the fractional frame. Geometry becomes a pure relativity of scales. To summarize:
\begin{itemize}
\item[--] $T_\g$: the number of allowed presentations is two (initial- and final-point) in both flat and curved space. The deterministic view holds and the two presentations define inequivalent theories that, in principle, can be discriminated by experiments sensitive enough to detect a deviation from standard physics. In alternative, the stochastic view holds exactly and the presentation problem is replaced by an uncertainty on distance and time measurements.
\item[--] $T_q$: the number of allowed presentations is infinite (a one-parameter family) in flat space and is reduced to two in the presence of gravity. The deterministic view holds and the two presentations define inequivalent theories. However, one can also adopt the stochastic view as an approximation and regard the two presentations as an intrinsic uncertainty effect.
\item[--] $T_v$: the number of allowed presentations is infinite in flat space and is reduced to two in the presence of gravity. The deterministic view holds and the two presentations define inequivalent theories. There is no stochastic view.
\item[--] $T_1$: the number of allowed presentations could be reduced to two only in the presence of gravity, provided multiscale local inertial frames existed. The deterministic view holds and different presentations (two or infinitely many) define inequivalent theories. There is no stochastic view.
\end{itemize}

\bigskip

\llabel{28} \textbf{\emph{To show that the presentation problem signals an inconsistency, let us just confine ourselves to classical physics. The fundamental principle governing classical dynamics is that the classical trajectories minimize a quantity that we call the action. While, as we go from one frame to another, the action may look different written in terms of the fields, it is the same quantity that we must calculate in every frame. In other words, once we decide that the action looks a certain way in a given frame, in any other frame its functional form must completely be determined via the usual Lorentz field transformations. This property just follows from the requirement of the invariance of the action. This functional form will not be preserved in multifractional theories, as the nonscalar $v(x)$ changes from one frame to another. So, in essence, one has only one opportunity to choose a unique spacetime point in the universe, and once chosen one does not have the luxury to keep changing it to suit one's needs just because one is conducting different experiments. That would mean that one is changing the action depending upon what experiment one is doing, when and where one is doing. Also, from the point of view of plain diffeomorphisms, the zeros or the singularities of $v(x)$ are special points which have an independent meaning, contrary to diffeo-invariant theories where a point acquires meaning only in relation to the happening of a physical event.}}\addcontentsline{toc}{subsection}{\lref{28} Does the presentation problem make the theory inconsistent?}

Tensor fields in multifractional spacetimes transform with different laws with respect to the standard case \cite{frc6,frc13}, and such laws replace usual Poincaré transformations as detailed in \lref{11}. Advocating arguments based on symmetries that cannot be valid for actions with factorizable measures can only mislead to dead ends. Moreover, when gravity is turned on the singularity point $\bar x$ in the measure is no longer a unique point in the universe. Rather, it is replicated at every local inertial frame (which exist both in $T_v$ and $T_q$), each with its own measure weight $v(x)$ attached. This realizes the intuitive characteristic of self-similar fractals that geometry is anomalous at any point of the set and with the same scaling law everywhere \cite{frc11,trtls}.

Although these arguments suffice, the core of this criticism affects the just-old version of multiscale theories where there was no superselection criterion for the choice of one of the four available presentations. The justification then was that such a choice is simply part of the definition of the theory. In \cite{trtls}, it was also suggested that the theory with fractional derivatives could realize a local notion of anomalous geometry even in the absence of gravity. In that case, fractional calculus is shown to introduce a probabilistic character to spacetime: spacetime points $x$ become stochastic processes $X$; different presentations would simply amount to inequivalent prescriptions of integrodifferential calculus and, in turn, of stochastic integration.

With the advances made in \lref{27}, where we restricted the number of choices to two and removed any reference to a preferred point, we went a long way in giving a more satisfactory answer. The arbitrariness in the presentation has been reduced to only two options, and what was previously interpreted as an integrable singularity of the measure, to be found somewhere in the universe, corresponds in fact to local measurements with no spatial or time extension, $\ell^\mu=0$. This is the physical interpretation we were looking for.

%%%%%%%%%%%%%%%%%%%%%%%%%%%%%%%%%%%%%%%%%%%%%%%%%%%%%%%%%%%%%%%%%%%%%%%%%%%%%
%%%%%%%%%%%%%%%%%%%%%%%%%%%%%%%%%%%%%%%%%%%%%%%%%%%%%%%%%%%%%%%%%%%%%%%%%%%%%

\section{Field theory}\label{qft}

\llabel{29} \textbf{\emph{What is the action of the multifractional Standard Model?}}\addcontentsline{toc}{subsection}{\lref{29} What is the action of the multifractional Standard Model?}

Let $S_{\rm SM}=\int\rmd^Dq(x)\,\cL_{\rm SM}$ be the Standard-Model action, where we split the Lagrangian into an electroweak bosonic, an electroweak leptonic, a quark, a Yukawa, and a Higgs sector: $\cL_{\rm SM}=\cL_\text{ew-bos}+\cL_\text{ew-lep}+\cL_{\rm quark}+\cL_{\rm Yuk}+\cL_\Phi$. The Standard Model Lagrangian with ordinary derivatives is
\bs\label{StaMo}\ba
\cL_\text{ew-bos} &=&-\frac{1}{4}F^a_{\mu\nu}F_a^{\mu\nu}-\frac{1}{4}B_{\mu\nu}B^{\mu\nu}\,,\label{lym}\\
\cL_\text{ew-lep} &=&\rmi\overline{e_{\rm R}}\g^\mu\N_\mu e_{\rm R} +\rmi\overline{L}\g^\mu\N_\mu L\,,\label{lfree}\\
\cL_{\rm quark} &=&\rmi {\rm q}^{\dagger \a i} \bar \s^\mu (\N_\mu {\rm q})_{\a i} +\rmi \bar u^{\dagger}_\a \bar \s^\mu(\N_\mu\bar u)^\a +\rmi \bar d^{\dagger}_\a \bar \s^\mu(\N_\mu\bar d)^\a+\cL[t,b,c,s]\,,\label{lquarks}\\
\cL_{\rm Yuk} &=&-G_e\overline{L}\Phi \,e_{\rm R}+y'\epsilon^{ij}\Phi_i {\rm q}_{\a j}\bar u^\a-y''\Phi^{\dagger i} {\rm q}_{\a i}\bar d^\alpha+\text{H.c.}\,,\\
\cL_\Phi &=&-\left(\N_\mu\Phi\right)^\dagger\left(\N^\mu\Phi\right)+V(\Phi)\,,\label{hggk}\\
V(\Phi) &=& \frac{\la}{4}\left(\Phi^\dagger\Phi-\frac12 w^2\right)^2,\label{hggv}
\ea 
where the field strengths of the ${\rm SU}(2)$ and ${\rm U}(1)$ gauge fields $A_\mu^a$ and $B_\mu$ are
\ba
F^a_{\mu\nu} &=& \p_\mu A^a_\nu -\p_\nu A^a_\mu - g'\epsilon^a_{\ bc} A_\mu^b A_\nu^c\,,\label{fwei}\\
B_{\mu\nu}   &=& \p_\mu B_\nu -\p_\nu B_\mu\,,\label{bwei}
\ea
the gauge covariant derivatives are
\ba
\N_\mu L &=& \left(\p_\mu + \frac{\rmi}{2}g'\s_a A^a_\mu +\frac{\rmi}{2}g B_\mu\right)L\,,\label{covL}\\
\N_\mu e_{\rm R} &=& (\p_\mu +\rmi g B_\mu) e_{\rm R}\,,\\
(\N_\mu {\rm q})_{\a i} &=& \p_\mu  {\rm q}_{\a i} + \rmi g_s C^a_\mu (\la^a)_\a^{\ \b} {\rm q}_{\b i}+ \frac{\rmi}{2}g'{A}^a_\mu (\sigma_a)_{i}^{\  j} {\rm q}_{\a j}+\frac{\rmi}{6}g  B_\mu {\rm q}_{\a i}\,,\label{covQ}\\
(\N_\mu \bar u)^\a &=&\p_\mu  \bar u^\a + \rmi g_s C^a_\mu (\la^a)^\a_{\ \b} \bar u^\b -\frac{2\rmi}{3}g  B_\mu \bar u^\a\,,\label{covbarU}\\
(\N_\mu \bar d)^\a &=&\p_\mu  \bar d^\a + \rmi g_s C^a_\mu (\la^a)^\a_{\ \b} \bar d^\b +\frac{\rmi}{3}g  B_\mu \bar d^\a\,,\label{covbarD}\\
\N_\mu\Phi &=& \text{same as $\N_\mu L$}\,,\label{nab1}
\ea\es
the $\s_a$ are the $2\times 2$ Pauli matrices [generators of ${\rm SU}(2)$], $\g^\mu$ are the Dirac matrices, $\la^a$, $a=1,\ldots,8$, are the $3\times 3$ Gell-Mann matrices [generators of ${\rm SU}(3)$], $\bar \s^\mu = (\mathbbm{1},-\s^a)$, $C^a_\mu$ are the color gauge potentials, $L=\left(\begin{matrix} \nu_e\\ e_{\rm L}\\ \end{matrix}\right)$ is the left weak isospin doublet, $e_{\rm R}$ is the right isospin singlet, in $\cL_{\rm quark}$ we wrote only the first quark family $(u,d)$, ${\rm q}_i$, $i=1,2=u,d$ is a left-handed Weyl spinor under ${\rm SU}(2)$, $\bar u$ and $\bar d$ are antiquarks [singlets under ${\rm SU}(2)$], $\cL[t,b,c,s]$ is the Lagrangian for the other quarks, ``H.c.'' means Hermitian conjugate, $\Phi$ is the Higgs doublet, and $V(\Phi)$ is its potential. In standard Minkowski spacetime, $S_{\rm SM}=\int\rmd^Dx\,\cL_{\rm SM}$.
\begin{itemize}
\item[--] The Lagrangian $\cL_{\rm SM}$ in $T_1$ is the usual one \Eq{StaMo}. The couplings $g$, $g'$, $g_s$, $G_e$, $y'$, $y''$, $\la$, and $w$ are all constant.
\item[--] In $T_v$ \cite{frc13}, we have eqs.\ \Eq{StaMo} with $\p_\mu$ replaced by $\cD_\mu$ everywhere. The couplings $g$, $g'$, $g_s$, $G_e$, $y'$, $y''$, $\la$, and $w$ are all measure dependent with the following form:
\ba
C(x) &=&\sqrt{v(x)}\,C_0\,,\qquad C=g,g',g_s,G_e,y',y''\,,\nonumber\\ C_0&=&g_0,g_0',g_{s0},G_{e0},y_0',y_0''={\rm  const}\,,\label{gggyy}\\
\la(x) &=&v(x)\,\la_0\,,\qquad w(x)=\frac{w_0}{\sqrt{v(x)}}\,,\qquad \la_0,w_0={\rm  const}\,.\label{law}
\ea
\item[--] In $T_q$ \cite{frc12,frc13}, we have eqs.\ \Eq{StaMo} with $\p_\mu$ replaced by $\p/\p q^\mu(x^\mu)$ everywhere. The couplings $g$, $g'$, $g_s$, $G_e$, $y'$, $y''$, $\la$, and $w$ are all constant.
\item[--] The Standard Model in $T_\g$ has never been written down since it requires more study of the fundamentals of the theory. We do not know whether it can be defined simply by replacing $\p_\mu$ in \Eq{StaMo} with ${}_q\cD_\mu$ or $\mathbbm{D}_\mu$ everywhere.
\end{itemize}

\bigskip

\llabel{30} \textbf{\emph{Why to extend a so well functioning Standard Model with a multiscale version of it?}}\addcontentsline{toc}{subsection}{\lref{30} Why to extend a so well functioning Standard Model with a multiscale version of it?}

As discussed in point \loref{04}, the main motivation of multifractional theories is not phenomenological, it is to address two fundamental problems of quantum gravity: the physical meaning and consequences of dimensional flow and whether it is possible to carry the quantization program in a perturbative framework. Once dimensional flow is implemented via the second flow-equation theorem in the measure of the theory, it affects virtually all sectors of physics, including that of fundamental quantum interactions. Then, in order to assess the viability of multifractional theories, it is mandatory to explore all such sectors, in particular the consequences of a multiscale spacetime on the theoretical and observational characteristics of QFT. The extension of the Standard Model is not an objective \emph{per se}; certainly, it is an occasion to place strong constraints on the free parameters of the measure, hence our interest in it.

\bigskip

\llabel{31} \textbf{\emph{In the theory with weighted derivatives, constants are promoted to fields, sometimes only time-dependent (for instance, the electric charge mentioned in question \lref{21}), sometimes not. What is the rationale behind these choices?}}\addcontentsline{toc}{subsection}{\lref{31} What is the origin of the spacetime-dependent couplings in the theory with weighted derivatives?}

There are four specifications to make from the start. First, all the couplings in gauge covariant derivatives and field interactions in the fractional-picture action of the Standard Model depend on the measure weight \Eq{vfacto}, which is a fixed profile of spacetime coordinates. Thus, according to eqs.\ \Eq{gggyy} and \Eq{law}, they depend on both time and space. Second, constants are not promoted to fields because the measure weight \Eq{vfacto} is not a scalar field. Third, the spacetime-dependent couplings \Eq{gggyy} and \Eq{law} are not \emph{ad hoc} but originate from gauge invariance and the requirement of being able to do free field theory. Fourth, one must distinguish between the couplings in the Lagrangian and observable couplings.

Let us clarify the origin of eqs.\ \Eq{gggyy} and \Eq{law}. Consider a generic Yang--Mills theory $S=\int\rmd^Dx\,v\,\cL$ with a gauge bosonic vector field $A_\mu^a$ (Abelian in the case of electromagnetism, non-Abelian in general) and fermionic matter $\Psi$ \cite{frc13}:
\be\label{lagyan}
\cL=-\frac{1}{2}{\rm tr}(\boldsymbol{\cF}_{\mu\nu}\boldsymbol{\cF}^{\mu\nu})+\rmi\overline{\Psi}\g_\mu\N_\mu\Psi-m\overline{\Psi}\Psi\,,
\ee
where $\boldsymbol{\cF}_{\mu\nu}:=\cF^a\!_{\mu\nu} t_a$, $F^a_{\mu\nu} =g^{-1}[\p_\mu(g A^a_\nu)-\p_\nu(g A^a_\mu)] - g f^a_{\ bc} A_\mu^b A_\nu^c$ is the field strength of $A$, $g=g(x)$ is the gauge coupling, and $t^a$ are the matrix representations of the Lie algebra $[t_a ,t_b]=\rmi f_{abc}t^c$ associated with the gauge group. The covariant derivative in \Eq{lagyan} is
\be \label{gcov}
\N_\mu=\cD_\mu +\rmi g  A_\mu^a t_a\,,
\ee
where $\cD_\mu$ is defined in \Eq{Ks}. \emph{A priori}, the coupling $g(x)$ can be spacetime dependent; to see what this dependence is, one defines the gauge-invariant matter current $J^\mu_a:=-g\,\overline{\Psi}\g^\mu t_a\Psi$, which is covariantly conserved:
\be\label{concu}
\N_\mu J^\mu_a=0\,.
\ee
Also, the Lagrangian density \Eq{lagyan} is invariant under a ${\rm U}(1)$ symmetry whose Noether current obeys the generalized conservation law $\check{\cD}_\mu(\overline{\Psi}\g^\mu\Psi)=0$, where $\check{\cD}_\mu=v^{-1}\p_\mu(v\,\cdot \,)$ (in the theory $T_v$, this type of derivative appears often in some conservation laws and in gravity \cite{frc11,frc13}). Since $\N_\mu J^\mu_a=0$ and $\check{\cD}_\mu(\overline{\Psi}\g^\mu\Psi)=0$ must agree when $f_{abc}=0$, this implies
\be \label{g}
g(x)=\sqrt{v(x)}g_0\,,
\ee
where $g_0$ is a constant. Therefore, all couplings in the fractional picture have the spacetime dependence given by eq.\ \Eq{g}, where $v(x)$ is determined by the second flow-equation theorem. In particular, the ${\rm U}(1)$ charge of electromagnetism in the fractional picture is $e(x)=\sqrt{v(x)}e_0$. The same dependence is found in Yukawa interactions. In the Higgs sector, the scalar potential is \Eq{hggv}, which gives a nonzero vacuum expectation value to the Higgs doublet. To obtain a Standard Model whose free sector is stable in the integer picture (a necessary requirement, if we want to have a manageable perturbation theory), both $\lambda$ and $w$ must acquire a specific dependence on the measure weight $v(x)$, given by eq.\ \Eq{law} \cite{frc13}. Then, the Higgs mass is the same in both the fractional and the integer picture.

None of the above couplings is a physical observable. In the case of weak interactions, all observable couplings (for example, the Fermi constant or the masses of the $W$ and $Z$ gauge bosons) are a combination of Lagrangian couplings, and it turns out that the measure dependence cancels out in such combinations \cite{frc13}. As a consequence, no exotic signatures are predicted in the weak sector alone. The electromagnetic sector is more interesting. The deformed conservation law for ${\rm U}(1)$ is a special case of eq.\ \Eq{concu}, $\cD_\mu J^\mu =0$, which leads to the nonconservation equation of the electric charge \cite{frc8}
\be\label{ech}
Q(t):=\int \rmd^{D-1}{\bf x}\,v({\bf x})\,J^0(t,{\bf x})\simeq \frac{e_0}{\sqrt{v_0(t)}}\,,
\ee
where $v({\bf x})$ is the spatial part of \Eq{vfacto} and the last approximated expression was found in ref.\ \cite{frc8}. Therefore, this particular observable coupling is only time-dependent because it comes from the usual definition of Noether charges. All the other Lagrangian couplings \Eq{gggyy} in the strong and weak sectors are spacetime-dependent but this property is not seen in the observable couplings of strong and weak interactions.

In the other multifractional theories where the Standard Model is obvious ($T_1$) or has been constructed ($T_q$ \cite{frc12,frc13}), there is no effective definition of spacetime-dependent couplings and they are all constant. The particle phenomenology is different from the case $T_v$; in particular, the weak sector of $T_q$ is nontrivial observationally.

\bigskip

\llabel{32} \textbf{\emph{Can multiscale effects be mimicked by more traditional extensions of the Standard Model such as effective field theories? There may be a strong call, from particle physicists, for some simplified exposition of the main ideas of multifractional theories. For instance, effective field theories speak the language in which most extensions of the Standard Model are usually formulated. It would be of big value to derive which higher-dimensional operators should be added to the Standard-Model Lagrangian to mimic multiscale effects.}}\addcontentsline{toc}{subsection}{\lref{32} Can multiscale effects be mimicked by more traditional extensions of the Standard Model such as effective field theories?}

To explain the QFT results in $T_v$ intuitively to a bigger circle of phenomenologists, it may be useful to make a link with more familiar formulations of physics beyond the Standard Model (this answer is taken from \cite{frc13}). What discussed in \lref{31} can be summarized by saying that the presence of an underlying multiscale geometry affects field theory in such a way that interaction terms (in gauge derivatives or in nonlinear potentials) acquire an explicit spacetime dependence of the form
\be\label{fff}
[1+f(x)]\phi^i\phi^j\cdots\,,
\ee
where $f(x)=f[v(x)]$ depends on the measure weight $v(x)$ and $\phi^i$ are some generic fields. Terms such as \Eq{fff} have a naive interpretation of ``having promoted coupling constants to fields'' and, in some sense, some of the physical effects we encounter are similar to those in models with varying couplings. Another possibility to mimic effects of the form \Eq{fff} is to add higher-dimensional operators to a traditional Lagrangian. For instance, in a scalar-field theory one would have
\be\nonumber
V(\phi)\to[1+f(x)]V(\phi)\sim [1+\phi^m+\phi^n+\cdots]V(\phi)
\ee
for some exponents $m$ and $n$, and one would fall into the context of effective field theories.

These are only superficial analogies not capturing the real nature of the multiscale paradigm. The most evident departure is that $v(x)$ is not a scalar field and none of the above interpretations based on ordinary field theories has any such premade, nontrivial integrodifferential structure. Since $v(x)$ [hence $f(x)$] is fixed by the geometry, it cannot be interpreted as a field and the higher-order-operators comparison dies as soon as one writes down the classical or quantum dynamics [classically, one does not vary with respect to $v(x)$; at the quantum level, $v(x)$ does not propagate]. The varying-coupling analogy is also of limited utility in the long run, since it does not explain why only certain couplings, but not others, depend on spacetime.

\bigskip

\llabel{33} \textbf{\emph{Is field theory unitary?}}\addcontentsline{toc}{subsection}{\lref{33} Is field theory unitary?}

No, it is not, but it does not lead to problems. To accept this paradoxical answer, we should examine its grounds. In a general multiscale geometry, the usual symmetries are deformed and Noether currents are modified accordingly. In the theory $T_v$, these currents obey conservation laws such as \Eq{concu}, where the gradient operator encodes the multiscale nature of the geometry. In the theory $T_q$, one has conservation laws with $q$-derivatives, $\N_{q^\mu(x^\mu)}J^\mu_a=0$. Thus, once written according to the differential structure typical of the theory, there is a notion of current conservation that implies unitarity. More precisely, both $T_v$ and $T_q$ admit an integer picture where we have a standard unitary QFT, which is necessary and sufficient to compute the QFT observables of the theory. On the other hand, however, the conservation laws with nonstandard derivatives are equivalent, in $T_v$ and $T_q$ \cite{frc8,frc13}, to nonconservation laws with standard derivatives, which means that these theories in the fractional picture, where the weighted and $q$-derivatives are written as standard derivatives multiplied by measure factors, are classically \emph{dissipative}, i.e., nonunitary at the quantum level \cite{frc2,frc6}. Therefore, although the auxiliary QFT developed in the integer picture is unitary, the QFT in the fractional picture is nonunitary. Also the study of quantum mechanics indicate that unitarity is violated but in a controllable way \cite{frc5}.

The model $T_1$ has no integer picture and the system is manifestly nonconservative. It was in this context, similar to the more general case of multiscale field theories with nonfactorizable measures, that violation of unitarity was first predicted \cite{fra2}. However, even if field theory can never be trivialized, nonconservation laws can be interpreted as governing an exchange of probability densities between the multiscale world and its $D$-dimensional topological bulk \cite{fra2}. Again, nonunitarity is there but under check.

Presumably, in the theory $T_\g$ there will be conservation laws of the form ``$\p^\a_\mu J^\mu=0$,'' where the gradient is made of multifractional derivatives such as \Eq{dissum}, \Eq{capuq}, or \Eq{capuq2}. Then, conservation in terms of first-order ordinary gradients will appear only in the IR asymptotic regime (it cannot be exact: fractional derivatives reduce to ordinary ones only asymptotically). Or, nonconservation equations as in the $T_1$ case could appear.

\bigskip

\llabel{34} \textbf{\emph{What is the propagator in multifractional theories?}}\addcontentsline{toc}{subsection}{\lref{34} What is the propagator in multifractional theories?}

We give the example of a real massive scalar in flat space, which captures all the main features of propagators. Also, we omit the causal prescription of the propagator and consider a generic Green function solving $(\cK_x-m^2)\,G(x,x')=\de_q(x,x')$, where $m$ is the mass of the scalar and $\de_q(x,x')$ is the equivalent of the Dirac distribution in a multifractional geometry. In general, the structure of the Green function is
\be
G(x,x')=\int\frac{\rmd^Dp(k)}{(2\pi)^D}\,\bE(k,x)\bE^*(k,x')\,\tilde G(k)\,,\qquad \tilde G(k)=-\frac{1}{-\tilde\cK(k)+m^2}\,,
\ee
where $\rmd^D p(k)=\prod_\mu\rmd p^\mu(k^\mu)=:\rmd^Dk\,w(k)$ is the measure in momentum space [$w(k)=\prod_\mu w_\mu(k^\mu)$ is the measure weight], $k^\mu$ are the momentum coordinates in the fractional frame, and $\bE(k,x)$ are the ``plane waves'' of the theory, i.e., the eigenfunctions of the Laplace--Beltrami operator $\cK$: $\cK_x \bE(k,x)=\tilde\cK(k)\,\bE(k,x)$.
\begin{itemize}
\item[--] In $T_1$, the fact that $\cK^\dagger\neq\cK=\B$ implies that the action \Eq{Sphi} is physically inequivalent to an action with kinetic term $-(1/2)\p_\mu\phi\p^\mu\phi$. In the first case, the Green function in momentum space is
\be\label{prop1}
\tilde G_1(k)=-\frac{1}{k^2+m^2}\,,
\ee
where $k^2=k_\mu k^\mu=-(k^0)^2+|{\bf k}|^2$. There are two poles at ${\rm Re} k^0=\pm\sqrt{m^2+|{\bf k}|^2}$ and the usual interpretation of fields as particles. In the case of the kinetic term $-(1/2)\p_\mu\phi\p^\mu\phi$, the structure of $\tilde G(k)$ is completely different and branch cuts may arise for $\a=2/D$ (this conclusion is reached by adapting the findings of ref.\ \cite{fra2} for $\tilde T_1$ to the factorizable measure of $T_1$). These problems disappear in $T_v$, the natural upgrade of $T_1$.
\item[--] In $T_v$, the Dirac distribution is $\de_q(x,x')=\de(x-x')/\sqrt{v(x)v(x')}$, the plane waves are the weighted phases $\bE(k,x)=\exp(\rmi x_\mu k^\mu)/\sqrt{w(k)v(x)}$ \cite{frc3}, $\tilde\cK(k)=-k^2$, and the Green function in momentum space is eq.\ \Eq{prop1} \cite{frc2,frc6}. Again, we have two mass poles and fields are associated with particles.
\item[--] In $T_q$, the momentum measure is eq.\ \Eq{mompk}. The delta distribution is $\de_q(x,x')=\de[q(x)-q(x')]$, plane waves are $\bE(k,x)=\exp[\rmi q_\mu(x_\mu) p^\mu(k^\mu)]$, and the Green function is attractively simple in geometric coordinates:
\be\label{prop2}
\tilde G_q(k)=-\frac{1}{p^2(k)+m^2}\,,\qquad p^2(k):=\sum_\mu [p^\mu(k^\mu)]^2=k^2+\dots\,.
\ee
The usual poles are replaced by branch points, which cannot be determined analytically in general.
\item[--] In $T_\g$, we have not calculated the full multiscale propagator yet, but we can guess its general structure at any plateau of $\dh$, where (up to weight factors) $\cK_\g\sim\p^{2\g}$ is a fractional derivative and $\cK_\g \bE(k,x)\sim |k|^{2\g} \bE(k,x)$ for each direction and up to a constant (to be determined by the type of derivative \cite{frc1,frc4}) \cite{frc1,frc2,frc4}. Then, the Green function is something of the form (mass term rescaled) \cite{frc2}
\be\label{prop3}
\tilde G_\g(k)\sim-\frac{1}{F^{2\g}(k)+m^{2\g}}\,,\qquad F^{2\g}(k):=\sum_\mu |k^\mu|^{2\g}\,,
\ee
which, for $2\g\not\in\mathbbm{N}$, has a branch point at ${\rm Re} k^0=\pm(m^{2\g}+\sum_i |k^i|^{2\g})^{1/(2\g)}$ and a branch cut corresponding to a continuum of modes of rest mass $>m$.
\end{itemize}

\bigskip

\llabel{35} \textbf{\emph{The multiscale idea is quite exotic because it involves a nonstandard algebra of derivatives, and it may be difficult to understand its consequences for a field theory. What is the physics behind perturbation theory?}}\addcontentsline{toc}{subsection}{\lref{35} What is the physics behind perturbation theory?}

This question is difficult to answer because QFT is yet unknown in $T_\g$ (and in the less interesting case $T_1$), while in $T_v$ and $T_q$ it is under full control only in the integer picture. We do not know much about the physical interpretation, i.e., about what happens in the fractional picture. The following descriptions are an orientative start.
\begin{itemize}
\item[--] The absence of symmetries and of a self-adjoint Laplacian has fatally stalled progress in the case of $T_1$. An example of the problems one may incur into is in the form of the propagator, which changes with the prescription made on the kinetic term (see the previous question). Therefore, we directly move to its upgrade $T_v$.
\item[--] In the theory with weighted derivatives $T_v$, we have point particles but a perturbative treatment of their interactions does not follow conventional Feynman rules. The main problem is that ordinary momentum is not conserved, as remarked in \lref{11} and \lref{33}. Vertices in anomalous geometries do not combine like delta distributions as in ordinary QFT, since the Dirac delta is smeared to a sort of landscape of volcanoes [one for each term $n$ in eq.\ \Eq{meag}]. Each external momentum brings a distribution $\sim |k|^{-\b}$ (where $\b$ depends on $\a_n$) which does not combine into a vertex distribution $\sim |k_{\rm tot}|^{-\b}$.
\item[--] In the theory with $q$-derivatives $T_q$, we do not even have a notion of particle in the fractional picture, due to the form of \Eq{prop2}. Once recast the system into the integer picture both in position and in momentum space, we have effective particle fields in an effective ordinary QFT [mass poles at $p^2=-m^2$ in eq.\ \Eq{prop2}], and everything goes through smoothly. But in the physical frame, none of that holds. As in the case of $T_v$, it seems that quantum interactions are heavily altered by taking place in a multiscale anomalous geometry, which dissipates energy and momentum into the embedding bulk. In other words, quantum physics cannot be described by the nonadaptive measurements units of the fractional picture but, as soon as we consider adaptive units [i.e., multiscale coordinates $q(x)$] and move to the integer picture, a standard QFT emerges. The resulting ``observables'' must be reconverted to nonadaptive measurements, which is all we have in the real world. The surprising thing is that this procedure works and the final physical observables are well defined. It may be that some deep mechanism is in action such that the scale hierarchy of the geometry and the measurement of quantum phenomena by a macroscopic apparatus of size $s$ affect each other in some yet poorly understood way. In some still mysterious sense, the presence of yet another scale $s\gg\ell$ in the system, determined by the measurement apparatus and represented by the final conversion from the integer to the fractional picture, alters the multiscale hierarchy in quantum interactions and tames it to a finite result. The appearance of such a scale in a recent comparison of the multifractional paradigm (with $\a_\mu=1/2$) with quantum-gravity uncertainties \cite{CaRo2a,CaRo2b} may be especially informative.
\item[--] In the case of the theory $T_\g$, the branch cut in eq.\ \Eq{prop3} signals the presence of an infinite number of unstable quasiparticles for which we do not have a representation by Feynman diagrams. We hereby recast the propagator \Eq{prop3} explicitly as such a superposition of pseudoparticle modes. Ignoring the index $\mu$ everywhere and taking $k>0$, we have
\be\label{quas}
-\frac{1}{k^{2\g}+m^{2\g}}=-\int_0^{k}\rmd \k\frac{f(\k)}{\k^2+m^2}\,,\qquad f(\k)=2\g\k^{2\g-1}\frac{\k^2+m^2}{(\k^{2\g}+m^{2\g})^2}\,.
\ee
This continuum of quasiparticles of mass $>m$ is equivalent to the superposition of massive particle modes of momenta $\k$ smaller than $k$, weighted by the distribution $f(\k)$. The momentum distribution is plotted in figure \ref{fig3} for some values of $1/2\leq\g<1$. For $\g=1/2$ and $m\neq 0$, $f(0)=1$ and $f(\k)$ tends to 1 asymptotically at large $\k$; at $\k=m$, there is a global minimum. This case does not correspond to a continuum of quasiparticles since the propagator has a simple pole at $k=-m$ in this case. For $1/2<\g<1$, $f(\k)$ vanishes both at $\k=0$ and asymptotically at large $\k$, with in between a local maximum at some $0<\k<m$ and a minimum at $\k=m$. As $\g$ increases, the maximum gets closer to the minimum until the latter disappears at some critical value $\g=\g_*$; for $\g>\g_*$, the distribution has a global maximum at $\k=m$. In the massless limit for $1/2<\g<1$, the monotonic profile $f(\k)= 2\g\k^{1-2\g}$ diverges at $\k=0$ and vanishes asymptotically at large $\k$. Therefore, for massless fields the main contribution in $f(\k)$ comes from the $\k=k$ mode, while for massive fields it comes from the branch point $\k=m$ for sufficiently large fractional exponent $\g$.
\begin{figure}
\centering
\includegraphics[width=8.4cm]{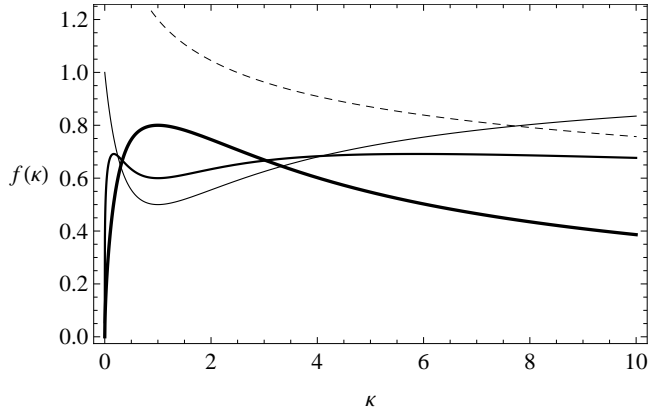}
\caption{\label{fig3} The distribution $f(\k)$ in eq.\ \Eq{quas} for $m=1$ and some values of $\g$: with increasing thickness, $\g=0.5, 0.6, 0.8$. The dashed curve is the $m=0$ case for $\g=0.6$.}
\end{figure}

This rewriting of the fractional Green function in terms of a superposition of ordinary propagators clarifies the physical interpretation of field theory in $T_\g$ and it could help in the construction of perturbative QFT therein.
\end{itemize}

\bigskip

\llabel{35b} \textbf{\emph{Does Lorentz violation lead to a fine tuning in loop corrections, as predicted by the general argument of Collins et al.\ \cite{CPSUV,CPS}?}}\addcontentsline{toc}{subsection}{\lref{35b} Does Lorentz violation lead to a fine tuning in loop corrections?}

In scenarios with modified dispersion relations breaking Lorentz symmetry, the extra terms lead to large fine tunings at the quantum level \cite{CPSUV,CPS}. More precisely, loop corrections to the propagator generally lead to Lorentz violations several orders of magnitude larger than the tree-level estimate, unless the bare parameters of the model are fine-tuned. Thus, even if one starts with a classical theory with tiny Lorentz-symmetry violations, one may end up with an observationally unacceptable enhancement of order of percent.

Usually, this happens in models where the dispersion relation acquires terms which dominate at small scales, as for instance in Lifshitz-type field theories \cite{IRS} and, presumably, in Ho\v{r}ava--Lifshitz gravity. There may exist quantum-gravity models which can bypass that argument \cite{GRP} (but see ref.\ \cite{Pol11}), and it was checked in ref.\ \cite{frc9} that also the multifractional theories $T_v$ and $T_q$ avoid this problem. $T_1$ and $T_\g$ have not been explored.

In both $T_v$ and $T_q$, the key reason is the existence of an integer frame hosting an ordinary, formally Lorentz invariant field theory. After field and coupling redefinitions (in $T_v$) or after moving to geometric coordinates (in $T_q$), loop calculations in the integer frame disclose no bad news. In the case of $T_v$, there also is an explicit calculation in the integer frame with nontrivial measure-dependent interactions \cite{frc9}. The Dyson series for the full quantum propagator $\tilde G$ of a scalar field in momentum space can be formally resummed to $\tilde G = \tilde G_1+A \tilde G_1+A(A \tilde G_1)+\dots=[1-A]^{-1}\tilde G_1$, where $\tilde G_1$ is given by eq.\ \Eq{prop1}, $A:=\tilde G_1\p^2(\tilde\Pi\p^2\,\cdot\,)$, and $\tilde\Pi(k^2) \sim (k^2+m^2) \ln (k^2/m^2)$ is the self-energy function for large $|k^2|$. In a coupling expansion up to quadratic order and keeping only the first two terms of the Dyson series, In the large-$k$ limit one has $\tilde G \sim 1/(k^2-C\ln k^2/k^2)$, where $C$ is a constant. Thus, in the ultraviolet ($k\to\infty$) the correction term is subdominant with respect to the usual one, and the fine-tuning problem is avoided \cite{CPSUV,CPS}. These modifications to the propagator are not introduced by hand, they are derived from the theory. Thus, they also bypass some other general arguments related to Collins et al.'s Lorentz violations \cite{Pol11}.

\bigskip

\llabel{36} \textbf{\emph{What about CPT symmetry?}}\addcontentsline{toc}{subsection}{\lref{36} What about CPT symmetry?}

Having discussed the transformation properties of the fields and the violation of local Poincaré symmetries in \lref{11}, we consider discrete Lorentz transformations: charge conjugation C, parity P, and time reversal T. The requirement of having a positive-semidefinite measure weight implies that the geometric coordinates are odd under reflection $q^\mu(-x^\mu)=-q^\mu(x^\mu)$. Since the measure weight \Eq{vfacto} is even in the coordinates, classical multifractional theories are invariant under parity and time-reversal transformations PT (see \cite{frc2} for $T_\g$ and \cite{frc13} for $T_v$ and $T_q$; the case of $T_1$ is obvious). The presence of measure weights in the action does not affect the charge properties of spinors \cite{frc13}, so that also C is preserved classically.

Since QFT is performed in the integer picture, where $T_v$ and $T_q$ look the same as the ordinary Standard Model, the fate of CPT symmetry at the quantum level is the same as in the usual case, although differences in quantitative predictions may arise by the mechanisms detailed in \lref{32}.

%%%%%%%%%%%%%%%%%%%%%%%%%%%%%%%%%%%%%%%%%%%%%%%%%%%%%%%%%%%%%%%%%%%%%%%%%%%%%
%%%%%%%%%%%%%%%%%%%%%%%%%%%%%%%%%%%%%%%%%%%%%%%%%%%%%%%%%%%%%%%%%%%%%%%%%%%%%

\section{Classical gravity and cosmology}\label{cosmo}

\llabel{37} \textbf{\emph{What is the gravitational action?}}\addcontentsline{toc}{subsection}{\lref{37} What is the gravitational action?}

\begin{itemize}
\item[--] In $T_1$, the action of gravity is \cite{fra2,frc11}
\be
S_1[g,\phi^i] =\frac{1}{2\kappa^2}\int\rmd^Dx\,v\,\sqrt{-g}\,\left[R-\om(v)\p_\mu v\p^\mu v-U(v)\right]+S_1[\phi^i]\,,\label{Sg2}
\ee
where $\om$ and $U$ are functions of the weight $v$, $R$ is the ordinary Ricci scalar and $S[v,\phi^i]$ is the matter contribution minimally coupled with the metric. Apart from the dependence on the measure, the system is nonautonomous (i.e., the Lagrangian depends explicitly on the coordinates) unless $\om=0=U$. Even setting $\om=0=U$, the gravitational sector is not the Einstein--Hilbert action, due to the presence of $v$ in the measure.
The equations of motion are different from those in an ordinary scalar-tensor theory, since $v$ is not a scalar field and the action is not varied with respect to it.
\item[--] In $T_v$, the weighted derivatives in eq.\ \Eq{Ks} are not the only ones appearing in the action of this theory. Derivatives with more general weights
\be\label{bD}
{}_\b\cD:=\frac{1}{v^\b}\p(v^\b\,\cdot\,)
\ee
are necessary when tensor fields of rank greater than 1 enter the dynamics. In the case of gravity, eq.\ \Eq{bD} is used to define the metric connection
\be\label{leci2}
{}^\b\G^\rho_{\mu\nu}:= \tfrac12 g^{\rho\s}\left({}_\b\cD_{\mu} g_{\nu\s}+{}_\b\cD_{\nu} g_{\mu\s}-{}_\b\cD_\s g_{\mu\nu}\right)\,.
\ee
It turns out that the only nontrivial covariant derivative among several consistent possibilities is $\N_\s^- g_{\mu\nu} := \p_\s g_{\mu\nu}-{}^\b\G_{\s\mu}^\tau g_{\tau\nu}-{}^\b\G_{\s\nu}^\tau g_{\mu\tau}$. Covariant conservation $\N^-_\s g_{\mu\nu}=0$ of the metric translates into the Weyl-integrable condition $\N_\s g_{\mu\nu}=(\b\p_\s\ln v)\,g_{\mu\nu}$ with respect to the standard covariant derivative. Defining the fractional Riemann tensor
\be
\cR^\rho_{~\mu\s\nu}:= \p_\s {}^\b\G^\rho_{\mu\nu}-\p_\nu {}^\b\G^\rho_{\mu\s}+{}^\b\G^\tau_{\mu\nu}{}^\b\G^\rho_{\s\tau}-{}^\b\G^\tau_{\mu\s}{}^\b\G^\rho_{\nu\tau}\label{rite2}
\ee
and the curvature invariants $\cR_{\mu\nu}:= \cR^\rho_{~\mu\rho\nu}$ and $\cR:= g^{\mu\nu}\cR_{\mu\nu}$, the gravitational action is eq.\ \Eq{eha} \cite{frc11}. For $\b=0$, one recovers the case \Eq{Sg2}. In general, one can obtain a relatively simple integer frame (\emph{not} equivalent to standard general relativity) only when the gauge invariance of Weyl-integrable spacetimes is implemented (exactly if $\om=0=U$, approximately if $\om$ or $U$ are nonvanishing), which results in fixing $\b=2/(D-2)$. In $D=4$, $\b=1$ and the metric is a bilinear field density of weight $-1$.
\item[--] In $T_q$, the metric connection and the Riemann tensor are defined from the ordinary expressions, with the replacement \Eq{xq}:
\ba
{}^q\G^\rho_{\mu\nu} &:=& \tfrac12 g^{\rho\s}\left(\frac{1}{v_\mu}\p_{\mu} g_{\nu\s}+\frac{1}{v_\nu}\p_{\nu} g_{\mu\s}-\frac{1}{v_\s}\p_\s g_{\mu\nu}\right)\,,\label{leciq}\\
{}^q R^\rho_{~\mu\s\nu} &:=& \frac{1}{v_\s}\p_\s {}^q\G^\rho_{\mu\nu}-\frac{1}{v_\nu}\p_\nu {}^q\G^\rho_{\mu\s}+{}^q\G^\tau_{\mu\nu}\,{}^q\G^\rho_{\s\tau}-{}^q\G^\tau_{\mu\s}\,{}^q\G^\rho_{\nu\tau}\,,\label{riemq}
\ea
plus the curvature invariants ${}^q R_{\mu\nu}:= {}^q R^\rho_{~\mu\rho\nu}$ and ${}^q R:= g^{\mu\nu} {}^q R_{\mu\nu}$. The action of gravity and matter is
\be\label{Sgq}
S_q[g,\phi^i] =\frac{1}{2\kappa^2}\int\rmd^Dx\,v\,\sqrt{-g}\,({}^q R-2\Lambda)+S_q[\phi^i]\,,
\ee
where in $S_q[\phi^i]$ the metric is minimally coupled.
\item[--] Gravity with multifractional derivatives is still under construction.
\end{itemize}

\bigskip

\llabel{38} \textbf{\emph{What are the main features of cosmological dynamics in multifractional spacetimes?}}\addcontentsline{toc}{subsection}{\lref{38} What are the main features of cosmological dynamics in multifractional spacetimes?}

Despite the full dynamical equations having been laid down already, cosmological solutions have not been discussed in detail. The little we know shows signs of an exotic evolution. Here we write only the $D=4$ Friedmann equations ($00$ component of Einstein equations and the trace equation) for a homogeneous and isotropic background evolving with scale factor $a(t)$ and Hubble parameter $H=\dot a/a$. These expressions were derived in ref.\ \cite{frc11} from the full Einstein equations of the theories \Eq{Sg2}, \Eq{eha}, and \Eq{Sgq}.
\begin{itemize}
\item[--] In $T_1$, the Friedmann equations with curvature $\textsc{k}=0,\pm1$ and a perfect fluid with energy density $\rho$ and pressure $P$ are
\ba
H^2&=&\frac{\k^2}{3}\rho-\frac{\textsc{k}}{a^2}-H\frac{\dot v}{v}+f_1(v,\dot v)\,,\label{00fe}\\
\frac{\ddot a}{a}&=&H^2+\dot H=-\frac{\k^2}{6}(\rho+3P)+H\frac{\dot v}{v}+f_2(v,\dot v)\,,\label{01fe}
\ea
where possible measure-dependent terms have been collected into two contributions $f_1$ and $f_2$. The dynamics of this model has not been studied beyond a preliminary inspection \cite{fra2}.\footnote{The homogenous cosmology of $\tilde T_1$ is the same of $T_1$, since the two models have the same type of derivatives and they differ only in the factorizability of the measure.}
\item[--] In $T_v$, one has a non-Riemannian geometry ($g_{\mu\nu}$ not conserved) that strongly resembles a Weyl-integrable spacetime. With some manipulation, the action \Eq{eha} looks like that of a scalar-tensor theory, with the difference that the scalar field is here replaced by a function of the measure weight $v$. After a conformal transformation to a frame (the integer picture) where the metric is conserved, the Friedmann equations read
\ba
H^2&=&\frac{\k^2}{3}\,\rho+\frac{\Om}{2}\frac{\dot v^2}{v^2}+\frac{U(v)}{6v}-\frac{\textsc{k}}{a^2}\,,\label{friw}\\
\frac{\ddot a}{a}&=&-\frac{\k^2}{6}\,(\rho+3P)+\frac{U(v)}{6v}\,,\label{friw2}
%\frac{\textsc{k}}{a^2}+2H^2+\dot{H}=\frac{\k^2}{6}\,(\rho-3P)+\frac{\Om}{2}\frac{\dot v^2}{v^2}+\frac{U(v)}{3v}\,,
\ea
where $U(v)$ is a ``potential'' term for $v$ determined by the geometry (in general, solutions require $U\neq 0$) and $\Om=-3/2+f(v)$, where $f$ is a function of $v$ that, just like $f_{1,2}$ in eqs.\ \Eq{00fe} and \Eq{01fe}, is not necessary usually and can be set to zero.
\item[--] In $T_q$, the dynamics is
\ba
\frac{H^2}{v^2}&=&\frac{\k^2}{3}\,\rho-\frac{\textsc{k}}{a^2}\,,\label{fri}\\
\frac{\ddot a}{a}&=&-\frac{\k^2}{6}\,v^2(\rho+3P)+H\frac{\dot v}{v}\,.\label{ra}
\ea
A simple power-law solution $a(t)=[q^0(t)]^p$ with a binomial measure illustrates the typical evolution \cite{frc11}. The log oscillations of the measure give rise to a cyclic universe characterized by epochs of contraction and expansion, which progressively evolve to a monotonic scale factor at times $t\gg t_*$. The average or coarse-grained scale factor is given by the zero-mode contribution only, i.e., setting $F_\om=1$ in the measure. At early times $t\lesssim t_*$, the coarse-grained particle horizon expands faster than in standard cosmology. In this theory, we also know what happens when inhomogeneities are included \cite{frc14} (see question \lref{40}).
\item[--] The cosmology of $T_\g$ is unknown.
\end{itemize}
In none of the theories the evolution in the presence of radiation and dust matter has been considered yet and it would be important to check whether multifractional cosmologies are realistic. The extreme rigidity of the dynamics, where the evolution of curvature is governed by that of the measure, should make all these cosmological models easily verifiable.

\bigskip

\llabel{39} \textbf{\emph{Can you get acceleration from geometry without slow-rolling fields?}}\addcontentsline{toc}{subsection}{\lref{39} Can you get acceleration from geometry without slow-rolling fields?}

Yes, you can. In $T_1$, the term $H\dot v/v+f_2$ in the right-hand side of eq.\ \Eq{01fe} can give a positive contribution (averaging over log oscillations). In $T_v$, the term $\propto\Om(\dot v/v)^2<0$ in the right-hand side of eq.\ \Eq{friw} acts like the kinetic term of a phantom field (without having the theoretical problems associated with it), while the term $\propto U/v$ is like a potential or a time-varying cosmological constant and, since $U>0$ for self-consistency of the solutions (it is not imposed by hand \cite{frc11}), it gives a positive contribution to the right-hand side of eq.\ \Eq{friw2}. Phantoms typically trigger super-acceleration; the bouncing vacuum solution found in ref.\ \cite{frc11} confirms this expectation.

The theory $T_q$ is less transparent. Since $v\simeq 1+|t/t_*|^{\a-1}$, for an expanding universe one has $H\dot v/v\propto \dot v \propto (\a-1)|t/t_*|^{\a-2}<0$ and the right-hand side of eq.\ \Eq{ra} can vanish for an equation of state $w=P/\rho<-1/3$. Thus, it would seem that one needs a strong slow roll to get acceleration. However, measure factors $1/v^2<1$ are hidden in $\rho$ and $P$, inside the kinetic term of fields, and they suppress it. By this mechanism, potentials can dominate even if fields are not in very-slow roll.

\bigskip

\llabel{40} \textbf{\emph{Can you explain inflation with this mechanism?}}\addcontentsline{toc}{subsection}{\lref{40} Can you explain inflation with this mechanism?}

Not in $T_q$, because the flatness problem is not solved \cite{frc11}. However, the slow-roll condition is relaxed. In standard inflation, the primordial spectrum of scalar and tensor perturbations is described, as a first approximation, by the power spectrum $\cP_{\rm s,t}=\cA_{\rm s,t}(k/k_0)^n$, where $k=|{\bf k}|$ is the comoving wavenumber, $k_0$ is an experiment-dependent pivot scale and $n=n_{\rm s}-1,\,n_{\rm t}$ are, respectively, the scalar and tensor spectral index. In the theory $T_q$ with a binomial measure, this power law is deformed by the multifractional geometry according to the following expression \cite{frc14}: 
\ba
\cP_{\rm s,t}(k) &\simeq&\cA_{\rm s,t}\left(\frac{k}{k_0}\frac{\a+\left|\frac{k_0}{k_*}\right|^{1-\a}}{\a+\left|\frac{k}{k_*}\right|^{1-\a}}\right)^n\left[1+ A n\cos\left(\om\ln\frac{k_\infty}{k}\right)+B n\sin\left(\om\ln\frac{k_\infty}{k}\right)\right.\nonumber\\
&&\qquad\qquad\qquad\qquad\qquad \left.
~~-A n\cos\left(\om\ln\frac{k_\infty}{k_0}\right)-B n\sin\left(\om\ln\frac{k_\infty}{k_0}\right)\right].\label{Pmusc}
\ea
In the limit $k_*\ll k< k_\infty$ and averaging on log oscillations, one gets an effective power law $\cP_{\rm s,t}(k)\sim (k/k_0)^{n_{\rm eff}}$, where $n_{\rm eff}=\a n$. In particular, the effective spectral index of the primordial scalar spectrum is
\be
n_{\rm eff}-1 \simeq \a(n_{\rm s}-1)
\ee
asymptotically. If $\a$ is sufficiently small, one can soften the slow-roll condition \cite{frc11} and get viable inflation, even when $n_{\rm s}$ deviates from 1 by more than $10\%$ \cite{frc14}. One can see this intuitively by noting that the factor $1/v^2$ in the left-hand side of eq.\ \Eq{fri} can match a nontrivial time evolution of the right-hand side even when $H$ is approximately constant, while in standard cosmology a quasi-de Sitter evolution requires a matter energy density $\rho>-3P$.

However, one still needs a scalar field in slow roll in order to have a shrinking horizon during acceleration. For the cosmological toy model $T_1$ and the theory $T_v$, the Friedmann equations are known \cite{frc11} but they have not been studied, nor have cosmological perturbations been considered. Nothing is known about the cosmology of $T_\g$.

\bigskip

\llabel{41} \textbf{\emph{Can you explain dark energy with this mechanism?}}\addcontentsline{toc}{subsection}{\lref{41} Can you explain dark energy with this mechanism?}

We do not know, but work is in progress and preliminary results are encouraging. A cosmological constant term of purely geometric origin arises in $T_v$, both in homogeneous cosmological solutions \cite{frc11} and in black holes \cite{frc15}; however, it is not clear whether it can act as dark energy in a fully realistic evolution of the universe.

\bigskip

\llabel{42} \textbf{\emph{Are the big-bang and black-hole singularities resolved?}}\addcontentsline{toc}{subsection}{\lref{42} Are the big-bang and black-hole singularities resolved?}

The answer depends on the theory. We do not know in the case of $T_1$ and $T_\g$, but there are some results for the other two theories. 

In $T_v$, the vacuum solution $a(t)$ of the dynamics \Eq{friw}--\Eq{friw2} with $\textsc{k}=0$ is a \emph{bouncing} universe that tends to de Sitter asymptotically in the future \cite{frc11}. If one could show that general stable solutions with $\rho\neq 0$ have the same feature, there would be a concrete possibility to solve the big-bang problem in this theory. Regarding black holes, it was recently shown that spherically-symmetric solutions to the Einstein equations are of Schwarzschild--de Sitter type, hence the pointwise singularity at the center persists \cite{frc15}. Thus, the fate of singularities in $T_v$ is not clear.

In $T_q$, an original reinterpretation of the big-bang problem was proposed \cite{frc11}. Since a shift $q^\mu(x^\mu)\to q^\mu(x^\mu)+x^\mu_{\rm bb}$ does not change the measure, an arbitrary constant $x^\mu_{\rm bb}$ can be added which would leave the gravitational action formally unchanged but would regularize the scale factor $a[q(t)]\to a[q(t)+t_{\rm bb}]$ at $t=0$. In the light of the second flow-equation theorem \cite{first}, we can now exclude this shift: the constant vector $x^\mu_{\rm bb}$ can be assimilated to the presentation label $\bar x^\mu$, but we already have fixed that in the final- or initial-point prescriptions in the deterministic view of the theory and in the $T_{\g=\a}\cong T_q$ approximation (see \lref{27}). Also, the arguments below eq.\ \Eq{infip} clearly show that what is really special is the null-distance configuration $\ell^\mu=0$, not the coordinate point $x^\mu=0$. Therefore, the shift regularization cannot be implemented consistently in the theory. The same effect could be achieved without any shift in the geometric coordinates, setting $\a=0$; then, the constant term would come from the modulation factor in the measure $q(t)=t+F_\om(t)$ \cite{frc11}. This geometric configuration has not been considered much in the past, since it does not correspond to a traditional dimensional flow where the spacetime dimension changes at large scale excursions $\De\ell$: in this case, the dimension is constant in average but it is modulated by log oscillations. An alternative that capitalizes on the stochastic view of \cite{CaRo2a,CaRo2b} is that, due to the intrinsic microscopic uncertainty in the geometry, we cannot probe the zero scale of the big bang, which is thus screened by this most peculiar effect. Notice, however, that this mechanism does not work in the case of black holes: the singularity oscillates between a point and a ring topology (the two extrema of the initial- and final-point presentations) without ever disappearing \cite{frc15}. We leave all these possibilities open for future study.

%%%%%%%%%%%%%%%%%%%%%%%%%%%%%%%%%%%%%%%%%%%%%%%%%%%%%%%%%%%%%%%%%%%%%%%%%%%%%
%%%%%%%%%%%%%%%%%%%%%%%%%%%%%%%%%%%%%%%%%%%%%%%%%%%%%%%%%%%%%%%%%%%%%%%%%%%%%

\section{Quantum gravity}\label{qg}

\llabel{43} \textbf{\emph{Is dimensional flow really so important in quantum gravity, where there may not even be a notion of spacetime? The claim that one of the most striking phenomena we come across the landscape of quantum-gravity models is dimensional flow might be true in some abstract sense. However, some would say that the very concept of spacetime might not make sense and thus the theory of multifractional spacetimes might not be of any interest to them. Thus, multifractional models are addressed to a very particular sub-community of the overall quantum-gravity community.}}\addcontentsline{toc}{subsection}{\lref{43} Is dimensional flow really so important in quantum gravity, where there may not even be a notion of spacetime?}

Some theories of quantum gravity do not admit a notion of spacetime at the fundamental level. The most clear example of that is the group of GFT-spin foams-LQG, where geometry emerges from a combinatorial structure (e.g., ref.\ \cite{Ori09}). Even in CDT, where the path integral over geometries is regularized by a discretization procedure and the continuum limit is eventually taken, a smooth spacetime arises only in the so-called phase C in the phase diagram of the theory, while all the other phases correspond to non-Riemannian geometries (mutually disconnected lumps of space in the branched-polymeric phase A, large-volume configurations of vanishing time duration in phase B, and signature changes in phase ${\rm D}={\rm C}_{\rm b}$ \cite{AJL5,AmJ,Amb16}). Nevertheless, the Hausdorff, spectral, and walk dimensions are indicators valid also in non-Riemannian geometries, as discussed in question \loref{01} and showed in refs.\ \cite{COT2,COT3} for the GFT-spin foams-LQG group of theories and in the typical sets describing the non-Riemannian CDT phases \cite{AJL5,Dav92,JW,CW,DD,DJW1,DJW2}. Fractal geometry by itself is proof that we do not need a smooth manifold to have dimensional flow \cite{Fal03}.

Whether one sees them as independent theories of geometry or as effective models describing the flow of other theories in certain regimes, multifractional spacetimes are not addressed to a restricted audience. They do not lack personality since they are based on a characteristic paradigm, they are a top-down approach from theory to experiments, and they offer their own predictions about physical observables. More popular quantum-gravity approaches on the market have better or clearer results about the UV finiteness, but in some cases the phenomenology and contact with experiment is still underdeveloped or even absent. The uniqueness argument in \loref{04} guarantees that multifractional theories have a certain degree of universality in dimensional flow, so that placing constraints on this proposal can help to assess the phenomenology of other theories with dimensional flow. If anything, one of the main messages of multifractional theories is that dimensional flow can be a testable property of exotic geometries, rather than an abstract property disconnected from physics.

\bigskip

\llabel{44} \textbf{\emph{Let me reformulate the question. Even accepting that dimensional flow exists for all quantum gravities, are dimensions really measurable?}}\addcontentsline{toc}{subsection}{\lref{44} Are dimensions really measurable?}

If $dh$, $\ds$, and $\dw$ were not physical observables, dimensional flow would be only a mathematical feature useful to classify multiscale spacetimes. Some believe that these dimensions are not measurable and that they are just mathematical labels. Others recognize that the Hausdorff dimension is measurable but they do not acknowledge the same status for the spectral dimension; consequently (but this has never been said explicitly), also the walk dimension cannot be measured. Still others, like the author, are firm proponents of the measurability of all three dimensions.

That the spatial Hausdorff dimension is an observable is made clear by the following example \cite{trtls}. Consider an observer in a space with $\dh=D-1$ at large scales $\ell\gg\ell_*$ and $0<\dh<D-1$ at small scales $\ell\ll\ell_*$. They can make several balls of radius $\ell_1+\de \ell$ close to some average value $\ell_1\gg \ell_*$ (where $\de\ell\ll \ell_*$), submerge each ball in a container of water and measure the volume of displaced liquid, noting a distribution of volumes with average $\ell_1^{D-1}$ and width $\sim \ell_1^{D-2}\de \ell$. Making another set of balls of average radius $\de \ell<\ell_2\ll\ell_*$ with the same fluctuation $\de\ell$, they find an average volume $\ell_2^{\dh}$ and (for $D\geq 3$ and $\dh\geq 1$) a narrower distribution, since $1\ll (\ell_1/\ell_*)^{D-2} > (\ell_2/\ell_*)^{\dh-1}\ll 1$. The inequality may change direction for $\dh<1$ but, in any case, by comparing these dimensionless observables the experimenter realizes that they are living in a space with dimensional flow.

Ideally, the spectral and walk dimensions are measurable by placing a particle in a spacetime and let it diffuse. Literally. In practice, this procedure does not work when the scales we want to probe are much smaller than those covered by a molecule in Brownian motion. For that reason, and also to solve the negative-probabilities problem in quantum gravity, it may be better to adopt the QFT perspective that the diffusing probe is a quantum particle in a scattering process \cite{CMNa}. However, it is not yet clear how this would help since $\ds$ is the scaling of self-energy diagrams and, moreover, experiments with particle interactions cannot reach quantum-gravity scales. This does not mean that the spectral dimension is not a physical observable, since its relation (or even identification) with the dimension of momentum space (see \loref{01}) opens up several possibilities of measurement \cite{AAGM3}. 

When dealing with microscopic or very large scales, we cannot construct balls and submerge them in a liquid, or have ideal particles diffuse in spacetime, but appropriate experiments on high-energy physics or observations of cosmological scales can constrain both $\dh$ and $\ds$ with their characteristic tools. In the case of multifractional theories, this is illustrated by the many examples reported in section \ref{phen}.

\bigskip

\llabel{45} \textbf{\emph{There are many approaches to quantum gravity, some of which were listed in section \ref{intro}. Can you compare multifractional theories with these other scenarios?}}\addcontentsline{toc}{subsection}{\lref{45} Can you compare multifractional theories with other approaches to quantum gravity?}

We can make this comparison at five levels: (i) in the characteristics of dimensional flow, (ii) in other characteristics of the geometry, (iii) in terms of the UV properties of renormalizability or finiteness, (iv) in the way the multifractional paradigm, seen as an effective framework, captures the geometry of other theories, and (v) at the level of phenomenology and observational constraints.
\begin{itemize}
\item[(i)] The flow-equation theorems predict the general dimensional flow near the IR for any quantum gravity with nonfactorizable Laplacians [eq.\ \Eq{dhdsgen}] and for multifractional theories where Laplacians are factorizable in the coordinates [eq.\ \Eq{dhdsgen2}]. The coefficients $b$ and $c$ in eq.\ \Eq{dhdsgen} are determined by the dynamics of the theory, while $b_\mu$ and $c_\mu=1-\a_\mu$ in eq.\ \Eq{dhdsgen2} are free parameters with a restricted range (question \loref{07b}).
The spacetime dimensions in multifractional theories have been computed in \lref{13}. We compare the coefficients $b$ and $c$ in different theories of quantum gravity, expanding on the discussion of \cite{first}. The results are summarized in table \ref{tab3}.
\begin{table}[ht]
\begin{center}
\begin{tabular}{|llc|cc|cc|}\hline
\multicolumn{2}{|l}{}															 &  $D$  & $b_{\rm H}$    &$c_{\rm H}$& $b_{\rm S}$  	  & $c_{\rm S}$ \\\hline
%\multicolumn{2}{l}{Multifr.\ with $q$-der.} 			 &  $D$  &$-(\tfrac1\a-1)$& $1-\a$    & $\a-1$			    &	$1-\a$	  	\\
\multicolumn{2}{|l}{Asymptotic safety} 						 &  $4$  & 0				   		& ---		    &	$<0$					  &	$>0$ 		   	\\
\multicolumn{2}{|l}{CDT}							  					   &  $4$  & 0						  & ---				&	$<0$					  &	$2$			    \\
\multicolumn{2}{|l}{Near black holes}  						 &  $D$  & 0						  & ---				&$\tfrac{D+1}{2}$ &	$2$			    \\
\multicolumn{2}{|l}{Nonlocal gravity and string field theory} &  $D$  & 0						  & --- 			&	$<0$					  &	$2$			    \\
\multicolumn{2}{|l}{Fuzzy spacetimes}  						 &  $D$  & 0						  & --- 			&	$-D$					  &	$2$			    \\
\multicolumn{2}{|l}{Gravity with quantum particles} &  $3$  & 0						  & --- 			&$-\tfrac{21}{16}$&	$2$			    \\
$\k$-Minkowski & bicovariant $\N^2$, AN(3)			   &  $4$  & 0						  & --- 			&	$-2$    				&	$2$			    \\
$\k$-Minkowski & bicovariant $\N^2$, AN(2)				 &  $3$  & 0						  & --- 			&	$-\tfrac32$   	&	$2$			    \\
$\k$-Minkowski & bicrossproduct $\N^2$						 &  $4$  & 0						  & --- 			&	$1$     				&	$2$			    \\
$\k$-Minkowski & cyclic invariance (o.s.) 				 &  $D$  & $<0$ 				  & $1$       &	?								&	?						\\
\multicolumn{2}{|l}{Ho\v{r}ava--Lifshitz gravity}   &  $D$  & 0					    & ---       &	$<0$						&	$>0$ 				\\
\multicolumn{2}{|l}{GFT, spin foams, LQG (o.s.)}	   &$D(=4)$& $<0$  				  & $2$       &	$>0$						&	$2$		      \\\hline
\end{tabular}
\end{center}
\caption{\label{tab3}Parameters of the IR Hausdorff and spectral dimension of spacetime \Eq{dhdsgen} (subscript H and S, respectively) in quantum gravities. 
``Only space'' (o.s.) cases means that $\ell$ in eq.\ \Eq{dhdsgen} is a spatial scale (time dimension does not flow). Question marks indicate cases not studies in the literature.}
\end{table}

The Hausdorff dimension $\dh$ is the easiest to discuss: 
\begin{itemize}
\item[--] Asymptotic safety \cite{LaR5,RSnax,CES}, CDT \cite{AJL4,AJL5,BeH,SVW1}, spacetimes near black holes \cite{CaG,Mur12,ArCa1}, fuzzy spacetimes \cite{MoN}, and string field theory and nonlocal gravity \cite{CaMo1,Mod1} all have trivial dimensional flow in the Hausdorff dimension ($\dh=D$, where $D=4$ is some cases). 
\item[--] Noncommutative spacetimes usually have $\dh=D$ \cite{ArAl1,Ben08,ArTr}, but in the case of $\kappa$-Minkowski with cyclic-invariant action \cite{AAAD} $b<0$ ($\dh$ increases from below) and $c=1$ \cite{ACOS,CaRo1}.
\item[--] Ho\v{r}ava--Lifshitz gravity has Lebesgue measure $\rmd t\,\rmd^{D-1}x$ but with a time coordinate with anomalous scaling $[t]=-z<-1$. One can reabsorb this scaling into the spatial gradients $\N^{2z}$ of the theory, so that $\dh=D$ \cite{fra7}.
\item[--] States of LQG and GFT describing general discrete quantum geometries display the kink profile of the binomial measure \Eq{meamu} without log oscillations \cite{COT3,Thu15} (see figure 6 of ref.\ \cite{COT3}). In the analytic example of the lattice $\cC_\infty=\mathbb{Z}^{D-1}$, the Hausdorff dimension reads $\dh-1=\ell[\psi(\ell+D-1)-\psi(\ell)]$, where $\psi$ is the digamma function: $\dh=2+O(\ell)$ in the UV ($\ell$ is measured in units of the lattice spacing), while in the IR $\dh=D-(D-1)(D-2)/(2\ell)+O(\ell^{-2})$, giving $b<0$ and $c=1$.
\end{itemize}
Concerning the spectral dimension $\ds$ near the IR:
\begin{itemize}
\item[--] In asymptotic safety, $\ell$ is the IR cutoff governing the renormalization-group equation of the metric \cite{LaR5,RSnax,CES}. The multiscale profile of the spectral dimension is calculated analytically at each plateau and numerically in transition regions. The author is unaware of any semianalytic approximation giving $b$ and $c$ in \Eq{dhdsgen}. 
\item[--] In Ho\v{r}ava--Lifshitz gravity, $\ds\simeq1+(D-1)/z<D$ in the UV and $\ds\simeq D$ in the IR \cite{Hor3}, so that $b<0$. No semianalytic profile connecting the UV to the IR has been computed, so that we cannot say much about $c$ apart that it is positive. Using anomalous transport theory, it should be possible to find such profile with the multiscale tools of \cite{frc4}.
\item[--] The rest of the models have $c=2$, without exception. In CDT, $b<0$ is found numerically \cite{AJL4,BeH,SVW1}. In a nonlocal field-theory model near a black hole, $b=(D+1)/2$ \cite{ArCa1}. In fuzzy spacetimes, $b=-D$ \cite{MoN}. In nonlocal gravity with $\rme^\B$ operators as in string field theory, $b<0$ (one can show that $b=-36$ in $D=4$) \cite{CaMo1}. In LQG and GFT, one can check numerically that $b>0$ for all the classes of states inspected, that is, dimensional flow occurs from a UV with low dimensionality, reaches a local maximum $>D$, and then drops down to the IR limit from above \cite{COT3,Thu15}. Effective bottom-up approaches to LQG confirm dimensional flow to an UV spectral dimension smaller than $D$, although they do not make an analysis of quantum states \cite{Ron16,MiTr}.
\item[--] To date, the spectral dimension for $\kappa$-Minkowski with cyclic-invariant measure has not been calculated. The other noncommutative examples, all with $c=2$, are the following: in $D=3$ Einstein gravity with quantized relativistic particles, $b=-21/16$ \cite{ArAl1}; in Euclidean $\k$-Minkowski space with bicovariant Laplacian and AN(3) momentum group manifold, $D=4$ and $b=-2$ \cite{Ben08,ArTr}; with AN(2) momentum group manifold, $D=3$ and $b=-3/2$ \cite{ArTr}; with bicrossproduct Laplacian, $D=4$ and $b=1$ \cite{ArTr}. The bicovariant-Laplacian results are compatible with an independent calculation in generic $D$, where $b<0$ and $c=2$ \cite{AnHa1,AnHa2}.
\end{itemize}
In none of these cases, except hints in the GFT-spin foams-LQG group \cite{COT2,ThSt}, complex dimensions preluding to log oscillations have been detected. In the cases with discrete structures, this may be due to technical limitations in the way the spectral dimension has been computed, while in asymptotic safety the cutoff identification or the truncation of the action may play a role.
\item[(ii)] Asymptotic safety, phase C of CDT (after sending the triangulation size to zero), nonlocal gravity, string theory, and Ho\v{r}ava--Lifshitz gravity are defined on a continuum and spacetime, no matter how irregular it looks like at small scales due to quantum effects, can be described by a fundamental or effective metric $g_{\mu\nu}$. Phases A, B, and D of CDT do not correspond to metric manifolds but they admit a continuum description. $\k$-Minkowski and other noncommutative spacetimes are defined in a continuous embedding, but noncommutativity of the coordinates makes the spacetime structure highly non-Riemannian. GFT, spin foams, and LQG are all defined on pre-geometric structures such as group manifolds and combinatorial graphs; a continuous-spacetime approximation is reached in certain regimes (not only limited to the obvious semiclassical limit). In LQG and spin foams, the continuum limit is subject to a number of delicate technicalities, while in GFT its realization is perhaps more transparent \cite{GOS1,GOS2,GO}. Multifractional spacetimes are fundamentally discrete in the sense that there is a DSI at ultrasmall scales (see \lref{18}). This symmetry is not exact and at larger scales it gives way to a continuum. This transition happens via a natural coarse-graining procedure on the harmonic structure of the geometry \cite{fra4,frc2}.
\item[(iii)] In nonperturbative approaches such as asymptotic safety, CDT, LQG, and spin foams, UV finiteness is achieved by other means than perturbative renormalizability. In asymptotic safety, via the functional renormalization approach \cite{Nie06,NiR,CPR,Lit11}. In CDT \cite{lol08,AJL8,Fousp,AGJL4} and spin foams \cite{Ori01,Per03,Rov10,Per13}, via the well-definiteness of the path integral of (pre)geometries. In LQG, via canonical quantization of the gravitational constraints on a Hilbert space of (pre)geometric states (the spin networks) \cite{rov07,thi01}. GFT includes spin foams and LQG but it is a Lagrangian field theory on a group manifold; therefore, its renormalization properties can be tested either perturbatively (which constrains the forms of the kinetic term allowed by renormalizability) \cite{GeBo,BeG12,CORi1,CORi2} or nonperturbatively via the functional renormalization approach \cite{BBGO,BGMO1,BGMO2,BeLa}. Also the other major theories discussed in this review are based on perturbative field-theory renormalization, although in very different forms: examples are perturbative superstring theory (genus-expansion series of Riemann surfaces) \cite{GS85a,Mar86,VerVe,D'HP1,D'HP2,D'HP3,D'HP4,D'HP5,D'HP6,D'HP7,D'HP8}, noncommutative field theory (nonplanar graphs) \cite{GrWu1,GrWu2,GrWu3,RVTW,Ri07a,RVT,Ri07b,GMRT,MRT}, nonlocal gravity (traditional QFT but with nonlocal operators) \cite{Tom97,Mod1,MoRa,TaBM}, and Ho\v{r}ava--Lifshitz gravity (traditional QFT but with higher-order Laplacians) \cite{Hor09}. 

In nonperturbative formulations, UV finiteness is achievable but subject to a number of technical challenges or assumptions. For instance, in the functional renormalization approach used in asymptotic safety and in GFT a truncation of the effective action is performed. Still in GFT nonperturbative renormalization, all models considered so far are ``toy'' in the sense that they are with an Abelian group and without simplicity constraints (gravity requires a non-Abelian group and the implementation of simplicity constraints). In perturbative formulations, renormalization has been proven only at a finite order (as in perturbative string theory and Ho\v{r}ava--Lifshitz gravity), or at all orders but for a scalar field or other toy models (such is the case of GFT and noncommutative QFT), or modulo important technical or phenomenological issues (as in nonlocal gravity and Ho\v{r}ava--Lifshitz gravity).

The case of multifractional theories will be discussed in question \lref{47}.
\item[(iv)] Some quantum gravities have been connected directly with multifractional spacetimes.
\begin{itemize}
\item[--] The renormalization-group flow of asymptotic safety admits a complementary description in terms of a multifractional geometry \cite{fra7}, based on the observation that in the renormalization-group flow the physical momentum carries a scale dependence by the identification of the momenta $p_\textsc{as}$ with the cutoff scale $L$ of the renormalization-group flow. In the simplest case, $p_\textsc{as}(L)=1/L$. These momenta are scale-dependent and, by requiring the same dimensional flow of asymptotic safety, they can be matched by the geometric coordinates $p(k)$ in the momentum space of the theory $T_q$. Thus, in asymptotic safety physical rods are adaptive and momenta are implicitly scale-dependent, while in multifractional theories physical rods are nonadaptive and momenta are scale-independent, but one establishes a mapping by using geometric coordinates in position and momentum space, corresponding to adaptive ``mathematical'' rods and explicitly scale-dependent momenta.
This direction-by-direction mapping is exact (Laplacians are factorizable) and also explains the reason why these two theories predict the same spectral dimension of spatial slices, $\ds^{\rm space}\simeq 3/2$ in the deep UV, when $D=4$ and $\a=1/2$. This should be contrasted with nonfactorizable theories such as Ho\v{r}ava--Lifshitz gravity, for which $\ds^{\rm space}\simeq 1$ in the deep UV.
\item[--] The phase structure of CDT may find a counterpart in multifractional geometries \cite{frc2}. The branched polymers of phase A might be describable by a UV multifractional regime at scales $\ell_\infty<\ell\ll\ell_*$ where log oscillations modulate a highly nontrivial $\dh\simeq 2$ disconnected geometry. In phase B, the concepts of dimension, metric and volume seem not to play a major role, since a phase-B universe has no extension in the time direction and topology becomes important. This is akin to the most extreme limit of the multifractional measure, the so-called ``boundary-effect'' or ``near-boundary'' regime at scales $\ell\sim\ell_\infty$, where the binomial measure \Eq{meamu} (indices $\mu$ omitted here) is expanded at $|x/\ell_\infty|\sim 1$ and becomes $q(x)\sim \ln|x|$ \cite{fra4,frc2}. The name of this regime stems from its relation with an approximation of fractional derivatives near the extrema of integration in their definition and it signals a central role for topology, just like in phase B. This correspondence has not been formalized anywhere but it should not be hard to do so. It would be worth doing it not only for its simplicity, but also for the payback it brings. In particular, it immediately explains why random combs \cite{DJW1,AtGW,DJW2} cannot be associated with phase B: log oscillations are washed away in random structures.
\item[--] The anomalous scaling of the coordinates in Ho\v{r}ava--Lifshitz gravity can be easily interpreted in terms of binomial geometric coordinates \cite{fra7}. In these anisotropic critical systems \cite{Hor3,Hor09}, coordinates scale as $t\to \la^z t$ and ${\bf x}\to \la{\bf x}$ for constant $\la$, so that time and space directions have dimensions $[t]=-z$ and $[x^i]=-1$ in momentum units. This UV scaling is reproduced by an anisotropic multifractional model with $\a_0=1$ and $\a_i=1/z=1/(D-1)$. In particular, in four dimensions $\a_i=1/3$, the special value \Eq{a13} recently come to the fore \cite{CaRo2a,CaRo2b}. The correspondence of coordinates is $q^0(x^0)=x^0=x^0_{\rm HL}=t$, $q^i(x^i)=x^i_{\rm HL}$, and physical momenta are defined consequently, $p^0(k^0)=p^0_{\rm HL}$, $p^i(k^i)\sim p^i_{\rm HL}$. To get a multiscale geometry, one builds the total action with a hierarchy of differential Laplacian operators, from order $2z$ (UV) to $2$ (IR). Of course, symmetries and action differ in these two theories: while in Ho\v{r}ava--Lifshitz gravity the UV spectral dimension is anomalous due to the higher-order Laplacian, in multifractional theories it is so because of the nontrivial measure appearing in integrals and derivatives.
\item[--] Noncommutative and multifractional spacetimes enjoy different symmetries and are therefore physically inequivalent. Also, while we can devise noncommutative versions of multifractional theories and all noncommutative theories have nontrivial multiscale measures, we cannot interpret commutative multifractional theories as noncommutative theories. The ultimate cause of these discrepancies is the fact that noncommutative theories have nonfactorizable measures, while the measure of multifractional theories is factorizable \cite{CaRo1}. Nevertheless, these two classes of models have much in common, to the point where noncommutative geometry seems the natural candidate to generalize multifractional spacetimes to nonfactorizable measures \cite{CaRo1}. In particular, the spacetime algebra of $\k$-Minkowski spacetime is obtained by a noncommutative $q$-theory where geometric coordinates obey the Heisenberg algebra \cite{ACOS,CaRo1}, with a measure weight $v(x)$ reproducing the nontrivial measure found in the cyclic-invariant action of field theory on $\k$-Minkowski \cite{ACOS}. This correspondence is valid in the near-boundary regime discussed above and it yields eq.\ \Eq{infpl} as an important bonus: one can identify the scale in log oscillations with the Planck scale. Remarkably, the same identification is supported by a completely independent argument on distance uncertainties \cite{CaRo2a,CaRo2b}, but only for $\a=1/3$.
\item[--] Motivated by the contact points between multifractional and noncommutative spacetimes on one hand \cite{ACOS,CaRo2a,CaRo2b}, and the compatibility between the deformed Poincaré symmetries of $\kappa$-Minkowski spacetime and those of the effective-dynamics approach to LQG on the other hand \cite{hdaPa}, the constraint algebra of gravity in the multifractional theories $T_v$ and $T_q$ has been compared \cite{CaRo1} with the deformed algebra of LQG models of effective dynamics \cite{BoPa1,BoPa2,BBCGK,BBBD}. Although differences were expected for the reasons explained in the previous item, the types of deformation have been discussed in some detail \cite{CaRo1}. See question \lref{46}.
\item[--] A comparison of multifractional theories with models beyond general relativity at the border with quantum gravity, such as varying-$e$ models \cite{Bek82,BaMS,KiM}, VSL models \cite{Mag00,Mag03,Mag08b}, doubly special relativity \cite{DSR1,DSR2,DSR4,Mag02,JiS}, and fuzzy spacetimes \cite{MoN} can be found in refs.\ \cite{frc2,frc8} (see references therein for a more exhaustive bibliography).
\end{itemize}
Another but less direct connection is that the spacetime dimensions in asymptotic safety, Ho\v{r}ava--Lifshitz gravity, and GFT-spin foams-LQG have been reconsidered or found anew with the tools of anomalous transport theory \cite{CES,COT2,COT3}, which are heavily used in multifractional theories and have been proposed as a sharp instrument of analysis for quantum gravity in general \cite{fra6,frc4}.
\item[(v)] See question \lref{55}.
\end{itemize}

\bigskip

\llabel{46} \textbf{\emph{Some quantum gravities predict a deformation of the algebra of gravitational constraints. What is the constraint algebra for multifractional gravities?}}\addcontentsline{toc}{subsection}{\lref{46} What is the constraint algebra for gravity?}

It should not come as a surprise that the only available results are, once again, for $T_v$ and $T_q$ \cite{CaRo1}. We limit our attention to the classical algebra.

In the case of $T_v$ in $D=4$, the super-Hamiltonian constraint in the integer frame and in ADM variables can be written as $H[N] = H_0[N] + H_v[N]=\int \rmd^3x\,N(\cH_0+\sqrt{\tilde h}\mathcal{H}_v)$, where $N$ is the lapse function, $\tilde h$ is the determinant of the spatial metric, $\cH_0=\pi_{ij}\pi^{ij}/\sqrt{\tilde h}-\pi^2/(2\sqrt{\tilde h})-{}^{(3)}\tilde R\sqrt{\tilde h}$ is only metric dependent, $\pi_{ij}=\de S_v[\tilde g]/\de \dot{\tilde g}^{ij}$, and the density $\mathcal{H}_v$ is both metric and measure dependent. The diffeomorphism constraint is the usual one, $d[N^i]=-2\int\rmd^3x\,N^i \tilde h_{ij}d_{k}\pi^{kj}$, where $N^i$ is the shift vector. Since there are no dynamical degrees of freedom associated with $v$, there is no conjugate momentum $\pi_v$. Also, the $v$-dependent correction term $\cH_v$ is not affected by diffeomorphisms. Overall,
\ba
&&\{d[M^{k}],d[N^{j}]\}=d[\cL_{\vec{M}}N^{k}],\nonumber\\
&&\{d[N^{k}],H[M]\}=\{d[N^{k}],H_0[M]\}=H_0[\mathcal{L}_{\vec{N}}M],\label{hdav}\\
&&\{H[N],H[M]\}= \{H_0[N],H_0[M]\}=d[\tilde h^{jk}(N\partial_{j} M-M\partial_{j} N)],\nonumber
\ea
where $\cL$ is the Lie derivative. As claimed in question \lref{12}, standard diffeomorphism invariance is preserved in the integer frame of $T_v$ in the absence of matter; when interacting matter fields are present, diffeomorphism invariance is broken.

In the case of $T_q$, the algebra of first-class constraints is
\ba
&&\{d^q[M^{k}],d^q[N^{j}]\}=d^q\left[\frac{1}{v_j(x^j)}(M^j\partial_j N^{k}-N^j \partial_j M^k)\right],\nonumber\\
&&\{d^q[N^{k}],H^q[M]\}=H^q\left[\frac{1}{v_j(x^j)} N^j\partial_j M\right],\label{qhda}\\
&&\{H^q[N],H^q[M]\}=d^q\left[\frac{h^{jk}}{v_j(x^j)}(N\partial_{j} M-M\partial_{j} N)\right],\nonumber
\ea
where the index of the deformed measure weight $v_j$ is inert as usual. The constraints $H^q[N]$ and $d^q[N^{k}]$ generate time translations and spatial diffeomorphisms of the geometric coordinates $q^\mu(x^\mu)$, which means that these are not the usual time translation and diffeomorphisms.

A deformed constraint algebra appears in LQG when cancellation of quantum anomalies is imposed \cite{BoPa1,BoPa2,BBCGK,BBBD}. The only constraint deformed is the bracket of the super-Hamiltonian,
\be\label{hihb}
\{H[N],H[M]\} = d[\b h^{ij}(N\p_i M-M\p_j N)]\,,
\ee
where $\b$ is a function; in general relativity and in other quantization schemes of LQG \cite{AAN1,CGMM2}, $\b=+1$. From eqs.\ \Eq{hdav} and \Eq{qhda}, we can see that the constraint algebra of LQG, independently of the quantization scheme, differs from the algebras of $T_v$ and $T_q$. In the case of $T_v$, $\{H,H\}$ is untouched but deformations different from eq.\ \Eq{hihb} appear when matter is turned on. In the case of $T_q$, both $\{d,H\}$ and $\{d,d\}$ are modified (more precisely, the algebra is not deformed but the generators $d\to d^q$ and $H\to H^q$ are), contrary to what happens in LQG. Also, we cannot naively identify the LQG deformation function with $\beta = 1/v_i(x^i)$, since the left-hand side is a background-dependent function of the phase-space variables that can change sign, while the right-hand side is always positive and independent of the metric structure.

\bigskip

\llabel{47} \textbf{\emph{Are multifractional field theories renormalizable?}}\addcontentsline{toc}{subsection}{\lref{47} Are multifractional field theories renormalizable?}

This question is general but its implicit target is quantum gravity. A power-counting argument \cite{fra1,frc2} was at the origin of the multiscale paradigm. According to eq.\ \Eq{phik}, a scalar theory becomes super-renormalizable if $[\cK]=\dh$, that is to say, if the Laplace--Beltrami operator $\cK$ scales as a momentum to the power of the Hausdorff dimension of spacetime. For a polynomial potential $V=\sum_{n=0}^N \s_n\phi^n$, the coupling $\s_N$ of the highest power has engineering dimension $[\s_N]=\dh-N(\dh-[\cK])/2$ and the theory is power-counting renormalizable if $[\s_N]\geq 0$, i.e.,
\ba
&&N\leq \frac{2\dh}{\dh-[\cK]}\qquad {\rm if}\quad [\cK]<\dh\,,\label{n}\\
&& N\leq+\infty \qquad\qquad {\rm if}\quad [\cK]\geq\dh\,.\label{n2}
\ea
When $[\cK]>\dh$, the theory is super-renormalizable. Concentrating on eq.\ \Eq{n2}, if $\dh=D$, we need higher-order derivative operators, which introduce ghosts (Stelle gravity is a masterpiece example of this \cite{Ste77,Ste78}). If $\dh\simeq D\a$ in the UV, then we need either a second-order $\cK$ for $\a=2/D$ (as in the original multiscale proposal $\tilde T_1$ \cite{fra1,fra2,fra3}, in $T_1$, and in $T_v$) or a $\cK$ with anomalous scaling for general $\a<1$ (as in $T_q$ and $T_\g$). In the second case, however, if $\g=\a$ one has $[\B_q]=2\a=[\cK_\a]$ in the UV and, from eq.\ \Eq{n}, one obtains the usual condition $N\leq 2D/(D-2)$. Therefore, the theories $T_q$ and $T_{\g=\a}$ are not more renormalizable than in standard spacetime. If $\g\neq\a$, we have $[\cK]\geq\dh$ only if
\be
\g\geq \frac{D\a}{2}\,.
\ee
The limiting case is $\a=2/D$, where $\g=1$ and one recovers either $T_1$ or $T_v$. In $D=4$, having $\g<1$ and asking for power-counting renormalizability corresponds to having a non-normed spacetime. However, the condition for a norm was found in the absence of log oscillations \cite{frc1} and the latter disrupt the standard properties of spacetime anyway. Moreover, the presence of an intrinsic distance uncertainty in the deep UV of $T_\g$ (question \lref{27}) further indicates that having a norm is bound to become, sooner or later in the UV, an obsolete requirement. Cognitive estrangement is thus generally expected in the extreme regimes of multifractional spacetimes. The question is whether it is due to physically acceptable mechanisms.

Therefore, the power-counting argument gives good news for $T_1$ and $T_v$, bad news for $T_q$, and unclear news for $T_\g$. To check whether renormalizability is actually improved (or not) on a multifractional spacetime, one must go beyond the power-counting argument and employ either perturbative or nonperturbative QFT techniques. The only clear results we have so far are perturbative and only for $T_v$ and $T_q$ in the deterministic view. Here we review them and provide a new insight in $T_q$ and $T_\g$. The bottom line is that we have no news for $T_1$ (but, as we said, we do not care too much about that, since the upgrade of $T_1$ is $T_v$), bad news for $T_v$ (against the power-counting argument), bad news for $T_q$ in the deterministic view (in line with the power-counting argument), and intriguing news both for $T_q$ in the stochastic view and for $T_\g$ in either view.

In the theory with weighted derivatives, the degree of divergence of Feynman graphs in a scalar field theory does not improve with respect to standard QFT \cite{frc9}. An easy argument showing that the renormalizability of this theory is basically the same as that of the standard theory is the following. In the fractional frame, the measure in the momentum integration in loops is $\rmd^Dk\,w(k)$, where the weight $w(k)$ is such that the scaling dimension of the measure is smaller than $D$. However, when coupled with the full expression with two fractional phases $\bE(k,x)=\rme^{\rmi k\cdot x}/\sqrt{w(k)v(x)}$ (such as in propagators), the latter include two factors $w^{-1/2}$, which cancel the weight in the measure. Thus, the degree of divergence of momentum integrals remains the same as in the integer field theory. The actual degree of divergence of some diagrams differ with respect to the power-counting argument \cite{frc9} but essentially agrees with its main conclusion. Yet another, more intuitive way to understand this point is to notice that the free multifractional propagator in position space is of the form
\be\label{Gv}
G_v(x,y)=\frac{G_1(y-x)}{\sqrt{v(y-\bar x)v(x-\bar x)}}
\ee
for any factorizable positive semidefinite measure $v$ in presentation $\bar x$ \cite{frc6} (see question \lref{34}). Therefore, the divergence of $G_v(x,y)$ at coincident points $x\sim y$ is solely determined by the usual propagator $G_1(x-y)$ and not by the prefactor $\sim 1/v(y)$. 

The theory with $q$-derivatives in the deterministic view does not work, either. Its basic renormalization properties can be inferred from position space, according to the following scaling argument. In the rest of the answer, we omit spacetime indices and also avoid cumbersome expressions in geometric polar coordinates; a rigorous calculation could easily fill the gaps in this heuristic reasoning without major surprises. The free propagator is $G_q(x,y)=G_1[q(y)-q(x)]$ and its behaviour at $x\sim y$ is the same as the standard theory. For instance, in the massless case 
\be\label{G0}
G_q(x,y)\sim \frac{1}{|q(y)-q(x)|^{D-2}}\sim \frac{1}{|v(y-\bar x)(y-x)|^{D-2}}
\ee
upon Taylor expanding around $x=y$, and at coincident points inverse powers of $q(x)-q(y)$ diverge as inverse powers of $x-y$. Here $|q(y)-q(x)|=\sqrt{\sum_\mu [q^\mu(y^\mu)-q^\mu(x^\mu)]^2}$.

The only points where these arguments fail are those corresponding to the measure singularity at $y=\bar x=x$, where the above expressions vanish. The main conclusion is not modified in the deterministic view, but something interesting may happen in the stochastic view. As said in \lref{27}, we can adopt this view in $T_q$ when regarded as an approximation of $T_{\g=\a}$. We can see here how by computing the Green function both in $T_\g$ and in $T_q$; for $T_\g$, we only sketch a back-of-the-envelope calculation. The second flow-equation theorem selects the initial-point and the final-point presentation as special among all the others, and in $T_\g$ one can always choose either presentation thanks to translation invariance. Therefore, the propagator will be of the form $G_\g(x,y)=G_\g(y-x)$. Calling $r=|y-x|$, recall that the Fourier transform $\cF$ of a power law $r^\b$ is proportional to $k^{-(\b+1)}$. In $D$-dimensional Euclidean space, in polar coordinates we have $k^{-2}\propto\cF[r]=\cF[r^{D-1} r^{2-D}]$. The factor $r^{D-1}$ is the Jacobian in polar coordinates, which leads to $G_1(r)\sim r^{2-D}$ as the Green function in position space. Similarly, from the propagator \Eq{prop3} we get $k^{-2\g}\propto\cF[r^{2\g-1}]=\cF[r^{D\a-1} r^{2\g-D\a}]$, and identifying $r^{D\a-1}$ with the Jacobian in a space with UV Hausdorff dimension $\dh\simeq D\a$, we get the free propagator $G_\g(r)\sim r^{2\g-D\a}$ in the theory $T_\g$ at the plateau $\dh\simeq\a$, which can be generalized to the whole dimensional flow and to the presence of log oscillation. When $\g=\a$,
\be\label{G0g}
G_\a(x-y)\sim \frac{1}{|q(y-x)|^{D-2}}\stackrel{{\rm UV}}{\sim} \frac{1}{|y-x|^{\a(D-2)}F_\om^{D-2}(y-x)}\,,
\ee
where we have taken an isotropic binomial measure to illustrate the typical UV behaviour. On the other hand, we cannot use the initial- or final-point presentations in $T_q$ because we cannot conveniently fix $\bar x$ case by case. However, if we did, from eq.\ \Eq{G0} we would obtain exactly the same behaviour as in $T_{\g=\a}$:
\be\label{G02}
G_q(x,y)\simeq G_1[q(y-x)-q(0)]\simeq G_\a(x-y)\,.
\ee
Thus, eq.\ \Eq{G0g} is the typical Green function of the theory $T_\g$ with fractional derivatives of order $\a$, approximated by the theory $T_q$ with $q$-derivatives. Let us discuss its main properties, beginning with the deterministic view. For $F_\om=1$ (coarse-grained or no log oscillations), the singularity of the propagator (or of the Newtonian potential, to cite another example) is softened but, in accordance with the power-counting argument, not removed. Nevertheless, in the limit $\a\to 0$, we reach the $\a=0$ geometric configuration already met in \lref{42} and the propagator (or the potential) tends to a constant. This phenomenon is very similar to what found in nonlocal theories and is related to asymptotic freedom \cite{BMS,SFC}. It signals the possibility that interactions, including gravity, become weak in the deep UV. The limit $\a_\mu\to 0$ cannot be reached in $T_\g$ if we require spacetime to be normed (question \loref{07b}), but if we regard $\a$ in the Hausdorff dimension in eq.\ \Eq{phik} as the average fractional charge we can get $\a=0$ by setting some of the charges $\a_\mu$ to negative values. As said in \loref{07b}, these geometries are strange (or even unphysical) because they have no norm along some or all directions \cite{frc1} and, in general, the dimension of time or of spatial slices become negative.

However, this is not the end of the story. If $F_\om\neq 1$ is nontrivial, then at scales $\sim\ell_\infty$ the divergence becomes of the form $\sim(\ln 1)^{2-D}$ for any $\a$, since $q(y-x)\sim\ln(y-x)$ in that case; this is the near-boundary regime described in \lref{45}. Going at even smaller scales, $G_\g$ diverges periodically at the zeros of $F_\om(y-x)$. This behaviour, induced by the discreteness of the geometry at these scales, is totally different from what we would expect in a traditional resolution of singularities or in a renormalization scheme in a continuum. In the absence of a better name and of an explanation, we call this a \emph{DSI divergence} or DSI singularity.\footnote{In \cite{frc11}, the DSI approach to the big bang was compared at first sight to the BKL singularity. A quantitative comparison is still missing.} Notice that this possibility is realized only if the amplitudes of the log oscillations are large enough. The most negative contribution to $F_\om$ is given by an angle of $5\pi/4$, where $F_\om=1-\sqrt{2}(A+B)$. Assuming $A=B$, $F_\om$ vanishes for as small an amplitude as
\be
A=B=\frac{1}{2\sqrt{2}}\approx 0.35\,,
\ee
which is not excluded by CMB observations \cite{frc14} (see question \lref{49}).

In the stochastic view, the propagator is
\be
G_\a\sim G_q\stackrel{{\rm UV}}{\sim} \frac{1}{|r(1\pm\cX)|^{D-2}},
\ee
where $\cX\sim |\ell_*/r|^{1-\a}$ is an adaptation to polar coordinates of the correction \Eq{XT}. The DSI oscillations are just a blurring out of spacetime below the intrinsic uncertainty $\cX$, which grows as we approach the singular point $r=0$. This stochastic noise is the authentic texture of spacetime at these scales and it screens the observer from singularities: the point $r=0$ cannot be reached physically. Since we cannot measure lengths smaller than $r\cX$, the existence of a norm at these scales is irrelevant and we can contemplate exponents $\a<1/2$.

So, are infinities tamed in $T_\g$ or not? We do not know, but the quest for an answer promises to be stimulating both in the deterministic view (where we have the mysterious DSI singularity or a very exotic non-normed geometry) and in the stochastic view just described. 

%%%%%%%%%%%%%%%%%%%%%%%%%%%%%%%%%%%%%%%%%%%%%%%%%%%%%%%%%%%%%%%%%%%%%%%%%%%%%
%%%%%%%%%%%%%%%%%%%%%%%%%%%%%%%%%%%%%%%%%%%%%%%%%%%%%%%%%%%%%%%%%%%%%%%%%%%%%

\section{Observations}\label{phen}

\llabel{48} \textbf{\emph{Can a multifractal observer be aware of being in a multifractal spacetime?}}\addcontentsline{toc}{subsection}{\lref{48} Can a multifractal observer be aware of being in a multifractal spacetime?}

Yes, they can. An observer can recognize whether the underlying geometry is standard or multiscale (in particular, multifractal or multifractional) by measuring dimensionless quantities such as the ratio of two observables of the same kind \cite{trtls}. We saw an example of this procedure in question \lref{44} for the measurement of volumes. Another instance is the following. Consider $T_q$ in $D=1+1$ dimensions and suppose that two nonrelativistic objects a and b of very different size move with velocities $V_{x,{\rm a}}=\De x_{\rm a}/\De t$ and $V_{x,{\rm b}}=\De x_{\rm b}/\De t$ in the fractional picture. In the integer picture, one can compute the geometric velocity
\be
V_q=\frac{\De q(x)}{\De q(t)}=\frac{\De x|1\pm\cX|}{\De t|1\pm\cX^0|}=V_x \left|\frac{1\pm\cX}{1\pm\cX^0}\right|\,,
\ee
where we used eq.\ \Eq{Dex1}. Clearly, the ratio of the velocities of a and b will be different in a multifractional spacetime with respect to an ordinary spacetime, $V_{x,{\rm a}}/V_{x,{\rm b}}\neq V_{q,{\rm a}}/V_{q,{\rm b}}$, and a discrimination between the two spaces is possible when we measure the ratios of several objects, or when a and b are related to each other and some physical law predicts the value of such ratio.

This naive example is obviously inapplicable to the real world where, if multiscale geometry were true, one would find exotic effects at the scales of relativistic quantum physics or smaller. However, the main mechanism can be adapted to more realistic experiments.

\bigskip

\llabel{49} \textbf{\emph{Have these theories been constrained by observations? What are the constraints?}}\addcontentsline{toc}{subsection}{\lref{49} Have these theories been constrained by observations? What are the constraints?}

Yes. The multifractional theories $T_v$ and $T_q$ with a binomial measure have been confronted with experiments and observations, and bounds have been placed on the scale $\ell_*$, on the fractional exponents $\a_\mu=\a_0,\a$, and on the amplitudes $A$ and $B$ of log oscillations. The first datum that was considered was the variation of the fine-structure constant $\a_\textsc{qed}$ in quasars, but the bound on $T_v$ thus found was poor \cite{frc8}. The construction of the multifractional Standard Model permitted to use known constraints on electroweak interactions, including the estimate of the muon lifetime and of $\a_\textsc{qed}$, and the Lamb shift \cite{frc12,frc13}. Astrophysical processes such as the first black-hole merger observed by LIGO and gamma-ray bursts (GRB) from distant objects placed the strongest bounds on $\ell_*$ \cite{qGW}, while the main contribution of cosmology comes from the CMB black-body and inflationary spectra \cite{frc14}. The latter do not constrain $\ell_*$ efficiently but do allow to constrain the fractional charge (hence, the dimension of spacetime) and the log oscillations. Tables \ref{tab4}--\ref{tab7} summarize these results.\footnote{In the line ``CMB black-body spectrum '' in table \ref{tab5}, we correct a typo of table 2 of ref.\ \cite{frc14}; compare eq.\ (3.17) therein.}
\begin{table}[ht]
\begin{center}
\begin{tabular}{|l|ccc|c|c|}\hline
$T_v$ ($\a_0,\a\ll 1/2$)  							          & $t_*$ (s)        & $\ell_*$ (m) & $E_*$ (GeV)     & $A,B$ & source      \\\hline
Muon lifetime     																& --- 						 & --- 					& --- 						& --- 	& \cite{frc13}\\
Lamb shift       																  & ${<10^{-23}}$& $<10^{-14}$  & $>10^{-2}$      & --- 	& \cite{frc13}\\
Measurements of $\a_\textsc{qed}$ 							  & ${<10^{-26}}$& $<10^{-18}$  & $>10^{1}$       & ---		& \cite{frc13}\\
$\Delta\a_\textsc{qed}/{\a_\textsc{qed}}$ quasars	& ${<10^{11}}$ & $<10^{20}$   & $>10^{-37}$     & ---   & \cite{frc14}\\
Gravitational waves 															& --- 						 & ---  				& --- 						& --- 	& \cite{qGW}\\
Cherenkov radiation																& --- 						 & ---  				& --- 						& --- 	& this paper\\
GRBs 																						  & --- 						 & ---  				& --- 						& --- 	& \cite{qGW}\\
CMB black-body spectrum 													&   							 &   					  &   							& --- 	&   \\
CMB primordial spectra 													  &   							 &  					  &   							& 	  	&  \\\hline
\end{tabular}
\caption{\label{tab4}Absolute bounds on the hierarchy of multifractional spacetimes with weighted derivatives (obtained for $\a_0,\a\ll 1$). All figures are rounded. Items ``---'' are cases where the theory gives the standard result or where the experiments listed in the table are unable to place significant constraints. Empty cells are cases not explored yet.}
\end{center}
\end{table}

\begin{table}[ht]
\begin{center}
\begin{tabular}{|l|ccc|c|c|}\hline
$T_v$	($\a_0=1/2=\a$)							                & $t_*$ (s)     		& $\ell_*$ (m) & $E_*$ (GeV)     & $A,B$ & source      \\\hline
Muon lifetime     																& --- 							& --- 				 & --- 						 & --- 	 & \cite{frc13}\\
Lamb shift       																  & ${<10^{-29}}$ & $<10^{-20}$  & $>10^4$         & --- 	 & \cite{frc13}\\
Measurements of $\a_\textsc{qed}$ 								& ${<10^{-36}}$ & $<10^{-28}$  & $>10^{11}$    	 & ---	 & \cite{frc13}\\
${\Delta\a_\textsc{qed}}/{\a_\textsc{qed}}$ quasars	& ${<10^6}$ 	& $<10^{15}$   & $>10^{-32}$     & ---   & \cite{frc8}\\
Gravitational waves 															& --- 							& ---  				 & --- 						 & --- 	 & \cite{qGW}\\
Cherenkov radiation																& --- 						  & ---  			   & --- 						 & --- 	 & this paper\\
GRBs 				 																		  & --- 							& ---  				 & --- 						 & --- 	 & \cite{qGW}\\%\hline
CMB black-body spectrum 													& $<10^{-21}$	  		& $<10^{-12}$  & ${>10^{-4}}$& ---	 & \cite{frc14}  \\
CMB primordial spectra 													  &   							  &   					 &   							 &  	   & \\\hline
\end{tabular}
\caption{\label{tab5}Bounds on the hierarchy of multifractional spacetimes with weighted derivatives for $\a_0=1/2=\a$.}
\end{center}
\end{table}

\begin{table}[ht]
\begin{center}
\begin{tabular}{|l|ccc|c|c|}\hline
$T_q$	($\a_0,\a\ll 1/2$)						              & $t_*$ (s)         & $\ell_*$ (m) & $E_*$ (GeV)     & $A,B$ & source      \\\hline
Muon lifetime    																  & ${<10^{-13}}$ & $<10^{-5}$   & $> 10^{-12}$    & ---   & \cite{frc12}\\
Lamb shift        																& $<10^{-23}$       & $<10^{-15}$  & ${>10^{-2}}$& ---   & \cite{frc12}\\
Measurements of $\a_\textsc{qed}$ 								& ---        				& ---          & --- 						 & ---   & \cite{frc13}\\%\hline		
${\Delta\a_\textsc{qed}}/{\a_\textsc{qed}}$ quasars	& ---		 					& ---  				 & ---   					 & ---   & \cite{frc8}\\
Gravitational waves (pseudo)											& $<10^{-25}$       & $<10^{-17}$  & ${>10^{0}}$ & ---   & \cite{qGW}\\
Cherenkov radiation															  & $<10^{-60}$				& $<10^{-52}$  & ${>10^{36}}$& --- 	 & this paper\\
GRBs																						  &        					  &   					 &  							 & ---   &   \\%\hline
CMB black-body spectrum 													&  							    &   					 &  							 & --- 	 &   \\
CMB primordial spectra 													  & $<10^{12}$      &${<10^{20}}$& $>10^{-36}$	   & ---   & \cite{frc14}  \\\hline
\end{tabular}
\caption{\label{tab6}Absolute bounds on the hierarchy of multifractional spacetimes with $q$-derivatives (obtained for $\a_0,\a\ll 1$ in all cases but for the last one, where a likelihood analysis has been used). ``Pseudo'' indicates bounds obtainable in the stochastic view, where $T_q$ is regarded as an approximation of $T_{\g=\a}$.}
\end{center}
\end{table}

\begin{table}[ht]
\begin{center}
\begin{tabular}{|l|ccc|c|c|}\hline
$T_q$	($\a_0=1/2=\a$)							                & $t_*$ (s)     		& $\ell_*$ (m) & $E_*$ (GeV)     & $A,B$ & source      \\\hline
Muon lifetime    																  & ${<10^{-18}}$ & $<10^{-9}$   & $> 10^{-7}$     & ---   & \cite{frc12}\\
Lamb shift        																& $<10^{-27}$       & $<10^{-19}$  & ${>10^2}$   & ---   & \cite{frc12}\\
Measurements of $\a_\textsc{qed}$ 								& ---        				& ---          & --- 						 & ---   & \cite{frc13}\\%\hline
${\Delta\a_\textsc{qed}}/{\a_\textsc{qed}}$ quasars	& ---		 					& ---  				 & ---   					 & ---   & \cite{frc8}\\
Gravitational waves (pseudo)											& $<10^{-44}$       & $<10^{-35}$  & ${>10^{19}}$& ---   & \cite{qGW}\\
Cherenkov radiation																& $<10^{-81}$     	& $<10^{-73}$  & ${>10^{57}}$& --- 	 & this paper\\
GRBs 																							& 					        & 						 &  							 & ---   & \\%\hline
CMB black-body spectrum 													& $<10^{-26}$       & $<10^{-18}$  & ${>10^{2}}$& ---   & \cite{frc14} \\	
CMB primordial spectra 													  & ---               & ---				   & ---		  			 & ${<0.4}$ & \cite{frc14}  \\\hline
\end{tabular}
\caption{\label{tab7}Bounds on the hierarchy of multifractional spacetimes with $q$-derivatives for $\a_0=1/2=\a$. ``Pseudo'' indicates bounds obtainable in the stochastic view, where $T_q$ is regarded as an approximation of $T_{\g=\a}$.}
\end{center}
\end{table}

The results of \cite{first} stimulates us to review these bounds critically. They all arise from the binomial measure \Eq{meamu} with \Eq{binom2}, with or without log oscillations. However, the second flow-equation theorem does not really fix the coefficient $\ell_*/\a_\mu$ but, rather, it treats it as an arbitrary constant $\ell_* u_\mu$ not necessarily $\a_\mu$-dependent. The original motivation for the coefficient $\ell_*/\a_\mu$ is that the equations of motion and the physical observables in the theory $T_v$ depend only on the measure weight $v(x)=1+|x/\ell_*|^{\a-1}$ (log modulation is ignored), not on $q(x)$. The anomalous correction in $v(x)$ depends on the arbitrary scale $\ell_*$ and there is no need to introduce a new parameter $u_\mu$. However, in the theory $T_q$ having an $\a$-dependent or $\a$-independent constant $u_\mu$ can weaken some of the bounds in tables \ref{tab6} and \ref{tab7}. These new bounds with (index or label $\mu$ omitted everywhere)
\be\label{meau}
q(x)=x+\ell_* u\left|\frac{x}{\ell_*}\right|^{\a}\,,\qquad u=O(1)\,,
\ee
are shown in table \ref{tab8} and commented upon in \lref{64}. We will compute one of them explicitly in \lref{55}. As one can see by comparing tables \ref{tab6}--\ref{tab7} and \ref{tab8}, the new bounds are slightly weaker than the previous ones, except that from the CMB black-body spectrum which is almost unchanged.
\begin{table}[ht]
\begin{center}
\begin{tabular}{|l|ccc|c|}\hline
$T_q$	($\a_0,\a\ll 1/2$)						              & $t_*$ (s)         & $\ell_*$ (m) & $E_*$ (GeV)     & source    \\\hline
Muon lifetime    																  & ${<10^{-11}}$ & $<10^{-3}$   & $> 10^{-13}$    & this paper\\
Lamb shift        																& $<10^{-21}$       & $<10^{-13}$  & ${>10^{-4}}$& this paper\\
Gravitational waves (pseudo)											& $<10^{-22}$       & $<10^{-14}$  & ${>10^{-2}}$& this paper\\
Cherenkov radiation 														  & $<10^{-57}$				& $<10^{-49}$  & ${>10^{33}}$& this paper\\
GRBs 																						  & $<10^{-39}$       & $<10^{-30}$  & ${>10^{14}}$& \cite{qGW}\\%\hline
CMB black-body spectrum 													&   							  &   					 &  							 &   \\\hline
$T_q$	($\a_0=1/2=\a$)							                & $t_*$ (s)     		& $\ell_*$ (m) & $E_*$ (GeV)     & source      \\\hline
muon lifetime    																  & ${<10^{-17}}$ & $<10^{-8}$   & $> 10^{-8}$     & this paper\\
Lamb shift        																& $<10^{-26}$       & $<10^{-18}$  & ${>10^1}$   & this paper\\
Gravitational waves (pseudo)											& $<10^{-42}$       & $<10^{-33}$  & ${>10^{17}}$& this paper\\
Cherenkov radiation																& $<10^{-79}$     	& $<10^{-71}$  & ${>10^{55}}$& this paper\\
GRBs 			 																				& $<10^{-57}$       & $<10^{-48}$  & ${>10^{32}}$& \cite{qGW}\\%\hline
CMB black-body spectrum 													& $<10^{-26}$       & $<10^{-18}$  & ${>10^{1}}$& this paper\\\hline
\end{tabular}
\caption{\label{tab8}Absolute bounds (obtained for $\a_0,\a\ll 1$, upper part) and bounds for $\a_0=1/2=\a$ (lower part) on the hierarchy of multifractional spacetimes with $q$-derivatives with measure \Eq{meau}. The key formul\ae\ used to compute the constraints in the table are: for the muon lifetime, $t_*< (u\delta\tau/\tau_0^{\a_0})^{1/(1-\a_0)}$ replacing eq.\ (139) of \cite{frc13}, with $u=1$; for the Lamb shift, $E_*>\{u\delta E/[(2-\a_0)\De E]\}^{1/(\a_0-1)}|E_{2S}|$ replacing eq.\ (142) of \cite{frc13}, with $u=2-\a_0$; for gravitational waves, eq.\ \Eq{tslest2} with $u_\mu\propto C_\mu/(3-\a_\mu)$ and $2C_\mu=1$; for GRBs, eq.~\Eq{estilas} [$u=O(1)$]; for Cherenkov radiation, eq.~\Eq{chere2} with $2C_\mu=1$ [$u=O(1)$]; for the CMB black-body spectrum, a data fit with eq.\ (3.5) of \cite{frc14} with the factor $1/\a_0$ in the denominator replaced by $u=1$. ``Pseudo'' indicates bounds obtainable in the stochastic view, where $T_q$ is regarded as an approximation of $T_{\g=\a}$.}
\end{center}
\end{table}

\bigskip

\llabel{50} \textbf{\emph{Measurements of the anomalous magnetic moment of the electron tests QED to a much higher level of accuracy than the Lamb shift or the muon lifetime. Why not to use this datum?}}\addcontentsline{toc}{subsection}{\lref{50} Why not to use constraints on the anomalous magnetic moment of the electron?}

Indeed, the $g-2$ factor can constrain $T_v$ efficiently \cite{frc13}. From the triangular vertex in the integer picture, at one loop it is known that $g-2=\tilde\a_\textsc{qed}/\pi$. The fine-structure constant is measured with an accuracy of $\delta\a_\textsc{qed}/\a_\textsc{qed}\sim 10^{-10}$. Since, from eq.\ \Eq{ech}, the measured fine-structure constant in the fractional picture is (in $c=1=\hbar$ units) $\a_\textsc{qed}(t)\simeq Q^2(t)=\tilde\a_\textsc{qed}/v_0(t)$, for the binomial measure \Eq{meamu} with \Eq{binom2} the difference between the integer and fractional constant is $\Delta\a_\textsc{qed}=\a_\textsc{qed}(t)|t_*/t|^{1-\a_0}$. Demanding $\Delta\a_\textsc{qed}<\delta\a_\textsc{qed}$ and setting $t=t_\textsc{qed}=10^{-16}\,{\rm s}$, one obtains $t_*< 10^{-16-10/(1-\a_0)}\,{\rm s}$. The bounds from $\a_\textsc{qed}$ are reported in tables \ref{tab4} and \ref{tab5}, and they are several orders of magnitude stronger than the Lamb-shift bounds.

The theory $T_q$ is immune to similar constraints because it predicts the same $g-2$ factor and fine-structure constant as in the ordinary Standard Model. It is easy to understand why. The way the $q$-theory conveys multiscale effects to physical observables is via a transition from adaptive measurement units (integer picture) to nonadaptive ones (fractional picture). In the case of the Lamb shift, one borrows the standard QED result for the shift in the energy levels and applies it to the difference $\Delta p(E)$ between geometric energies; then, from $\Delta p(E)$ one extracts the actual Lamb shift $\Delta E$ and proceeds with the comparison with experiments \cite{frc12,frc13}. One could do essentially the same thing by looking at the hydrogen spectrum on a photographic plate, measuring the separation between two spectral lines; in either case, we are measuring dimensionful quantities. However, dimensionless quantities such as $\a_\textsc{qed}$ and $g-2$ are unaffected by having worked with composite momentum or position coordinates in the integer picture. Therefore, these fundamental\footnote{By fundamental, we mean that they are not obtained from the composition of other directly measurable quantities; see ref.\ \cite{trtls} and question \lref{48} for examples of nonfundamental observables that can discriminate the theory.} dimensionless observables remain the same in both frames of $T_q$. Curiously, this situation is complementary to the one for the muon lifetime, where $T_q$ is sensitive to changes in the geometry while $T_v$ is not \cite{frc13}.

\bigskip

\llabel{51} \textbf{\emph{What are the motivations and the gains of the bounds found for the multifractional Standard Models? None of the exotic realizations of the Standard Model contains any virtue with respect to the ordinary Standard Model. The constraints on the energy and length obtained appear to be irrelevant in view of the same quantities in renormalized QFT.}}\addcontentsline{toc}{subsection}{\lref{51} What are the motivations and the gains of the bounds found for the multifractional Standard Models?}

This criticism echoes question \lref{30} and we can only answer in the same way: the motivations of multifractional theories lie in quantum gravity (section \ref{moti}), not in the desire of modifying the celebrated Standard Model. If changing spacetime geometry carries consequences also for the fundamental particle interactions, then it becomes both interesting and necessary to verify whether these changes are compatible with extant experimental constraints.

\bigskip

\llabel{52} \textbf{\emph{Setting experimental limits on an \emph{ad hoc} proposal is not interesting. Setting limits on effective higher-dimensional operators is more systematic and model-independent, but it has already been done in the past.}}\addcontentsline{toc}{subsection}{\lref{52} Do these experimental limits come from an \emph{ad hoc} proposal?}

Sections \ref{moti} and \ref{qg} (questions \loref{04}--\loref{06}, \lref{43}--\lref{45}, and \lref{47}) bring a number of arguments on the fact that the multifractional proposal is neither \emph{ad hoc} nor sterile in its theoretical and phenomenological structure. On top of that, in \lref{32} we saw that expansions in higher-dimensional operators can mimic, and not even fully reproduce, only some aspects of the multifractional QFT phenomenology. 

\bigskip

\llabel{53} \textbf{\emph{Even granting that these theories are not ill defined and have some physical motivation, it is not possible to reach any conclusion about their phenomenology because they have not been developed rigorously. In particular, there is no top-down construction of a quantum field theory, let alone that of a Standard Model of electroweak and strong interactions.}}\addcontentsline{toc}{subsection}{\lref{53} Have these theories been developed rigorously to the point of being able to reach any robust conclusion about phenomenology?}

We hope that this review, and in particular section \ref{qft} (questions \lref{29}--\lref{36}), has convinced the reader that a top-down construction exists for the multifractional Standard Model in $T_v$ and $T_q$ \cite{frc13}, hence the phenomenology of these two theories comes directly from their foundations. We have not yet constructed the Standard Model for $T_\g$, but the resemblance of $T_q$ with $T_{\g=\a}$ (questions \lref{11}, \lref{34}, and \lref{47}) justifies the hope that the phenomenology of $T_\g$ be very similar to that found for $T_q$ in refs.\ \cite{frc12,qGW}.

\bigskip

\llabel{54} \textbf{\emph{Is it true that, in these theories, and despite the fact that the spacetime structure itself has been changed, it is assumed that only gravity is altered while the electromagnetic field is the usual one? What is the justification behind that?}}\addcontentsline{toc}{subsection}{\lref{54} Is it true that it is assumed that only gravity is altered while the electromagnetic field is the usual one?}

No, it is not true. The multiscale geometry of these spacetimes affect all fundamental interactions \cite{frc8,frc13}, including electromagnetism. The speed of light is modified accordingly in the theory $T_q$ \cite{trtls}, while in $T_v$ it is the usual $c=1$. See \lref{29} and \lref{31}. Amusingly, this question is somewhat ``complementary'' to \lref{24}.

\bigskip

\llabel{55} \textbf{\emph{Since, as claimed in \lref{10} and \lref{11}, multifractional theories lead to violations of Lorentz symmetries, then what are the constraints?}}\addcontentsline{toc}{subsection}{\lref{55} What are the constraints on violations of Lorentz symmetries?}

In general, all constraints coming from the Standard Model explicitly limit deviations from Lorentz invariance, but the strongest bounds to date are based on classical deformed dispersion relations (at the quantum level, we avoid problems; see \lref{35b}). The theory $T_\g$ has not been tested with these observations, due to its underdevelopment. We will not fill this gap here. The theory $T_v$ and the model $T_1$ have standard dispersion relations [see eqs.\ \Eq{prop1} and \Eq{Gv}] and do not predict any change in the propagation speed of particles. The remaining case is the theory $T_q$, for which constraints have been obtained from gravitational waves and GRBs \cite{qGW}. Let us begin with a general analysis of dispersion relations in $T_q$.

From eq.\ \Eq{prop2}, one finds the massive dispersion relation $[p^0(E)]^2=|{\bf p}|^2+m^2=\sum_i [p^i(k^i)]^2+m^2$, where $E=k^0$. This expression, valid for a scalar field, may be regarded as the general representative of dispersion relations for this theory. From now on, we drop the mass term, which plays no role in the main argument. Also, we approximate the dispersion relation for small multifractional corrections and combining the spatial momentum components into the absolute value $k=|{\bf k}|$. The latter approximation can be done in different ways that all give very similar results, modulo a prefactor $C_\mu$ in front of the correction which is always $O(1)$. Taking the binomial measure \Eq{binomp} with isotropic spatial hierarchy (which is all we need, according to \loref{07}), setting $k_i\simeq k/\sqrt{3}$ and defining $E_*=k_*$ [energy scale $E_*$ identified with the inverse of the time and length scales $t_*$ and $\ell_*=1/k_*$ in Planck units, eqs.\ \Eq{formu} and \Eq{infpl}], we get the full dispersion relation
\be
E^2\simeq k^2+2E_*^2\left[\frac{1}{\a_0}\left(\frac{k}{E_*}\right)^{3-\a_0} F_\om(k)-\frac{3}{\a}\left(\frac{k}{\sqrt{3} E_*}\right)^{3-\a} F_\om\left(\frac{k}{\sqrt{3}}\right)\right].\label{dire1}
\ee
This expression is simplified to
\be\label{dire}
E^2\simeq k^2+\frac{2E_*^2C_\mu}{3-\a_\mu}\left(\frac{k}{E_*}\right)^{3-\a_\mu}\,,
\ee
when log oscillations are averaged or absent ($F_\om=1$). For timelike fractal geometries ($\mu=0$, trivial measure in spatial directions) with ${\bf p}={\bf k}$, one has $C_0=(3-\a_0)/\a_0$; the correction is positive. For spacelike fractal geometries ($\mu=i$, trivial measure in the time-energy direction) with $p^0=E$, one has $C_i=-(3-\a)/[3^{(1-\a)/2}\a]$ and the correction is negative. Generic configurations with multifractional time and space directions can produce corrections of either sign, periodically suppressed by the log oscillations. The timelike and spacelike cases without oscillations are extreme representatives of this spectrum of possibilities, both corresponding to corrections with a unique sign and maximal amplitude.

Given a dispersion relation $E^2=E^2(k)$, the velocity of propagation of a wave front for a particle p is given by the group velocity
\be\label{gruve} 
V_{\rm p}:=\frac{\rmd E}{\rmd k}\,.
\ee
In this and the next question, we reserve the symbol $V$ for velocities and the reader should not confuse it with potentials. For the usual Lorentz-invariant dispersion relation $E^2=k^2+m^2$, in the small-mass limit one gets the difference $\De V:= V_{\rm p}-1\simeq -m^2/(2E^2)$ between the propagation speed of the particle and the speed of light. Plugging the timelike or spacelike approximations of eq.\ \Eq{dire1} into \Eq{gruve} and replacing $k\to E$ in the right-hand side consistently with the small-correction approximation, we get $\De V \simeq C_\mu(E/E_*)^{1-\a_\mu}$. This correction is less suppressed than those in usual modified dispersion relations in quantum gravity \cite{EMNan,ArCa2}, where $0<1-\a_\mu<1$ is replaced by some exponent $n\geq 1$. This determines a stronger and more sensitive bound on the characteristic energy $E_*$, via
\be\label{tslest}
E_* =\left|\frac{\De V}{C_\mu}\right|^{-\frac{1}{1-\a_\mu}}E\,.
\ee
We pause for a moment and highlight a caveat. The propagation speed \Eq{gruve} does not depend on the species of the particle. This is clear from eq.\ \Eq{dire1}, which is derived from the pole structure of a generic propagator in the massless limit, regardless of its tensorial structure. The effect of multiscale geometry on the propagation of particles is, thus, universal and the difference $\De V_{12}=V_1-V_2$ between the velocity of two species 1 and 2 is theoretically zero. In particular, the dispersion relation of photons acquire the same corrections \Eq{dire1} as any other particle and the speed of light is not $c=1$ \cite{trtls}. Therefore, in the deterministic view gravitons propagate at the speed of light and $\De V_{12}=0$. However, in the stochastic view the correction in the right-hand side of \Eq{dire1} represents a fluctuation of the geometry. Maximizing this fluctuation one finds eq.\ \Eq{dire} and, taking opposite signs for particles 1 and 2, one obtains eq.\ \Eq{tslest} with $\De V\to \De V_{12}/2$:
\be\label{tslest2}
E_* =\left|\frac{\De V_{12}}{2C_\mu}\right|^{-\frac{1}{1-\a_\mu}}E\,.
\ee
Little or nothing changes for phenomenology because the extra factor $1/2$ can modify the order of magnitude of $E_*$ at most by one.\footnote{Compare tables \ref{tab6} and \ref{tab7} with the numbers found at the end of ref.\ \cite{qGW} in the main body, where the factor $1/2$ is absent. The values in the table in ref.\ \cite{qGW} (also reported in ref.\ \cite{frc14}) use a different frequency peak.} If ignored, this delicate point may trigger question \lref{54}, since in eq.\ \Eq{tslest} $\De V$ is the difference between the particle propagation velocity and a constant speed of light $c=1$.

From eq.\ \Eq{tslest2} and similar others, one usually extracts two types of bounds, an ``absolute'' one giving the most conservative estimate of multifractional effects (typically obtained for $\a_0,\a\ll 1/2$ or zero) and one for a specific choice of $\a_0$ or $\a$, as in tables \ref{tab6}--\ref{tab8}. Here we consider the bounds on the propagation speed of gravitational waves from the LIGO observation of the black-hole merger GW150914 \cite{Abb16}. Following \cite{YYP}, we take the gravitational-wave signal to peak at frequencies $f=\om/(2\pi)\sim100\,{\rm Hz}$, corresponding to $\om\approx 630\,{\rm Hz}$, an energy $E=\hbar\om\approx 4.1\times 10^{-13}\,{\rm eV}$, and a velocity difference
\be\label{bov}
|\De V_{12}|<4.2\times 10^{-20}\,.
\ee
The bounds for $C_\mu$ fixed as in the text below eq.\ \Eq{dire} are shown in the line ``Gravitational waves (pseudo)'' of tables \ref{tab6} and \ref{tab7} (there is no detectable difference between the timelike and the spacelike cases), while for $2C_\mu=1$ they are in table \ref{tab8}. Bounds on $E_*$ are converted to bounds on $t_*$ and $\ell_*$ via eqs.\ \Eq{formu} and \Eq{infpl}.

The bounds from photon time delays in GRBs are more severe but obtained in a more heuristic way \cite{qGW}. The difference in the velocities of two photons with different energies emitted in a GRB at the same time is $|\De V_{12}|\propto (E_2^{1-\a_\mu}-E_1^{1-\a_\mu})/E_*^{1-\a_\mu}$. Taking $E_2\gg E_1$ (highly-energetic photons), one gets eq.\ \Eq{dire} with $\De V\to\De V_{12}$ (and no $1/2$ factor). Letting $d$ be the luminosity distance between the source and us and $\De t=t_1-t_2$ the time delay in the arrival of the photons, we also have $1\gg \De V_{12}\sim d/t_1-d/t_2\simeq d\De t/t_1^2\simeq V_2^2\De t/d\sim \De t/d$. The observed sources of bright GRBs are in the range of redshift $z=0.16-3.37$ (i.e., \cite{BAJPP}), corresponding to $d\sim 10^{25}-10^{27}\,{\rm m}$. For typical photon emissions, $\De t\sim 10^{-2}-10^{-1}\,{\rm s}$, so that $\De V_{12}\sim 10^{-20}-10^{-18}$. Taking $E_2\sim 10^{-4}\,{\rm GeV}$ and the most conservative value $\De V_{12}\sim 10^{-18}$, we get
\be\label{estilas}
E_*>10^{-4+\frac{18}{1-\a_\mu}}\,{\rm GeV}\,.
\ee
This bound \cite{qGW}, shown in table \ref{tab8}, is much tighter with respect to the other constraints, even discounting a few orders of magnitude with respect to a rigorous estimate.

\bigskip

\llabel{56} \textbf{\emph{But there are much stricter constraints in particle physics, for instance those of refs.\ \cite{CoGl2,KoMe}. One derives limits on coefficients of effective operators, which are typically more stringent than those quoted above. Even for Lorentz-invariant operators, current limits are mostly in the TeV range or higher.}}\addcontentsline{toc}{subsection}{\lref{56} Are there stricter constraints in particle physics?}

Good point. The constraints reviewed in ref.\ \cite{CoGl2} are on the difference $\De V$ of the maximal attainable velocity of (i) photons and electrons (from photon decay) \cite{CoGl1}, (ii) muons and electrons (from muon decay) \cite{CoSr}, (iii) muon and electron neutrinos (from neutrino oscillations) \cite{Bru86}, (iv) neutral kaons K-long and K-short \cite{HMSa}, (v) photons and atoms \cite{LJHRF}, and (vi) photons and cosmic-ray protons (via vacuum Cherenkov radiation) \cite{CoGl1}. Lorentz-violating effects combined with CPT violation were discussed in ref.\ \cite{KoMe}.

Just like the propagation speed, the maximal attainable velocity of a particle is independent of its species in multifractional theories at the classical level, but the constraints (i)--(v) are calculated in quantum field theory and they are nontrivial also in $T_v$ (only when charged particles are involved, since the theory is nontrivial only in the QED sector \cite{frc13}) and in the deterministic view of $T_q$ and $T_\g$. Even at the classical level, the microscopic stochastic fluctuations of geometry in $T_q$ and $T_\g$ can induce a relative excursion between velocities which cannot exceed the experimental bounds. This mechanism is very different from the Lorentz violation from CPT-even renormalizable rotationally invariant interactions in ordinary spacetime \cite{CoGl2}. 

The bound from Cherenkov radiation (vi) is stronger than the others but it requires energies much greater than those accessible in colliders. To see whether we can use it to constrain multifractional theories, let us first review its origin in ordinary spacetimes with modified dynamics. Primary cosmic rays (i.e., originated outside the Solar System) are made of protons and atomic nuclei; ultra-high-energy cosmic rays (UHECR) carry energies greater that $10^{18}\,\text{eV}$. Cosmic rays with energies above $10^{19}\,\text{eV}$ have been observed systematically \cite{Aab14,Auger}, but isolated events associated with primary protons of energy $E_{\rm UHECR}\approx 1-3\times 10^{20}\,\text{eV}$ have also been detected \cite{Lin63,Bir94,Hay94}. Assume to be in a spacetime where the speed of light $c_x$ is smaller than the usual $c$ (the reason of the symbol $c_x$ will become clear soon). A proton travelling faster than light would rapidly release energy via photon emission, $p\to p+\g$, until its speed drops below luminal. While travelling a distance $V_p t$ with velocity $V_p$, the particle produces a shock wave of photons travelling at speed $c_x$. At time $t$, the electromagnetic wave produced at $t=0$ has traveled a distance $c_x t$ and the angle of the shock wave with respect to the proton trajectory has $|\cos\theta|=(c_x t)/(V_p t)=c_x/V_p\leq1$. The threshold for the production of Cherenkov radiation is thus reached when the particle travels at the same speed of the wave front, $V_p=c_x$. Therefore, restoring $c=1$ units temporarily, from the special-relativistic energy of the proton $E=m_p c^2/\sqrt{1-(V_p/c)^2}$, where $m_p c^2\approx 938.28\,\text{MeV}$ is the proton rest mass, one gets the threshold energy $E_{\rm min}=m_p c^2/\sqrt{1-(c_x/c)^2}$. Since superluminal UHECRs must have become subluminal well before reaching us, their energy must be smaller than the threshold energy, $E_{\rm UHECR}<E_{\rm min}$. Taking $E_{\rm UHECR}\approx 10^{11}\,\text{GeV}$, one gets the bound (back to $c=1$ units) \cite{CoGl1}
\be\label{chere}
1-c_x^2<\left(\frac{m_p}{E_{\rm UHECR}}\right)^2\approx 10^{-22}\,.
\ee

Vacuum Cherenkov radiation can be realized in Lorentz-violating extensions of the Standard Model \cite{LePo,Alt07,AnTa} and we now ask whether it happens also in multifractional theories. For $T_v$, the answer is negative. As in the case of gravitational waves examined in question \lref{55}, the theory $T_v$ is left unscathed because the speed of light is $c_x/c=1$ there, and the bound \Eq{chere} has nothing to say. The case of $T_q$ is more interesting. As pointed out in ref.\ \cite{trtls}, in the fractional frame particles can travel at speed slightly higher than light, and vacuum Cherenkov radiation can occur. We can make a crude estimate of the effect from eq.\ \Eq{chere}. To measure the maximal departure $\De c=c_q-c_x$ of the speed of light $c_x$ in the fractional frame from the standard speed of light $c_q=c=1$ (the geometric velocity of photons in the integer frame), we combine eqs.\ \Eq{tslest} [not \Eq{tslest2}; see the discussion above] and \Eq{chere}, noting that $1-c_x^2=(1+c_x)(1-c_x)\simeq 2\De c$ when $\De c$ is small:
\be\label{chere2}
E_* >\left|\frac{1}{2C_\mu}\left(\frac{m_p}{E_{\rm UHECR}}\right)^2\right|^{-\frac{1}{1-\a_\mu}}E_{\rm UHECR}\,.
\ee
If we use the binomial measure \Eq{meamu} with $\a$-dependent coefficients, then $\De c\propto -C_\mu>0$ in a spacelike fractal geometry and $C_\mu=C_i=-(3-\a)/[3^{(1-\a)/2}\a]$. For this choice, one finds the absolute and $\a=1/2$ bounds of tables \ref{tab6} and \ref{tab7}, respectively. For a generic $2C_\mu=1$, one gets
\be\label{chere3}
E_* >10^{11+\frac{22}{1-\a_\mu}}\,{\rm GeV}\,,
\ee
and the weaker bounds reported in table \ref{tab8}. Comparing eq.\ \Eq{chere3} with \Eq{estilas}, we see two factors that improve the GRB bound. One is in the velocity difference \Eq{chere}, which is 4 orders of magnitude smaller than in the GRB case. The other, and most important, is the reference energy $E_{\rm UHECR}$, 15 orders of magnitude larger than that of typical GRB photons. It is no wonder that the values reported in the ``Cherenkov radiation'' line of tables \ref{tab6}--\ref{tab8} are much tighter than those from GRBs. 

\bigskip

\llabel{57} \textbf{\emph{Is the dispersion relation \Eq{dire}, which is claimed to affect the propagation of gravitons, photons or other particles, physical? It was derived from the propagator \Eq{prop2}, which has the conventional form in terms of the $p$'s. However, any dispersion relation in which one mixes momentum components in two or more coordinates, or where one calls ``$p(k)$'' momentum $p$, will take an unconventional form without having unconventional physics.}}\addcontentsline{toc}{subsection}{\lref{57} Is the theory with \texorpdfstring{$q$}{}-derivatives trivial? (v)}

This is questions \lref{15}, \lref{16}, and \lref{22}--\lref{25} disguised in another form. Once we choose the time and length units of our devices as the scaling units of the fractional coordinates $x^\mu$ in position space, we also automatically fix the momentum and energy units as the scaling units of the fractional coordinates $k^\mu$ in momentum space:
\be
[k^\mu]=1\,.
\ee
In the case of the theory with $q$-derivatives, the measure \Eq{rdp} in momentum space is fixed uniquely by eq.\ \Eq{mompk}, so that the momentum-space analogue of eq.\ \Eq{xq},
\be\label{kp}
k^\mu\to p^\mu(k^\mu)\,,
\ee
is not a change of coordinates but a mapping from the fractional frame where observables are computed and the integer frame where the theory looks simpler. In particular, the propagator \Eq{prop2} is a highly nontrivial and rigid function of $k^\mu$, even if it has the usual form in terms of $p=p(k)$. All these properties are determined by the symmetries of the theory. We can obtain any dispersion relation without unconventional physics only in a theory admitting eq.\ \Eq{kp} as a coordinate transformation leaving physical observables invariant. This is not the case of $T_q$, as we discussed at length in section \ref{frafi}.

\bigskip

\llabel{58} \textbf{\emph{Even if the replacements $x\to q(x)$ and $k\to p(k)$ were somehow physical, they are not done at the required level of rigor. In particular, one would need to follow a first-principle approach where one starts with a field action and performs the well-known procedure to get the Hamiltonian density. Until such a rigorous analysis is done, it is not justified to assume that the symbols that are used such as $p$, $k$, $E$, and so on, have the meaning of momentum and energy.}}\addcontentsline{toc}{subsection}{\lref{58} Is the phenomenology of the theory with \texorpdfstring{$q$}{}-derivatives robust? (i)}

A first-principle approach is followed. Quantum-gravity motivations aside, we have a spacetime measure dictated by the second flow-equation theorem (question \loref{04}) and a momentum space measure determined by that automatically (question \loref{07b}). We have a field action, both for the Standard Model \cite{frc12,frc13} and for gravity \cite{frc11} (questions \lref{29} and \lref{37}; the general structure of field actions in $T_q$ are discussed in \lref{11} and \lref{22}). The Hamiltonian analysis could not be easier than in $T_q$: it follows all the steps of the standard case with the replacements $x\to q(x)$ and $k\to p(k)$, and it is not necessary to repeat it here in detail.\footnote{In \cite{CaRo1}, the algebra of first-class constraints of gravity plus matter for the theory with $q$-derivatives has been written down (question \lref{46}). Instances of Hamiltonian analyses of $T_1$, $T_v$, and other multiscale theories can be found in refs.\ \cite{fra2,frc5,frc6,frc10,CaRo1}. Equation \Eq{hpo} is an example in $T_v$.} The example of a classical real scalar field in flat space will suffice. From the action \Eq{qS}, one obtains the momentum $\Pi_\phi=\p\cL/\p q^0(t)=\p_{q(t)}\phi$, the super-Hamiltonian density $\cH$, and the supermomentum density $\cH_i$:
\be
\cH=\Pi_\phi \p_{q(t)}\phi-\cL=\frac12 \Pi_\phi^2+\frac12\sum_{i=1}^{D-1}[\p_{q^i(x^i)}\phi]^2+V(\phi)\,,\qquad \cH_i=\Pi_\phi\p_{q^i(x^i)}\phi\,.
\ee
The Hamiltonian is $H=\int\rmd^{D-1}q({\bf x})\,\cH$, where one integrates only on spatial coordinates. For $V(\phi)=m^2\phi^2/2$ and using the Fourier transform
\be
\phi(x)=\int\frac{\rmd^Dp(k)}{(2\pi)^D}\,\rme^{\rmi \eta_{\mu\nu}p^\mu(k^\mu) q^\nu(x^\nu)}\phi_k\,,
\ee
it is not difficult to quantize canonically and to identify $H$ as the charge conserved under fractional time translations. At the classical level, $p^0(E)$ is the geometric energy in the integer picture and, hence, $k^0=E$ is the energy in the fractional picture. All of this stems from the fact that $p^\mu(k^\mu)$ is Fourier conjugate to $q^\mu(x^\mu)$.

\bigskip

\llabel{59} \textbf{\emph{Granting that a given action describes this framework, it is a fact that there would not be two types of momenta $p$ and $k$ for the same field (for instance, gravity), as they appear in the modified dispersion relation \Eq{dire}. Therefore, at the level in which the theory currently stands, it is impossible to claim that one can make contact with experiments and observations.}}\addcontentsline{toc}{subsection}{\lref{59} Is the phenomenology of the theory with \texorpdfstring{$q$}{}-derivatives robust? (ii)}

As said in \lref{57}, there is only one momentum for a field, which is $k^\mu$. The geometric momentum $p^\mu=p^\mu(k^\mu)$ is only a convenient tool to cast the theory $T_q$ in the integer picture.

\bigskip

\llabel{60} \textbf{\emph{Experimental constraints of multifractional models are typically based on equations which show an extreme sensitivity to the value chosen for the parameters $\a_0$ or $\a$. Does this indicate that the domain of validity of these formul\ae\ is limited and that a more refined analysis is required?}}\addcontentsline{toc}{subsection}{\lref{60} Does the extreme sensitivity to the value of \texorpdfstring{$\a_\mu$}{} of the formul\ae\ used for experimental constraints indicate that their domain of validity is limited and that a more refined analysis is required?}

Tables \ref{tab4}--\ref{tab8} show that the bounds on the scales of the binomial measure can change by a few orders of magnitude when varying the fractional exponents $\a_\mu$ in the range $[0,1)$; the results for values close to zero and for $\a_\mu=1/2$ are compared. This sensitivity on a fundamental parameter of the theory with a clear-cut geometric interpretation should not be regarded as a drawback. In fact, this feature is an invaluable bonus: it guarantees that these theories can be easily falsified. Already the estimates from GRBs are an example of this: they exclude the values $\geq 1/2$ for $T_q$ in the absence of log oscillations and they limit the parameter space of this theory in an unprecedented way, the characteristic energy of the momentum measure being pushed very close to grand-unification and Planck scales. 

A key difference with respect to other quantum-gravity-inspired dispersion-relation bounds \cite{EMNan,ArCa2} is that our constraints are obtained directly from a full theory, without invoking any generic assumption encoding uncontrolled effects in heuristic umbrella constants. We do have free parameters but they are fundamental, intrinsic to the theory. In this respect, our approach is less qualitative, more rigid and, therefore, more sensitive to the strength of the observational constraints \cite{qGW}.

\bigskip

\llabel{61} \textbf{\emph{Are there constraints from tests of the equivalence principle?}}\addcontentsline{toc}{subsection}{\lref{61} Are there constraints from tests of the equivalence principle?}

Not yet, but it is an interesting question.

\bigskip

\llabel{62} \textbf{\emph{Are there constraints on the dimension of spacetime?}}\addcontentsline{toc}{subsection}{\lref{62} Are there constraints on the dimension of spacetime?}

Yes, there are for the theory with $q$-derivatives. A likelihood analysis of the primordial CMB scalar spectrum excludes portions in the parameter space of $T_q$, due to the fact that CMB data disfavor the logarithmic oscillations of the spectrum \Eq{Pmusc}. The marginalized likelihood for the spatial fractional exponent $\a$, when $N$ in eq.\ \Eq{omN} is fixed, indicates that $\a\lesssim 10^{-1},10^{-0.2},10^{-0.25}$ at the 95\% confidence level for, respectively, $N=2,3,4$. From eqs.\ \Eq{dhuv2} and \Eq{dsuv},
\be\label{dhst}
\begin{matrix}
N=2:\qquad & \ds^{\rm \,space}=\dh^{\rm \,space}\lesssim 0.3\qquad\text{(UV)}\,,\\
N=3:\qquad & \ds^{\rm \,space}=\dh^{\rm \,space}\lesssim 1.9\qquad\text{(UV)}\,,\\
N=4:\qquad & \ds^{\rm \,space}=\dh^{\rm \,space}\lesssim 1.7\qquad\text{(UV)}\,.
\end{matrix}
\ee
Higher $N$ should give similar constraints. This result is somewhat surprising, as it forces an \emph{upper} bound on the dimension of space in the UV. Therefore, the primordial universe is very well described by the standard inflationary model in a smooth spacetime with four topological dimensions but, as soon as one assumes that spacetime geometry undergoes dimensional flow, this flow must be nontrivial to fit data. During this flow, the effective dimension of space is reduced at least by 1 ($N=3$ case) in the UV.

There are no analogous results for the other multifractional theories, although it is possible that eq.\ \Eq{dhst} could apply also to the case with fractional derivatives thanks to the $T_{\g=\a}\cong T_q$ approximation.

\bigskip

\llabel{63} \textbf{\emph{If the length scales of these theories are so small, how is it possible to test them at cosmological scales? Modifications to gravity are strongly suppressed during inflation. The reason is that the ratio between the inflationary energy density and Planck density (at which classical gravity is believed to break down) is very small, $\rho_{\rm infl}/\rho_{\rm Pl}\sim (\ell_\Pl H)^2 \sim 10^{-8}$, where we estimated the typical energy scale during inflation to be about the grand-unification scale, $H\sim 10^{15}\,\mbox{GeV}$. Thus, quantum corrections or corrections from exotic geometries are expected to be well below any reasonable experimental sensitivity threshold.}}\addcontentsline{toc}{subsection}{\lref{63} If the length scales of these theories are so small, how is it possible to test them at cosmological scales?}

This type of argument holds only when corrections to general relativity are limited to higher-order curvature corrections. As is known in quantum gravity (and, in particular, in string cosmology and in loop quantum cosmology), the effective dynamics of gravity in the early universe can be modified by far more sophisticated mechanisms than curvature corrections to the Einstein--Hilbert action. 

The case of multifractional spacetimes illustrates the point in a rather unique fashion. By definition of these theories, geometry is characterized by a hierarchy of fundamental scales. The main features of this configuration are exemplified to the bone by the binomial measure \Eq{meamu} with \Eq{binom2}. Here, we have two characteristic length scales $\ell_\infty\leq\ell_*$. At scales above $\ell_*$, spacetime looks smooth and the usual description of general relativity holds. However, when inspected at scales $\lesssim \ell_*$, in the deterministic view geometry changes properties smoothly and, in particular, the spacetime dimension decreases to some asymptotic value smaller than 4. If one further zooms in, at scales $\sim \ell_\infty$ a discrete symmetry emerges and the notion of smooth spacetime with well-defined dimensionality is lost. The length $\ell_\infty$ can be identified with the Planck length (see question \loref{07b}), while $\ell_*$ is constrained to be at least as small as the grand unification scale (table \ref{tab8}). Therefore, it might seem difficult that multifractional geometries could leave an observable imprint anywhere. However, primordial inflation expands Planckian scales to cosmological size. If geometry is modified at Planck scales, then we can expect that multiscale effects are magnified by the early-universe expansion up to the size of the visible sky. Such is indeed the case and CMB observations are capable of placing strong constraints on multifractional geometries \cite{frc14}. 

This cosmological mechanism is in action in most models of quantum gravity, but in the case of multifractional spacetimes there is also a subtler effect. Log oscillations are a manifestation of discrete UV symmetries and of the \emph{long-range} correlations typical of complex systems, anomalous stochastic processes (see, e.g., ref.\ \cite{Kel10} for a pedagogical review), and multifractals (via the so-called harmonic structure, reviewed in refs.\ \cite{frc1,frc2}). This long-range effect is clearly visible both in theoretical cosmology (where the oscillatory modulation of the scale factor dies out at scales much larger than $\ell_\infty$ and larger even than $\ell_*$ \cite{frc11}) and in observations, as we just remarked (see also \lref{62}). It is a most unusual phenomenon from the point of view of standard QFT, because it entails a symmetry (discrete scale invariance) that, despite being explicitly broken already near the UV, propagates to the IR and governs the physics at large scales.

\bigskip

\llabel{64} \textbf{\emph{How would the discrete spacetime at scales $\sim\ell_\infty$ look like to an observer?}}\addcontentsline{toc}{subsection}{\lref{64} How would the discrete spacetime at scales \texorpdfstring{$\sim\ell_\infty$}{} look like to an observer?}

If spacetime is discrete at scales $\sim \ell_\infty$, then we could picture it as a totally disconnected set of points. How would an observer therein perceive this geometry? Certainly not as ``holes'' in the fabric of spacetime, since signals propagate only within the set; the holes picture would best suit an ideal observer living outside our universe, in the $D$-dimensional embedding space where the theory is defined. At the cosmological level, the visible effect of this spacetime geometry is a long-range logarithmic modulation of the power spectrum of primordial fluctuations \cite{frc14} as we discussed in the previous question. At the microscopic level, the stochastic view advanced here and in refs.\ \cite{CaRo2a,CaRo2b} predicts a fuzziness where measurements cannot be performed with arbitrary precision, and that get worse when trying to probe scales deeper in the UV.

\bigskip

\llabel{65} \textbf{\emph{Are multifractional theories ruled out?}}\addcontentsline{toc}{subsection}{\lref{65} Are multifractional theories ruled out?}

A multifractional theory is ruled out observationally if the length scale $\ell_*$ in the binomial measure is much smaller than the Planck scale, $\ell_*\ll \lp$. In momentum space, this corresponds to $E_*\gg\ep$.
\begin{itemize}
\item The phenomenology of $T_1$ has not been studied and we cannot say much about it. The spectral dimension of $T_1$ is the same as $T_v$ because the diffusion equation in these theories is one the adjoint of the other \cite{frc7}. Therefore, the observable consequences of their dimensional flow should be about the same. This is a non-issue, since $T_1$ was useful as a first exploration of the multifractional paradigm but nowadays it has been replaced by the more rigorous $T_v$.
\item The most conservative bounds on $T_v$ (table \ref{tab4}) are very weak because the theory bypasses all the strongest tests. The $\a_\mu=1/2$ case is better constrained and measurements of the fine-structure constant require $E_*>10^{-8}\,\ep$.
\item Until now, the strongest bound on $T_q$ came from a crude estimate of the arrival time of photons with different energies emitted by GRBs \cite{qGW}. For $\a_\mu\ll 1/2$, this bound is $E_*>10^{-5}\,\ep$, while for $\a_\mu=1/2$ the theory is ruled out, since $E_*>10^{13}\,\ep$ (table \ref{tab8}). Inclusion of log oscillation could lead to an accidental erasure of corrections to dispersion relations, but not without fine tuning \cite{qGW}. The only chance to avoid the GRB bound would be to disprove the estimate reported in \lref{55} by a precise calculation. However, the constraints from emission of Cherenkov radiation by cosmic rays, which are several orders of magnitude stronger, rely only on the multifractional modification of special relativity, and they look much harder to evade. All these constraints are valid in the deterministic view and could be avoided by invoking the stochastic view and considering the possibility that stochastic fluctuations cancel out when integrated along the photon or cosmic-ray paths \cite{CaRo2a,CaRo2b}. In fact, eq.\ \Eq{tslest2} assumed that the two particles for which one is measuring the velocity difference experience maximal and opposite fluctuations. On the other hand, in average the effect could be just zero and all constraints from gravitational waves, GRBs, and UHECRs would evaporate.
\item Like in the case of $T_1$, we do not have direct calculations of physical observables in $T_\g$ and we conjectured that the $T_{\g=\a}\cong T_q$ approximation allows one to apply the constraints found for $T_q$ to $T_\g$. In that case, what said for $T_q$ would hold also here: the stochastic view bypasses all the strongest tests (gravitational waves, GRBs, and UHECRs) only if stochastic fluctuations are averaged out, while the deterministic view is constrained much more severely. There are three possible ways in which the theory $T_\g$ can be rescued: (i) giving up the deterministic view and adopting only the stochastic view, which is more justified here than in $T_q$ (where it is a juxtaposed approximation when meant in the sense of \cite{trtls}); (ii) finding that, despite their similarities, $T_q$ and $T_\g$ are essentially different in some key physical consequences and that some bounds do not apply after a closer scrutiny; (iii) finding that the fractional derivatives in $T_\g$ must or can be taken with an order $\g$ smaller than the fractional exponent $\a$ in the measure. Case (ii) is particularly interesting. As discussed in \loref{07b}, the value $\a_\mu=1/2$ is special according to some rigorous arguments advanced for $T_\g$, which is closely similar to $T_q$ when $\g=\a$. However, these two theories are mathematically different and arguments rigorously valid for $T_\g$ can be taken only as suggestions in $T_q$, and we do not expect that any theoretical argument in the future will fix $\a_\mu$ uniquely for $T_q$ (or $T_v$). Vice versa, the $\a$-dependent observational constraints obtained for $T_q$ are robust for that theory, but only indicative for the yet-unexplored case of $T_\g$. In particular, we cannot conclude that GRBs rule out $T_\g$ just because they rule out $T_q$ for the range $\a_\mu\geq 1/2$ for which $T_\g$ is normed. However, the UHECR bound of $T_q$ is strong for all $0\leq \a_\mu<1$ and it could be avoided in $T_\g$ only with a radical departure from $T_q$ in special relativity. Again, the explicit construction of $T_\g$ and calculations of its predictions will settle the question.
\end{itemize}

%%%%%%%%%%%%%%%%%%%%%%%%%%%%%%%%%%%%%%%%%%%%%%%%%%%%%%%%%%%%%%%%%%%%%%%%%%%%%
%%%%%%%%%%%%%%%%%%%%%%%%%%%%%%%%%%%%%%%%%%%%%%%%%%%%%%%%%%%%%%%%%%%%%%%%%%%%%

\section{Perspective}\label{concl}

\llabel{66} \textbf{\emph{In a nutshell, what are the main virtues of multifractional theories?}}\addcontentsline{toc}{subsection}{\lref{66} In a nutshell, what are the main virtues of multifractional theories?}

\begin{itemize}
\item[--] \emph{They are a novel paradigm} because, contrary to many other effective models, what one modifies here is not the dynamics but the integrodifferential structure describing how we measure the geometry. Dynamics is modified as a byproduct of having a spacetime that can be conveniently treated with multidisciplinary tools of fractal geometry, anomalous transport theory, and complex systems. This framework is different from much of the mainstream in theoretical physics, quantum gravity, and cosmology, and some researchers find this intellectually stimulating.
\item[--] \emph{They are simple} without being simplistic. It is the first attempt to control the most generic profile of dimensional flow in a purely analytic way. All the usual techniques employed in quantum field theory and classical gravity can be adapted, with caution. This allowed us to extract the first serious experimental constraints \cite{frc12} not long since the original proposal \cite{fra4}.
\item[--] \emph{Their phenomenology is rich} and spreads across all scales, from elementary particle interactions to cosmology. It is also rigid enough to allow to exclude large portions of the parameter space.
\item[--] \emph{They have much potential} yet untapped, especially regarding observational constraints and major open issues in cosmology and quantum gravity (see question \lref{69}).
\end{itemize}

\bigskip

\llabel{67} \textbf{\emph{And their problems?}}\addcontentsline{toc}{subsection}{\lref{67} And their problems?}

\begin{itemize}
\item[--] The novelty of the paradigm carries some difficulties such as the breaking of symmetries (but the emergence of others \dots) in the UV and the consequent need to choose a frame in position space. This is unattractive for someone accustomed to work in Lorentz-invariant theories, not only because Lorentz invariance is a powerful theoretical asset making life simpler, but also because preferred frames are usually more difficult to justify scientifically and epistemologically, and can be much trickier when it comes to extract physical observables. These are not the first models of gravity and matter breaking Lorentz invariance, and surely they will not be the last; however, their foundations are so different with respect to other, more conventional proposals that it is natural to find resistance. Many of the questions collected here were actually raised during interactions between the author and colleagues. Some of these questions had already been answered in the literature at the moment of their formulation, while others triggered more thinking. One of the goals of this work was to gather all these issues in one basket and address them in a unified systematic way.
\item[--] The most interesting among the proposals, the theory with multifractional derivatives, has not been developed much. A top priority will be to make it progress.
\end{itemize}

\bigskip

\llabel{68} \textbf{\emph{What is the agenda for the future?}}\addcontentsline{toc}{subsection}{\lref{68} What is the agenda for the future?}

The recent proposal of the stochastic view \cite{CaRo2a,CaRo2b} confirms that dimensional flow is a solid manifestation of quantum gravity, while the original motivation of the multifractional paradigm was to quantize gravity successfully precisely because of dimensional flow \cite{fra1}. This is the usual dualism of multifractional spacetimes viewed as effective models or as fundamental theories. Both possibilities are viable and can be pursued in parallel and in several distinct ways. In order of importance:
\begin{enumerate}
\item To complete the formulation of the theory with multifractional derivatives, starting from a coherent and useful definition of multiscale fractional calculus (question \lref{11}), the construction of perturbative QFT thereon, the study of its renormalizability, and the study of its cosmology. Quantizing gravity consistently will be one of the main goals.
\item To verify the viability of the $T_{\g=\a}\cong T_q$ approximation explicitly by extracting experimental constraints directly from $T_\g$. If these bounds turned out to be close to those obtained in $T_q$, then the $T_{\g=\a}\cong T_q$ approximation would be confirmed and one could use the simple $T_q$ setting to explore more features of $T_\g$ in advance. If, on the contrary, the direct bounds on $T_\g$ departed from those on $T_q$, we would have to treat these theories separately.
\item To study the late-time cosmology of all theories, in order to check whether we can explain late-time acceleration with multiscale geometry (question \lref{41}).
\item To check whether the $\a=0$ configuration can help to address the big-bang problem (question \lref{42}).
\item To study the role of complex dimensions and degenerate geometries, their theoretical viability, and their physical consequences (question \lref{14}).
\item To investigate the relation between the near-boundary regime of multifractional spacetimes and phase B of CDT \cite{frc2} (question \lref{45}).
\item The multiscale model $\tilde T_1$ of refs.\ \cite{fra1,fra2,fra3} has a Lorentz-invariant measure $\rmd^Dx\,v(s)$ but its Laplace--Beltrami operator is not self-adjoint. For this reason, it was abandoned in favor of the multifractional paradigm \cite{fra4}. However, $\tilde T_1$ is an example of geometry obeying the first flow-equation theorem that could be used for phenomenology, without the ambition of defining a theory with a rigorous top-down construction.
\end{enumerate}
The status of each theory, together with the discontinued model $T_1$, is summarized in table \ref{tab9}.
\begin{table}[ht!]
\centering
\begin{adjustwidth}{-1.7cm}{}
\footnotesize
\begin{tabular}{|l|cccc|}\hline
                           & $T_1$	               				& $T_v$       			& $T_q$ 		          & $T_\g$  												 \\\hline
Calculus									 & {\ding{51}}           				& {\ding{51}} 			& {\ding{51}}        & {\ding{51}}? 										 \\
													 & \cite{fra2,fra3,frc2} 				& \cite{frc2} 			& \cite{frc2}        & \cite{frc1,frc2,frc4}, this paper \\\hline
Momentum transform         & {\ding{51}}  				 				& {\ding{51}} 			& {\ding{51}}        & {\ding{51}}?                      \\
													 & derivable from ref.\ \cite{fra2}	& \cite{frc3} 			& \cite{frc11,frc14} &                                   \\\hline
Self-adjoint Laplacian 	   & {\ding{55}}									& {\ding{51}}       & {\ding{51}}				 & {\ding{51}}											 \\
and momentum operator      & \cite{frc3,frc5}							& \cite{frc3}	      & \cite{frc11}			 & \cite{frc2,frc4}, this paper			 \\\hline
Spectral dimension         & {\ding{51}}           				& {\ding{51}} 			& {\ding{51}}        & {\ding{51}}?											 \\
													 & \cite{frc7}           				& \cite{frc7} 			& \cite{frc7}        & \cite{frc4}											 \\\hline%\hline
Classical mechanics        & {\ding{51}}  				 				& {\ding{51}} 			& {\ding{51}} 			 &  																 \\
													 & 							 								& \cite{frc5} 			& \cite{frc10} 			 &   																 \\\hline
Quantum mechanics          & {\ding{55}}  				 				& {\ding{51}} 			& {\ding{51}} 			 &  																 \\
													 & implicit in ref.\ \cite{frc5} & \cite{frc5} 			&  						 			 &   																 \\\hline%\hline
Scalar field theory        & {\ding{51}}           				& {\ding{51}} 			& {\ding{51}}        & {\ding{51}}? 										 \\
													 & \cite{fra2,fra3}     				& \cite{frc6} 			& \cite{frc11}			 & \cite{frc2}, this paper					 \\\hline
Standard Model						 & {\ding{51}} 				 				  & {\ding{51}} 			& {\ding{51}} 			 &      														 \\
													 & this paper		 				 				& \cite{frc8,frc13} & \cite{frc12,frc13} &      														 \\\hline
Power-counting 				     & \ding{51}             				& {\ding{51}}       & {\ding{55}}        & depends on norm                   \\
renormalizability					 & \cite{fra1,fra2}, this paper & this paper        & this paper         & this paper						             \\\hline
Perturbative   				     & ---           								& {\ding{55}}       & \multicolumn{2}{c|}{depends on parameters and on view?}\\
renormalizability					 & 									      & \cite{frc9}, this paper & this paper         & this paper						             \\\hline%\hline
Gravity and cosmological   & {\ding{51}}                  & {\ding{51}}       & {\ding{51}}        &  																 \\
equations									 & \cite{frc11}                 & \cite{frc11}      & \cite{frc11}       &  																 \\\hline
Early-universe dynamics    & ---                          & {\ding{51}}?      & {\ding{51}}        &  																 \\
													 &                              & \cite{frc11}      & \cite{frc11,frc14} &  																 \\\hline
Dark energy						     &                              & 						      & 						       &  																 \\
													 &                              & 						      & 									 &  																 \\\hline%\hline
Atomic and elementary      & ---  										    & {\ding{51}}       & {\ding{51}}        & 																	 \\
particle constraints			 & 													    & \cite{frc13}      & \cite{frc12,frc13} & 																	 \\\hline
Astrophysical constraints  & ---													& {\ding{55}}       & {\ding{51}}        & 																	 \\
													 & 												 & \cite{qGW}, this paper & \cite{qGW}, this paper & 															 \\\hline
Cosmological constraints   & ---													& {\ding{51}}       & {\ding{51}}        & 																	 \\
													 & 												      & \cite{frc14}      & \cite{frc14}       &     															 \\\hline
Ruled out?							   & ---													& No				        & Yes (in deterministic view) & 												 \\
													 & 											 & \cite{frc13}, this paper & \cite{qGW}, this paper & 															 \\\hline
\end{tabular}
\end{adjustwidth}
\caption{\label{tab9}Status of the multifractional model $T_1$ and of the three multifractional theories $T_{v,q,\g}$. Empty cells correspond to topics not studied yet. Items with a question mark ``?'' indicate partial results. If an nonempty item (with or without question mark) has no references given, the result is either obvious (no question mark) or easily doable (with question mark). The items ``---'' for $T_1$ are not of interest for the future since $T_1$ is a toy model replaced by $T_v$; however, one could still do some cosmological phenomenology with it.}
\end{table}

\bigskip

\llabel{69} \textbf{\emph{To conclude with a motivational appeal, why would I want to work on multifractional theories?}}\addcontentsline{toc}{subsection}{\lref{69} Why would I want to work on multifractional theories?}

Because they are based on a guiding principle whose implementation is gradually improving in rigorousness, their UV geometry is extremely interesting and affects all sectors in physics, they yield characteristic phenomenological predictions, they are rigid enough to be easily falsifiable by experiments, and they may contribute to the big-bang, the cosmological constant, and the quantum-gravity problems.

The guiding principle is the second flow-equation theorem \cite{first}, supported by multifractal geometry \cite{fra4,frc1,frc2} and motivated by quantum gravity (section \ref{moti}). In the UV, logarithmic oscillations can give rise to some esoteric form of propagation of quantum degrees of freedom (questions \lref{34}, \lref{47}, and \lref{63}) or else melt away in a stochastic structure not allowing for precise measurements (questions \lref{27} and \lref{47}). The effects of the multiscale geometry of these scenarios is not confined to the UV limit of gravity. On one hand, it propagates to large scales via the long-range modulation of log oscillations \cite{frc11,frc14}, which are a manifestation of microscopic discrete scale invariance \cite{frc3,frc2} (question \lref{63}). On the other hand, the nontrivial integrodifferential structure of multifractional theories modifies not just the gravitational sector [sections \ref{cosmo} and \ref{qg}; compare, in contrast, changes of the dynamics as in, say, $f(R)$ gravity] but also the Standard Model of particles (section \ref{qft}), thus opening up the possibility to constrain the theories with a great variety of experiments (section \ref{phen}). The wealth of bounds that have been obtained from atomic and particle physics, astrophysics, and cosmology are sensitive to the free parameters of the measure, in particular to the fractional exponents determining the dimension of spacetime. This property, together with the rigid theoretical structure of each proposal (especially $T_q$ and $T_\g$, the models with more symmetries), make multifractional scenarios easily falsifiable. Much still needs to be done in order to get control over $T_\g$ and the new developments on the stochastic view, but it can be done in a very reasonable time span. 

Finally, throughout this review-plus-plus we stumbled across many unsolved problems of modern theoretical physics, including the resolution of singularities such as the big bang or in black holes (question \lref{42}), the cosmological constant problem or the nature of dark energy (questions \lref{39} and \lref{41}), the nature of inflation (questions \lref{39} and \lref{40}), and the problem of quantum gravity (questions \loref{04}, \loref{06}, and \lref{47}). We cannot and do not claim that multifractional theories have the final answer to any of these topics, but they are contributing to the debate in an alternative way and there is a lot of potential to be uncovered from preliminary results.

We hope to report on, or to see news about, all this in the near future and, as paradoxical as it may sound, to come back with more frequently asked questions than now.

%%%%%%%%%%%%%%%%%%%%%%%%%%%%%%%%%%%%%%%%%%%%%%%%%%%%%%%%%%%%%%%%%%%%%%%%%%%%%
%%%%%%%%%%%%%%%%%%%%%%%%%%%%%%%%%%%%%%%%%%%%%%%%%%%%%%%%%%%%%%%%%%%%%%%%%%%%%

\section*{Acknowledgments}

The author is under a Ram\'on y Cajal contract and is supported by the I+D grant FIS2014-54800-C2-2-P. He thanks L.\ Modesto for the discussion that led to eq.\ \Eq{quas}, and M.\ Arzano, D.\ Oriti, and M.\ Ronco for other discussions.

\medskip

\noindent {\bf Open Access.} This article is distributed under the terms of the Creative Commons Attribution License (\href{https://creativecommons.org/licenses/by/4.0/}{CC-BY 4.0}), which permits any use, distribution and reproduction in any medium, provided the original author(s) and source are credited.

\end{document}